# Diversity-Oriented Synthesis of Polymers of Intrinsic Microporosity with Explicit Solid Solvation Cages for Lithium Ions

*Miranda J. Baran[1,2], Mark E. Carrington[3], Swagat Sahu[1], Artem Baskin[1], Junhua Song[3], Michael A. Baird[2], Simon J. Teat[4], Stephen M. Meckler[2], Chengyin Fu[3], David Prendergast[1,3], Brett A. Helms[1,3,5*]*

[1] The Joint Center for Energy Storage Research, 1 Cyclotron Road, Berkeley, CA 94720, USA

[2] Department of Chemistry, University of California, Berkeley, CA 94720, USA

[3] The Molecular Foundry, Lawrence Berkeley National Laboratory, 1 Cyclotron Road, Berkeley, CA 94720, USA

[4] Advanced Light Source, Lawrence Berkeley National Laboratory, 1 Cyclotron Road, Berkeley, CA 94720, USA

[5] Materials Sciences Division, Lawrence Berkeley National Laboratory, 1 Cyclotron Road, Berkeley, CA 94720, USA

Correspondence to: bahelms@lbl.gov

**Microporous polymers feature shape-persistent free volume elements (FVEs), which are permeated by small molecules and ions when used as membranes for chemical separations, water purification, fuel cells, and batteries.[1–3] It remains a significant challenge to identify FVEs with analyte specificity, due to difficulties in generating microporous polymer libraries with sufficient diversity for screening their properties. Here, we describe a diversity-oriented synthetic (DOS) strategy for microporous polymer membranes from which we identified those whose FVEs serve as solid solvation cages for lithium ions ($Li^+$). Key elements of our strategy included diversification of bis(catechol)-type monomers via multi-component Mannich reactions to introduce $Li^+$-coordinating functionality within individual FVEs, topology-enforcing polymerizations for generating macromolecular skeletal diversity for networking FVEs into different pore architectures, and several classes of on-polymer reactions for diversifying pore geometries and dielectric properties. Lead candidate membranes featuring such ion solvation cages exhibited both higher ionic conductivity and higher cation transference number than control membranes where FVEs were aspecific, which indicates conventional bounds for membrane permeability and selectivity for ion transport can be overcome.[4] Specifically, membranes whose solvation cages enhance $Li^+$ partitioning from the bulk electrolyte also benefit from having FVEs networked such that the number of solvent molecules bound to $Li^+$ is reduced, which increases $Li^+$ diffusivity in the membrane's pores. Such membranes showed promise as dendrite-suppressing anode–electrolyte interlayers in high-voltage lithium-metal batteries for electric mobility.**

Diversity-oriented synthesis introduces scaffold diversity in libraries of small molecules to expedite the identification of those that bind to a target host.[5–11] It follows, inversely, that the identification of a suitable host for a specific molecular species could be expedited were there synthetic tools available to architect negative space, e.g., in a microporous material, where pore dimensions need only be marginally larger than the encapsulated guest.[12] In principle, this could be carried out with regard to the topology of the microporous material's free volume, the types of chemical functionality displayed within it, and other diversity considerations. While reticular chemistry provides opportunities for library building with discrete microporous solids (e.g., particles consisting of microporous organic cages, metal–organic polyhedra, covalent organic frameworks, and metal–organic frameworks),[13–18] lacking is an analogous approach directed at the structure and topology of FVEs in amorphous microporous polymers and films thereof. As a result, it remains a formidable challenge to identify microporous polymer membranes with properly configured FVEs, whose analyte-specific interactions and network architecture might allow the membrane's transport properties to be tailored for a specific application.

Here, we report a diversity-oriented macromolecular synthesis by which FVEs in polymers of intrinsic microporosity (PIMs) are elaborated upon broadly to yield a library of polymers with highly differentiated architectural characteristics, which can be screened to identify those that allow them serve as hosts for specific analytes (**Fig. 1**). As a proof-of-concept, we designed and screened this library for PIMs with $Li^+$-coordinating and transporting properties, which would render them suitable for use in electrochemical devices where $Li^+$ is the working ion. PIMs whose FVEs were configured with explicit $Li^+$-coordinating motifs displayed in pores with specific topologies and length scales exhibited both higher ionic conductivities and higher $Li^+$ transference numbers than PIMs without such functionality, which is unusual given conventional performance trade-offs for analyte permeability and selectivity in polymer membranes.[12] We also show that PIMs featuring solid solvation cages for $Li^+$ are dendrite-suppressing lithium-ion conductors when implemented as an anode–electrolyte interlayer in high-voltage lithium-metal batteries.

To generate skeletal diversity, catechol serves as a common starting material from which known reactions yield spiro- or bicyclic bis(catechol)-type compounds **1** and **2**. Advancing further, in a forward synthetic manner (**Fig. 1a**), we found that spirobisindane **1** and dihyrdoethanoanthracene **2** are versatile substrates for functional group diversification via multi-component Mannich reactions[19,20] with formaldehyde and various secondary amines, e.g., yielding monomers **3**–**10** for polymerization with varying $Li^+$-coordinating functionality. While there are potentially four sites of aminomethylation for each skeletal class, we found that for each, generally only two sites were modified; aminomethylation proceeded in a regiospecific but not stereospecific manner, which was confirmed by single crystal X-ray analyses (**Supplementary Figs. 1–7**).

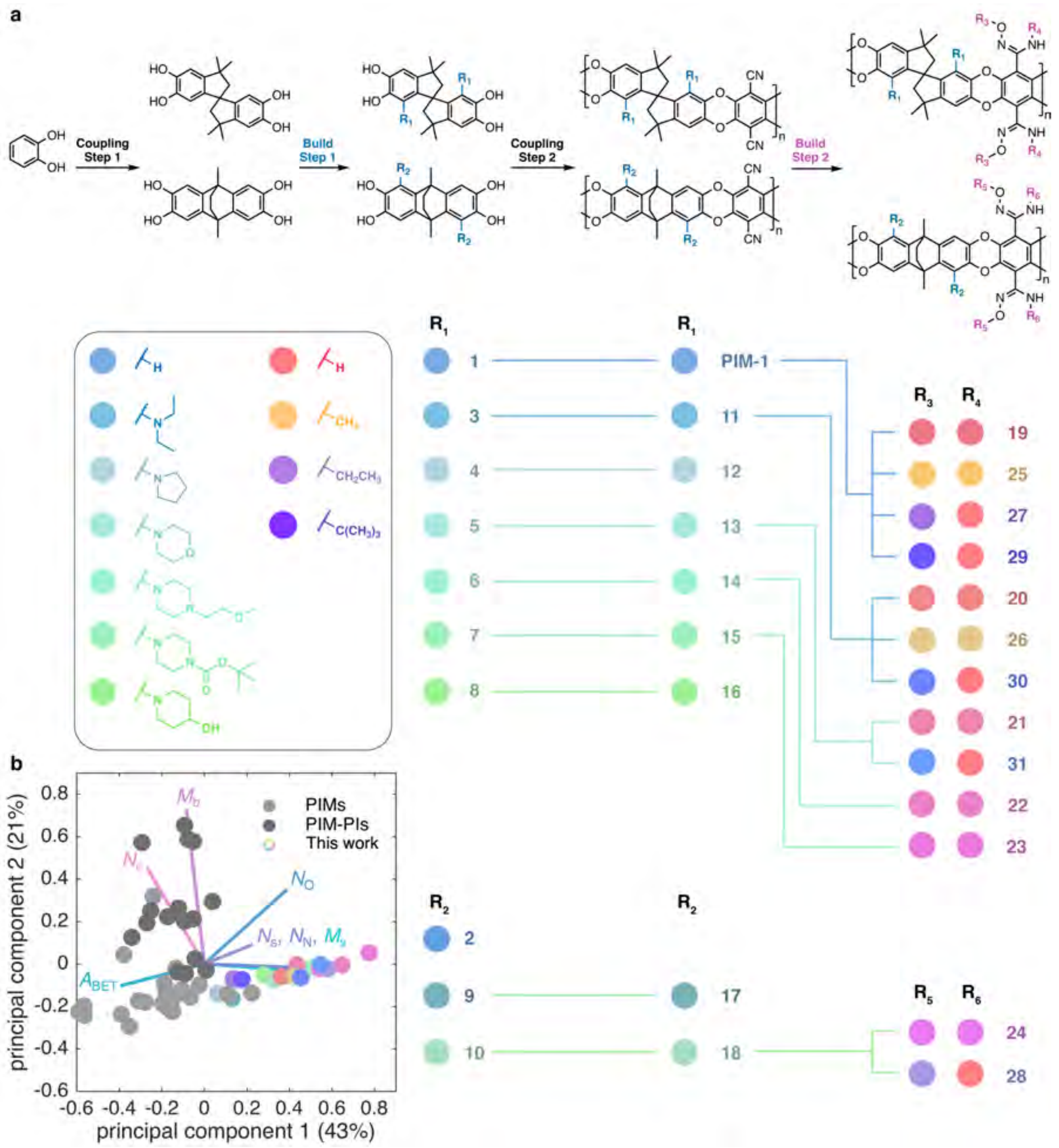

**Fig. 1: Diversity-Oriented Synthesis (DOS) of PIMs with Ion-Coordinating Functionality Embedded in Diverse Pore Network Architectures. a,** Forward synthesis of a diversity-oriented library of PIMs. **b,** Principle component analysis, which differentiates architectural attributes of the DOS PIM library from previously reported PIMs and PIM-Polyimide (PIM-PI) hybrids.

Having built skeletal diversity at the molecular level in bis(catechol) monomers, we next sought to do so at macromolecular length scales by coupling them together via base-mediated step-growth polymerization (**Fig. 1a**).[3,21] This allows the topology of interconnected FVEs within PIMs to vary alongside the $Li^+$-coordinating functionality. While spirobisindane **1** and dihyrdoethanoanthracene **2** polymerize with a variety of electrophiles to yield canonical PIMs,[3,21] aminomethylated monomers **3–10** are more complex and sterically hindered. In addition, PIMs with modified backbones often become insoluble during polymerization, which prevents them from reaching high molecular weights (*MW*). Nevertheless, polymerization of monomers **3–10** with tetrafluoroterephthalonitrile proceeded smoothly, affording aminomethylated ladder polymers **11–18** with *MW* of 19–146 kg $mol^{-1}$, BET surface areas of 11–505 $m^2$ $g^{-1}$, pore widths of 5.6–14.4 Å, and high solubility in a wide variety of common solvents (**Supplementary Fig. 8** and **Supplementary Tables 1** and **2**).

We next targeted the diversification of the dielectric environment of individual FVEs as well as their collective attributes as a network of interconnected pores by implementing several different on-polymer build steps. These characteristics should influence both the solvation structure and dynamics of $Li^+$ transport within the polymer. To do so, we initially targeted reactions promoting amidoxime formation[22–24] from benzonitriles along the polymer backbone to introduce a pluripotent functional group[25] for expanding the library with various polar-, non-polar, and pore-in-filling moieties. Nitrile-to-amidoxime interconversion was quantitative for all ladder polymers tested (**19–24**). From these common intermediates, we then developed several quantitative amidoxime-directed on-polymer reactions (**Fig. 1a**): including concurrent *N*- and *O*-monoalkylation under alkaline conditions using dimethyl sulfate (**25**, **26**), regiospecific *O*-alkylation under alkaline conditions using diethyl sulfate (**27**, **28**), and regiospecific *O*-alkylation under Lewis-acidic conditions[26] using di-*tert*-butyl dicarbonate (**29–31**). These further manipulations yielded polymers with *MW* of 12–222 kg mol$^{-1}$, BET surface areas of 10–454 m$^2$ g$^{-1}$, pore widths of 5.2–24.3 Å, and solubility in a wide variety of common solvents (**Supplementary Figs. 9 and 10** and **Supplementary Tables 1** and **2**).

To map the multi-dimensional structure space explored in this DOS PIM library, we used principal component analysis (PCA) (**Fig. 1b**). While PCA is common for assessing structure space for small-molecule DOS libraries, it has not been developed for differentiating polymers on the basis of their local microstructural characteristics within larger macromolecular architectures. To do so, we considered several features of relevance to polymers configured with hypothetical solvation cages for $Li^+$, including the molecular weight of all side chains per repeat unit ($M_s$), BET surface area ($A_{BET}$), number of oxygen atoms per repeat unit ($N_O$), number of nitrogen atoms per repeat unit ($N_N$), number of distinct types of side groups per repeat unit ($N_s$), molecular weight of polymer main chain per repeat unit ($M_b$), and the number of sites of contortion along the polymer main chain per repeat unit ($N_c$). PCA was then carried out on these parameters after normalization based on the respective sample means and sample standard deviations for each descriptor (**Supplementary Table 3**); the first two principal components accounted for 43% and 21% of the total variance within the data set, respectively (**Fig. 1b**). In comparing our DOS PIM library to previously reported PIMs and PIM-polyimide (PIM-PI) hybrids, we found that ours diverges on four key characteristics tied to the degree to which the polymers feature $Li^+$-coordinating *O*- and *N*-functionality as well as the diversity in their pairings along the polymer backbone, which in turn dictate aspects of both pore structure and dielectric environment due to the interconnectivity of FVEs. Further analysis of the DOS PIMs library by a second PCA incorporating macromolecular descriptors of diversity at longer length scales—including the number of peaks in the GIWAXS ($N_G$) assigned to various monomer reflections, the *d*-space value of the major GIWAXS peak ($L_G$), and the primary pore width of FVEs from NLDFT pore-size distributions based on $N_2$ isotherms ($L_{N2}$)—confirmed our monomer selections, ranges of chain conformations, and polymer packing in the condensed phase spanned a wide range (**Supplementary Fig. 11** and **Supplementary Table 4**). This suggested that the library could be well poised to identify PIMs with explicit solvation cages for $Li^+$ and thereby tunable properties for $Li^+$ transport.

To screen our library for PIMs whose FVEs might be solid solvation cages for $Li^+$, we initially determined the ionic conductivity ($\sigma$) for polymers infiltrated with a mixed carbonate electrolyte containing 1.0 M $LiPF_6$ as the supporting salt and compared these data relative to PIM-1, which features no explicit $Li^+$ solid solvation cage (**Fig. 2a**). PIMs **13, 14**, and **31** whose FVEs feature aminomethylation with $Li^+$-coordinating *N*- and *O*-functionality, exhibited the highest ionic conductivities at 25 ˚C—0.206 mS $cm^{-1}$, 0.195 mS $cm^{-1}$, and 0.192 mS $cm^{-1}$, respectively—which are nearly an order of magnitude higher than previously reported PIMs (e.g., PIM-1, $\sigma_{25^\circ C}$ = 0.038 mS $cm^{-1}$). For comparison, PIMs **13**, **14**, and **31** are almost as conductive as mesoporous polyolefin separators in-filled with electrolyte ($\sigma_{25^\circ C}$ = 0.413 mS $cm^{-1}$). Notably, PIMs **13**, **21**, and **31** each have *N*-alkylmorpholine as the side chain along their polymer backbones, but have markedly different $\sigma_{25^\circ C}$, highlighting the influential role played by local pore structure and dielectric environment set forth in the first and second build steps in the DOS. We also found that functional group elaboration in the second build step led to a large number of PIMs with high conductivity ($\sigma_{25^\circ C}$ = 0.063–0.191 mS $cm^{-1}$), but also low surface area due to their ultramicroporous character (**Supplementary Figs. 8–10**). This is due to shorter inter-chain spacing associated with hydrogen bonding when amidoximes are featured along the backbone and pore in-filling by various alkylated amidoxime-derived side-chains. This indicates that $\sigma$ in PIMs is enhanced by both the placement of $Li^+$-coordinating functionality within the pore network and the manifestation of an ultramicroporous polymer architecture that restricts the number of solvent molecules that can be bound to $Li^+$ within the pore network so that it might, in turn, have higher diffusivity therein.

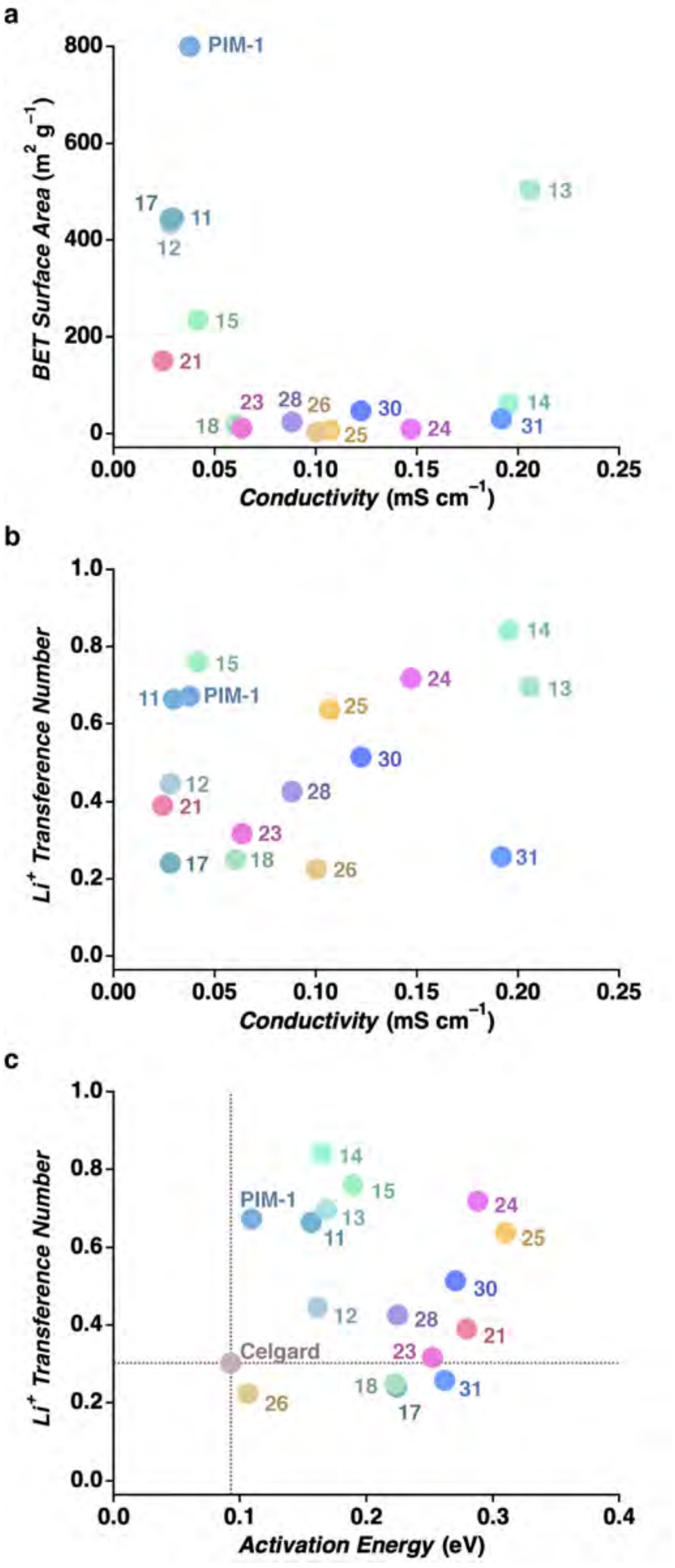

**Fig. 2: Structure–Transport relationships within the DOS PIMs library. a,** Conductivity vs. BET surface area, **b,** Conductivity vs. steady-state cation transference number, and **c,** Activation

energy vs. steady-state cation transference number. See **Supplementary Figs. 8–10** and **12–45** for $N_2$ isotherms and EIS spectra, respectively, and **Supplementary Table 5** for a summary of surface area, conductivity, transference number, and activation energy.

To explore the nature of $Li^+$-coordinating solid solvation cages in PIMs, we used multiscale computational modeling aimed at revealing the molecular structure of these cages and the energy landscapes for cage-to-cage $Li^+$ transport. Specifically, we used quantum mechanical calculations to analyze the stability of the various configurations and fragments of lead candidate PIM **13** (**Supplementary Figs. 46–52** and **Supplementary Tables 6–8**), and classical molecular dynamics simulations accompanied with free energy sampling[26,27] (**Supplementary Figs. 53–59** and **Supplementary Tables 9–10**) to reconstruct the dynamics of $Li^+$ (de)solvation in the condensed phase of PIM **13**, e.g., as $Li^+$ hops from one cage to another. We initially explored the free energy landscape of moving an isolated $Li^+$-ion from a cage to the bulk solvent and then along its motion between neighboring cages in PIM **13**, which corresponds to the low salt concentration limit and ignores anion interactions. The free energy profile (**Supplementary Fig. 58**) as a function of distance ($\boldsymbol{r}_1$) between the center of mass of a cage (**Supplementary Fig. 49**) and the cation exhibits a kinetic barrier of 0.29–0.31 eV for $Li^+$ cation dissociation from the cage and indicates that the cage-solvated state of the cation is 0.14–0.15 eV more stable (see more details **Supplementary Figs. 51–52**, and **Supplementary Tables 7–8**) than that of a fully solvated 4-fold (**Supplementary Fig. 47**) coordinated $Li^+$ in the bulk solvent. This suggests that in electrolytes with high concentrations of lithium salts (e.g., 1.0 M $LiPF_6$), the equilibrium $\mathrm{Li}^+_{\mathrm{cage}} \leftrightarrow \mathrm{Li}^+_{\mathrm{sol}} + \mathrm{cage}$ will be shifted to the left, as the dissociation constant for the ion in the cage is $K_d$ = 3.03 mM; in other words, nearly all of the cages will be occupied by $Li^+$-ions, and therefore the concentration of $Li^+$ in the membrane is significantly higher than in membranes without cages.

We next explored the kinetics of motion of $Li^+$ between cages. **Fig. 3a** shows the schematics of a plausible transport pathway, with $Li^+$ beginning and ending in neighboring cages along the polymer backbone. We used linear combinations $\boldsymbol{d} = \boldsymbol{r}_1 - \boldsymbol{r}_2$ and $\boldsymbol{s} = \boldsymbol{r}_1 + \boldsymbol{r}_2$ of distances $\mathbf{r}_1$ and $\mathbf{r}_2$ between $Li^+$ and the centers of mass for the two cages as representative collective variables (**Supplementary Figs. 55–57**). This guarantees that $Li^+$ is kept in close proximity to the polymer backbone (which is also favorable from a free energy standpoint) and does not diffuse extensively into the bulk solvent. In **Fig. 3d**, we show the 2D free energy landscape in these coordinates. The optimal, shortest pathway connecting two cages (**Fig. 3f,g**) is nearly symmetric and transitions through a partially solvated stage of the $Li^+$-ion (see **Fig. 3e**). The kinetic barriers are related to the free energy costs of dissociating $Li^+$ from the cage's *O*- and *N*-functionality (**Fig. 3c**). The free energy profile along the transport pathway (**Fig. 3b**) shows kinetic barriers of 0.24–0.27 eV. Due to steric constraints, $Li^+$ is only partially solvated along its pathway, and these configurations are 0.17–0.18 eV higher in free energy when compared to those when $Li^+$ is in either cage (**Fig. 3b,e**). These calculations compare favorably to the EIS data for activation barriers to ion transport in PIM **13** (**Fig. 2d**). It is worth noting that slightly lower kinetic barriers (0.24 eV versus 0.29 eV) along the cage-to-cage pathway are due to the correlated motion of the coordinating centers of adjacent cages that facilitates the ion transport. Taken together, our 1D and 2D free energy analysis suggests that a plausible mechanism of $Li^+$ transport is the hopping between vacant cages along the polymer backbone, rather than solvent-assisted cation diffusion.

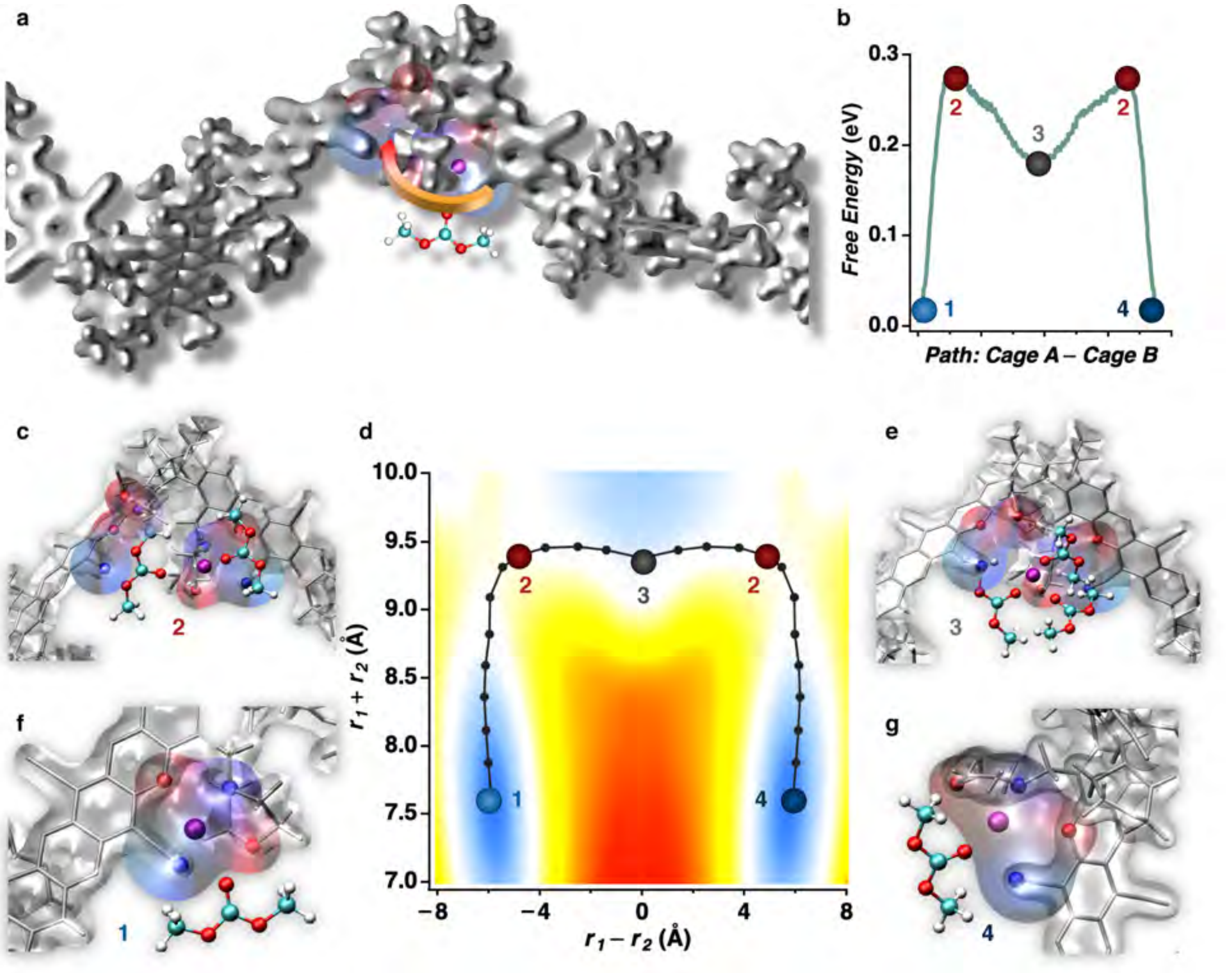

**Fig. 3: Molecular structure of $Li^+$ (PIM 13) solid solvation cages and free energy analysis of the cage-to-cage $Li^+$ ion transport. a**, Schematic of PIM **13** with two adjacent $Li^+$ solid solvation cages, and the direction of $Li^+$ transport, **b**, Free energy profile along the optimal path connecting the cages, **c**, Molecular configuration of $Li^+$ in an under-coordinated state, **d**, 2D free energy surface in $\mathbf{r_1} - \mathbf{r_2}$ and $\mathbf{r_1} + \mathbf{r_2}$ coordinates ($\mathbf{r_1}$ and $\mathbf{r_2}$ refer to the distances between $Li^+$ and the centers of masses of cage 1 and cage 2, respectively) with the optimal transport path, **e**, The molecular configuration of partly solvated $Li^+$, **f**, Molecular configuration of the $Li^+$ solid solvation cage (initial), **g**, Molecular configuration of the $Li^+$ solid solvation cage (final).

PIMs featuring $Li^+$ solid solvation cages are attractive as anode–electrolyte interlayers in Li–metal batteries, where they mediate $Li^+$ transport during Li metal plating and stripping while charging or discharging the battery, respectively.[28–30] Owing to the high $\sigma_{25C}$, high $t^+_{ss}$, and low $E_a$ for PIM **13**, Li–metal cells configured with PIM **13** as an interlayer should experience lower cell impedance and thereby access higher capacity than cells with other PIM interlayers (e.g., PIM-1), or in the absence of a PIM interlayer. To test this hypothesis, we assembled Li–NMC622 cells (cathode areal capacity = 1.44 mAh $cm^{-2}$) whose Li anodes either coated with PIM **13**, PIM-1 (positive control), or left bare (negative control). Cells were discharged galvanostatically at 1 mA $cm^{-2}$ and charged at 0.2 mA $cm^{-2}$ during cell cycling. Confirming our hypothesis, those configured with the PIM **13** interlayer accessed ~25% higher capacity than cells configured with PIM-1 or left bare. This gain in cathode utilization was sustained for over 400 cycles, after which 84% of the initial capacity of 1.01 mAh $cm^{-2}$ was retained. By comparison, cells configured with PIM-1 (positive control) initially access ~0.96 mAh $cm^{-2}$ before experiencing a high rate of capacity fade; cells featuring unmodified Li anodes access even lower capacity, ~0.86 mAh $cm^{-2}$. The modest gains initially by PIM-1 cells slowly diminishes over time, only further highlighting the advantages of implementing PIM **13** interlayers. Cathode utilization in cells configured with PIM 13 interlayers persisted across the cycling history: after 400 cycles, the cumulative gain in capacity was either 141% or 134% higher than the bare Li (negative control) and PIM-1 coated Li (positive control), respectively.

In addition to accessing higher capacity, cells implementing PIM **13** interlayers featured a charge–discharge overpotential of only 0.32 V, which was 0.19 V and 0.16 V lower than that of both positive (PIM-1) and negative controls (**Fig. 4e**). Furthermore, in high rate cycling up to 5 mA $cm^{-2}$, PIM **13** cells maintained both lower overpotentials and higher accessible capacities, compared to the controls. Such characteristics in Li–metal cells are desirable for batteries used in electric vehicles, trucks, and aircraft (**Supplementary Fig. 60**).[31–35] In this way, the advantages of PIM interlayers featuring explicit $Li^+$ solid solvation cages are clearly tied to their higher transference numbers and higher conductivity, which reduces polarization and the overpotentials for plating and stripping Li during cell cycling.

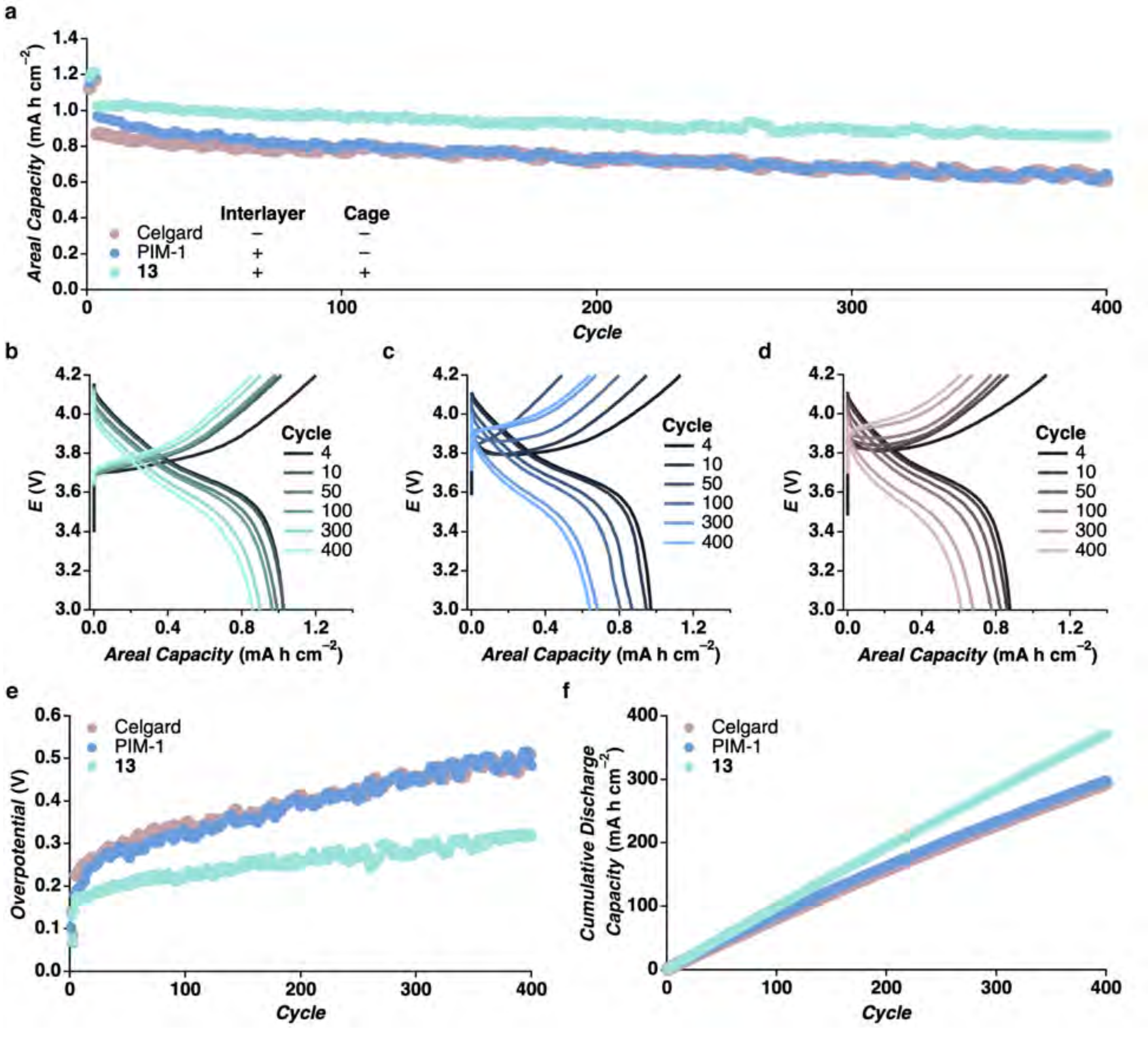

**Fig. 4: Li–NMC622 cell cycling. a** discharge capacity, **b–d** charge–discharge curves for PIM **13**, PIM-1, and Celgard respectively, **e** overpotential, and **f** cumulative discharge capacity.

Our introduction of $Li^+$-coordinating functionality into the pore networks of PIMs and subsequent identification of those whose FVEs are solvation cages for $Li^+$ have revealed a surprisingly simple way to overcome the conventional inverse relationship between conductivity and diffusion activation barrier for ion transport in PIM membranes. Specifically, the association of $Li^+$ with specific classes of *N*- and *O*-based ion-coordinating functionality on the polymer increases the $Li^+$ concentration in the polymer via partitioning from the bulk electrolyte. This solubility-enhanced ion transport in porous polymer membranes is rare, yet analogous to what has been observed for gas transport in microporous polymer membranes embedded with selective gas-binding metal–organic frameworks.[36] Furthermore, the architectural characteristics of our DOS PIM library makes possible exquisitely tunable ion transport selectivity. Both high $t^+_{ss}$ (> 0.8) and low $t^+_{ss}$ (< 0.2) can be intentionally selected from the DOS PIMs library and put to use in different classes of electrochemical devices. Thus, while the former are useful as Li-anode interlayers as we have demonstrated, the latter are potentially useful for fuel cells and redox-flow batteries, where high anion transference numbers are desirable.[3] DOS PIM libraries generated by these and future diversity-oriented synthetic approaches will greatly improve the prospects of identifying selective membranes for a wide variety of applications, and points to exciting new directions for macromolecular nanoionics.

**References**

[1] Slater, A. G. & Cooper, A. I. Function-led design of new porous materials. *Science* **348**, aaa8075 (2015).

[2] Koros, W. J. & Zhang, C. Materials for next-generation molecularly selective synthetic membranes. *Nat. Mater.* **16**, 289–297 (2017).

[3] Li, C., Meckler, S. M., Smith, Z. P., Bachman, J. E., Maserati, L., Long, J. R. & Helms, B. A. Engineered Transport in Microporous Materials and Membranes for Clean Energy Technologies. *Adv. Mater.* **30**, 1704953 (2018).

[4] Park, H. B., Kamcev, J., Robeson, L. M., Elimelech, M. & Freeman, B. D. Maximizing the right stuff: The trade-off between membrane permeability and selectivity. *Science* **356**, eaab0530 (2017).

[5] Schreiber, S. L. Target-oriented and diversity-oriented organic synthesis in drug discovery. *Science* **287**, 1964–1969 (2000).

[6] Burke, M. D., Berger, E. M. & Schreiber, S. L. Generating diverse skeletons of small molecules combinatorially. *Science* **302**, 613–618 (2003).

[7] Burke, M. D. & Schreiber, S. L. A planning strategy for diversity-oriented synthesis. *Angew. Chem., Int. Ed.* **43**, 46–58 (2004).

[8] Tan, D. S. Diversity-oriented synthesis: exploring the intersections between chemistry and biology. *Nat. Chem. Biol.* **1**, 74–84 (2005).

[9] Schreiber, S. L. Molecular diversity by design. *Nature* **457**, 153–154 (2009).

[10] Nielsen, T. E. & Schreiber, S. L. Diversity-oriented synthesis—Towards the optimal screening collection: a synthesis strategy. *Angew. Chem., Int. Ed.* **47**, 48–56 (2008).

[11] Galloway, W. R. J. D. *et al.* Diversity-oriented synthesis as a tool for the discovery of novel biologically active small molecules. *Nat. Commun.* **1**, 80 (2010).

[12] Haggarty, S. J. The principle of complementarity: chemical versus biological space. *Curr. Opin. Chem. Biol.* **9**, 296–303 (2005).

[13] Furukawa, H., Cordova, K. E., O'Keeffe, M. & Yaghi, O. M. The chemistry and applications of metal–organic frameworks. *Science* **341**, 1230444 (2013).

[14] Tranchemontagne, D. J., Ni, Z., O'Keeffe, M. & Yaghi, O. M. Reticular chemistry of metal–organic polyhedra. *Angew. Chem., Int. Ed.* 47, 5136–5147 (2008).

[15] Holst, J. R., Trewin, A. & Cooper, A. I. Porous organic molecules. *Nat. Chem.* **2**, 915–920 (2010).

[16] Diercks, C. S. & Yaghi, O. M. The atom, the molecule, and the covalent organic framework. *Science* **355**, eaal1585 (2017).

[17] Bisbey, R. P. & Dichtel, W. R. Covalent Organic Frameworks as a Platform for Multidimensional Polymerization. *ACS Cent. Sci.* **3**, 533–543 (2017).

[18] Cooper, A. I. Porous Molecular Solids and Liquids. *ACS Cent. Sci.* 3, 544–553 (2017).

[19] Mannich, C. & Krösche, W. Ueber ein Kondensationsprodukt aus Formaldehyd, Ammoniak und Antipyrin. *Archiv der Pharmazie* **250**, 647–667 (1912).

[20] Arend, M., Westermann, B. & Risch, N. Modern variants of the Mannich reaction. *Angew. Chem., Int. Ed.* **37**, 1045–1070 (1998).

[21] McKeown, N. B. & Budd, P. M. Polymers of intrinsic microporosity (PIMs): organic materials for membrane separations, heterogeneous catalysis and hydrogen storage. *Chem. Soc. Rev.* 35, 675–683 (2006).

[22] Patel, H. A. & Yavuz, C. T. Noninvasive functionalization of polymers of intrinsic microporosity for enhanced $CO_2$ capture. *Chem. Commun.* **48**, 9989–9991 (2012).

[23] Baran, M. J., Braten, M. N., Sahu, S., Baskin, A., Meckler, S. M., Li, L., Maserati, L., Carrington, M. E., Chiang, Y.-M., Prendergast, D. & Helms, B. A. Design Rules for Membranes from Polymers of Intrinsic Microporosity for Crossover-free Aqueous Electrochemical Devices. *Joule* **3**, 2968–2985 (2019).

[24] Tan, R., Wang, A., Malpass-Evans, R. *et al.* Hydrophilic microporous membranes for selective ion separation and flow-battery energy storage. *Nat. Mater.* (2019) doi:10.1038/s41563-019-0536-8

[25] Eloy, F. & Lenaers, R. The Chemistry of Amidoximes and Related Compounds. *Chem. Rev.* **62**, 155–183 (1962).

[26] Bartoli, G., Bosco, M., Locatelli, M., Marcantoni, E., Melchiorre, P. & Letizia Sambri, L. Unusual and Unexpected Reactivity of *t*-Butyl Dicarbonate ($Boc_2O$) with Alcohols in the Presence of Magnesium Perchlorate. A New and General Route to *t*-Butyl Ethers. *Org. Lett.* **7**, 427–430 (2005).

[26] Laio, A. & Gervasio, F. L. Metadynamics - A Method to Stimulate Rare Events and Reconstruct the Free Energy in Biophysics. *Rep. Prog. Phys*. **71**, 126601 (2008).

[27] Barducci, A., Bonomi, M. & Parrinello, M. Metadynamics. *Wiley Inter. Rev.: Comp. Mol. Sci*. **1**, 826−843 (2011).

[28] Li, C., Ward, A. L., Doris, S. E., Pascal, T. A., Prendergast, D. & Helms, B. A. A Polysulfide-Blocking Microporous Polymer Membrane Tailored for Hybrid Li–Sulfur Flow Batteries. *Nano Lett.* **15**, 5724–5729 (2015).

[29] Ward, A. L., Doris, S. E., Li, L., Hughes, M. A., Jr., Qu, X., Persson, K. A. & Helms, B. A. Materials Genomics Screens for Adaptive Ion Transport Behavior by Redox-Switchable Microporous Polymer Membranes in Lithium–Sulfur Batteries. *ACS Cent. Sci.* **3**, 399–406 (2017).

[30] Ma, L., Fu, C., Li, L., Mayilvahanan, K. S., Watkins, T., Perdue, B. R., Zavadil, K. R. & Helms, B. A. Nanoporous Polymer Films with High Cation Transference Number Stabilize Lithium Metal Anodes in Light-Weight Batteries for Electrified Transportation. *Nano Lett.* **19**, 1387–1394 (2019).

[31] Albertus, P., Babinec, S., Litzelman, S. *et al.* Status and challenges in enabling the lithium metal electrode for high-energy and low-cost rechargeable batteries. *Nat Energy* **3,** 16–21 (2018).

[32] Fredericks, W. L., Sripad, S., Bower, G. C. & Viswanathan, V. Performance Metrics Required of Next-Generation Batteries to Electrify Vertical Takeoff and Landing (VTOL) Aircraft. *ACS Energy Lett.* **3**, 2989-2994 (2018).

[33] Liu, J., Bao, Z., Cui, Y. *et al.* Pathways for practical high-energy long-cycling lithium metal batteries. *Nat Energy* **4**, 180–186 (2019).

[34] Viswanathan, V. & Knapp, B. M. Potential for electric aircraft. *Nat. Sustain.* **2**, 88–89 (2019).

[35] Sripad, S. & Viswanathan, V. Quantifying the Economic Case for Electric Semi-Trucks. *ACS Energy Lett.* **4**, 149–155 (2019).

[36] Bachman, J., Smith, Z., Li, T. *et al.* Enhanced ethylene separation and plasticization resistance in polymer membranes incorporating metal–organic framework nanocrystals. *Nat. Mater*. 15, 845–849 (2016).

**Acknowledgements**

This work was supported by the Joint Center for Energy Storage Research (JCESR), an Energy Innovation Hub funded by the U.S. Department of Energy, Office of Science, Office of Basic Energy Sciences. Portions of this work, including polymer synthesis and characterization, were carried out as a user project at the Molecular Foundry, which is supported by the Office of Science, Office of Basic Energy Sciences of the U.S. Department of Energy under contract no. DE-AC02-05CH11231. S.M.M. and M.E.C. acknowledge support from the Center for Gas Separations Relevant to Clean Energy Technologies, an Energy Frontier Research Center funded by the U.S.

Department of Energy, Office of Science, Basic Energy Sciences under award # DE-SC0001015 to carry out principal component analysis, gas sorption, and GIWAXS measurements. GIWAXS and single crystal measurements were carried out at the Advanced Light Source, which is a DOE Office of Science User Facility under contract no. DE-AC02-05CH11231. Computation was carried out at the National Energy Research Scientific Computing Center (NERSC), a U.S. Department of Energy Office of Science User Facility operated under the same contract. J.S. and C.F. acknowledge support from the Advanced Research Projects Agency-Energy Integration and Optimization of Novel Ion Conducting Solids (IONICS) program under Grant No. DE-AR0000774 to carry out Li–NMC622 cell cycling.

**Author contributions**

B.A.H. designed and directed the study. M.J.B., M.E.C., S.S., M.A.B., and S.M.M. synthesized and characterized the monomers and polymers. S.J.T. conducted single-crystal x-ray crystallography and analyzed the results. M.E.C. carried out principal component analysis on the polymer library's diversity metrics. M.J.B. and C.F. carried out ion transport experiments. J.S. conducted full cell cycling experiments. D.P. designed and directed the theoretical study. A.B. conducted the simulations. B.A.H. wrote the paper with contributions from all co-authors.

**Competing interests**

B.A.H., M.J.B., M.E.C., S.S., and S.M.M. are inventors on US provisional patent application 62/719,498 submitted by Lawrence Berkeley National Laboratory, which covers the DOS PIMs library as well as aspects of its use.

## Additional information

Supplementary information is available for this paper at XXX.

Reprints and permissions information are available at www.nature.com/reprints.

Correspondence and requests for materials should be addressed to B.A.H.

## Supplementary Methods

## Materials

Paraformaldehyde, morpholine, pyrrolidine, diethylamine, hydroxylamine, magnesium perchlorate, sodium hydroxide, lithium hydroxide monohydrate, dimethyl sulfate, diethyl sulfate, di-*tert*-butyl dicarbonate, dimethyl sulfoxide, *d*-chloroform, and $d_4$-*ortho*-dichlorobenzene were obtained from Sigma Aldrich. Hexanes, ethanol, toluene, methylene chloride, acetone, chloroform, methanol, dimethylformamide, tetrahydrofuran, sodium chloride, magnesium sulfate, and potassium carbonate were obtained from Fisher Scientific. 3,3,3',3'-tetramethyl-1,1'-spirobisindane-5,5',6,6'-tetraol was obtained from Alfa Aesar, tetrafluoroterephthalonitrile was obtained from Oakwood Chemicals, *tert*-butyl piperazine-1-carboxylate was obtained from Ark Pharm, 2-methoxyethyl-piperazine was obtained from Matrix Scientific, and $d_6$-dimethyl sulfoxide was obtained from Cambridge Isotope Laboratories. All chemicals were used as received without further purification. 9,10-dimethyl-9,10-dihydro-9,10-ethanoanthracene-2,3,6,7-tetraol was prepared according to literature procedure.[1]

## Instrumentation

**$^{1}$H and $^{13}$C Nuclear Magnetic Resonance (NMR) Spectroscopy**. $^{1}$H and $^{13}$C NMR spectra were recorded on Bruker Avance II at 500 MHz and 125 MHz, respectively. Chemical shifts are reported in δ (ppm) relative to the residual solvent peak ($CDCl_3$: 7.26 for $^{1}$H, 77.16 for $^{13}$C; $d_6$-DMSO: 2.50 for $^{1}$H, 39.51 for $^{13}$C; $d_4$-*o*-$C_6D_4Cl_2$, 120 °C: 7.22 and 6.95 for $^{1}$H). Splitting patterns are designated as s (singlet), br s (broad singlet), d (doublet), t (triplet), q (quartet), and m (multiplet).

**Electrospray Ionization Mass Spectrometry (ESI-MS)**. Spectra for compounds **3**–**6** were acquired on a Synapt G2 Q-TOF spectrometer. Spectra for compounds **7**–**10** were acquired on a Bruker MicroTOF spectrometer. ESI-MS was performed by the University of California, Berkeley QB3/Chemistry Mass Spectrometry Facility.

**Elemental Analysis (EA)**. EA was performed by the University of California, Berkeley College of Chemistry Microanalytical Facility.

**Single-Crystal X-ray Diffraction (XRD)**. Single crystals for compounds **3**–**7**, **9**, and **10** were selected and mounted on Mitegen® loops with Paratone oil and placed in an Oxford Cryosystems Cryostream 700 plus at $T$ = 100 K. Data were collected for **3**–**5** using a Bruker D8 diffractometer with APEXII CCD detector, with Mo $K_\alpha$ ($\lambda$ = 0.71073 Å), while for **6**, **7**, **9**, and **10** using beamline 12.2.1 at the Advanced Light Source with $\lambda$ = 0.7288 Å using a Bruker D8 diffractometer with a Bruker PHOTONII CPAD detector. Data reduction was performed and corrected for Lorentz and polarization effects using SAINT[2] v8.38a and were corrected for absorption effects using SADABS v2016/2.[3] Structure solutions were performed by SHELXT[4] using the direct method and were refined by least-square refinement against $F^2$ by SHELXL.[5]

**Size-Exclusion Chromatography (SEC)**. SEC using THF as the mobile phase was carried out with a Malvern Viscotek TDA 302 system calibrated with a 99 kDa monodisperse polystyrene standard. SEC using DMF (containing 0.2% *w/v* LiBr) as the mobile phase was carried out using a customized system consisting of a Shimadzu LC-20AD pump, Viscotek VE 3580 refractive index detector, and two mixed bed columns connected in series (Viscotek GMHHR-M). The system was operated at a temperature of 70 °C. Calibration on the system was performed with narrow poly(methyl methacrylate) standards (Polymer Laboratories) ranging from 620 g mol$^{-1}$ to 910,500 g mol$^{-1}$.

**Gas Adsorption and Desorption**. $N_2$ adsorption isotherms were collected at 77 K on a Micromeritics Tristar II 3020 gas sorption analyzer. $CO_2$ adsorption isotherms were collected on a Micromeritics ASAP 2020 gas sorption analyzer at 273 K. Nonlocal DFT (NLDFT) pore-size distributions from $N_2$ adsorption isotherms were calculated using SAIEUS software provided by Micromeritics using the 2D Heterogeneous Surface model. NLDFT pore-size distributions from $CO_2$ adsorption isotherms were calculated using the Micromeritics Carbon Slit model with high regularization. Samples were degassed under vacuum at 100–150 °C overnight prior to analysis.

**Grazing Incidence Wide-Angle X-ray Scattering (GIWAXS).** Samples for grazing-incidence wide-angle X-ray scattering (GIWAXS) were prepared by spin coating polymer inks onto silicon substrates. The substrates were previously cleaned by sonication at 40 °C in soap water, deionized water, acetone, and isopropyl alcohol before oven drying and UV/ozone treatment immediately prior to sample deposition. 11 through 16 were cast from chloroform solutions, and all other polymers were cast from solutions of NMP. Residual solvent was removed under vacuum at 90 °C for 24–48 h.

Scattering patterns were collected under a helium atmosphere at the Advanced Light Source beamline 7.3.3[6] of the Lawrence Berkeley National Laboratory. A photon energy of 10 keV, a sample-to-detector distance of 275 mm, and incident angles between 0.05 and 0.20 were used. The beamline was equipped with a Pilatus 2M detector, samples were calibrated against a silver behenate standard, and the data was processed using the Nika package for Igor Pro.[7]

Each GIWAXS pattern exhibited two or more distinct, broad scattering features typical of microporous polymers. The scattering vectors of these features and the corresponding real-space distances are detailed in **Supplementary Table 2**.

GIWAXS scattering patterns of microporous polymers are primarily informed by two types of structural features: 1) FVEs and concomitant semi-periodic regions of high and low electron density, which produce a broad feature whose peak intensity is at low $q$ and spans much of the GIWAXS range, and 2) repeating chemical moieties along the polymer chain (e.g., spiro centers), which produce smaller features superimposed on the broader peak.[8] Both effects are evident in the scattering patterns reported here. Separately, aging in thin films of PIMs is known to influence the overall intensity of the low-$q$ feature; thermal annealing can accelerate this aging effect.

In general, a strong peak is observed at $q$ = 0.6–0.7 Å$^{-1}$ in the scattering patterns of aminomethylated polymers, such as nitrile-containing PIMs **11**–**18**. Partially amplified by the broad low-$q$ feature, this peak is similar to a signature observed in previously reported PIMs with bulky appended functional groups.[8] A second feature in these PIMs at higher $q$-value (~1.3 Å$^{-1}$) results from scattering involving the appended group, as its intensity compared to the feature at $q$ = 0.6–0.7 Å$^{-1}$ qualitatively tracks with the size of the appended group. The lower-$q$ peak in PIM **15** is slightly shifted and a third peak, attributed to scattering associated with *tert*-butyl functionality, is present at $q$ = 0.95 $Å^{-1}$. These scattering patterns can be contrasted to those of polymers modified only at the nitrile group, such as PIMs **19**, **25**, **27**, and **29**. Capable of interchain hydrogen bonding, yet lacking a bulky aminomethylated functionalization, these generally do not exhibit a broad, low-$q$ feature or a peak at $q$ = 0.6–0.7 $Å^{-1}$.

When multiple modifications are present along the polymer backbone, complementary and competing effects are observed in the scattering patterns. Some aminomethylated polymers also bearing amidoxime groups, such as PIMs **20** and **21**, exhibit the peak at $q$ = 0.6–0.7 $Å^{-1}$, perhaps because the added bulk around the polymer backbone inhibits interchain hydrogen bond formation. Amidoxime-bearing polymers with even bulker chemical groups appended through the Mannich reaction exhibit complex GIWAXS patterns with many peaks, as seen with PIMs **22** and **23**. When the amidoxime groups on such polymers are alkylated, as in PIMs **26**, **30**, and **31**, the scattering patterns resemble those of PIMs **11–18**, although sometimes with shifted peak positions, highlighting the important role aminomethylation plays in dictating polymer chain packing in the condensed phase.

When plotted against NLDFT pore sizes obtained from $N_2$ and $CO_2$ isotherms, a good positive correlation ($R^2$ = 0.62) was achieved between the $d$-space value of the major GIWAXS peak and the maximum FVE obtained by gas porosimetry:

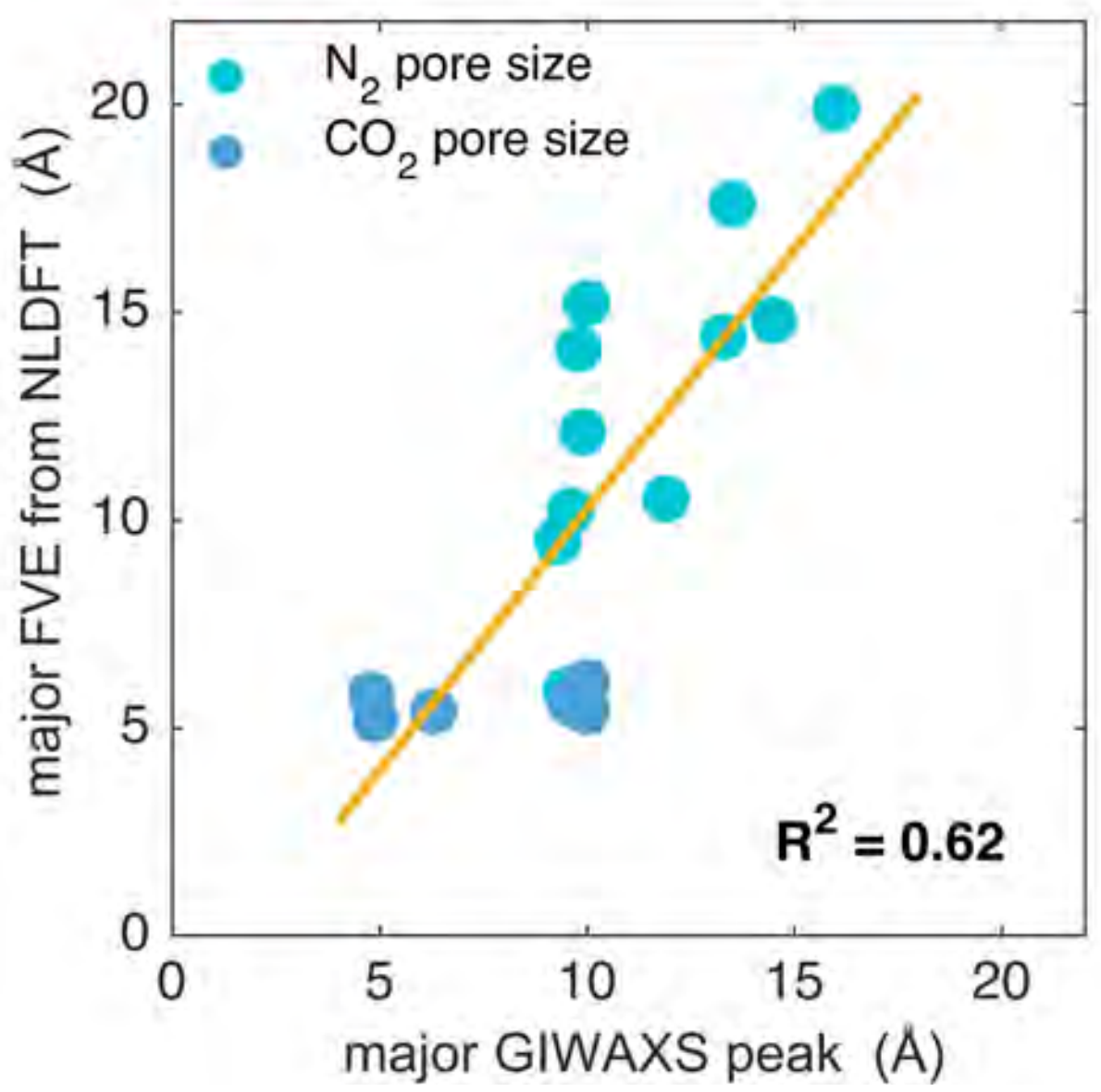

Collectively, these data suggest that the presence of complementary and competing effects in the scattering patterns of PIMs in our library showcase the library's diversity across multiple length scales, and that the major peak in the GIWAXS often provides an approximation of the size of accessible FVEs for PIMs in the condensed phase.

**Principal Component Analysis.** Principal component analysis was carried out in MATLAB® version R2015b using ten descriptors as defined below:

1. Molecular weight of side groups per repeat unit ($M_s$). The total molecular weight of components per repeat unit that do not comprise the polymer main chain. For example, in the case of **13**, the highlighted regions were designated as side groups:

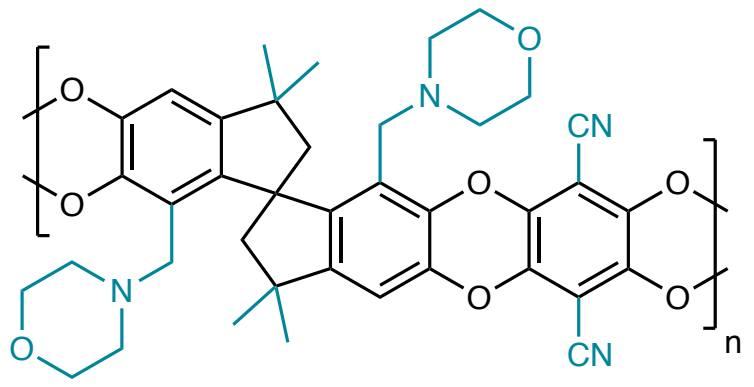

2. BET surface area ($A_{BET}$). The BET surface area as calculated from nitrogen adsorption experiments (see **Gas Adsorption and Desorption** above), or as reported in Supplementary References 9–11.
3. Number of oxygen atoms per repeat unit ($N_O$).
4. Number of nitrogen atoms per repeat unit ($N_N$).
5. Number of distinct types of side group per repeat unit ($N_s$). For example, in the case of **13**, there are three types of side group as highlighted below:

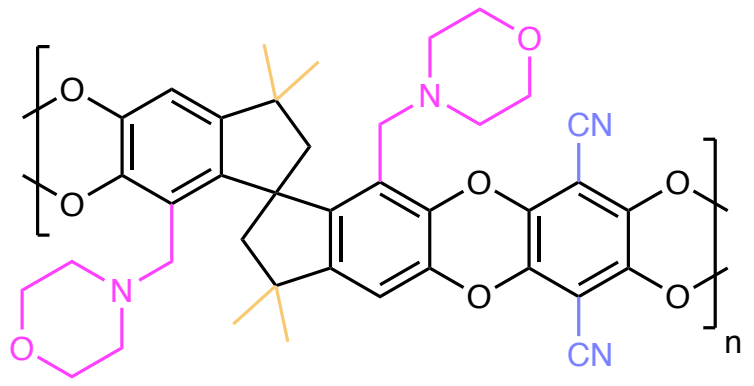

6. Molecular weight of polymer main chain per repeat unit ($M_b$). The total molecular weight of components per repeat unit that comprise the polymer main chain (backbone). For example, in the case of **13**, the highlighted regions were designated as main chain components:

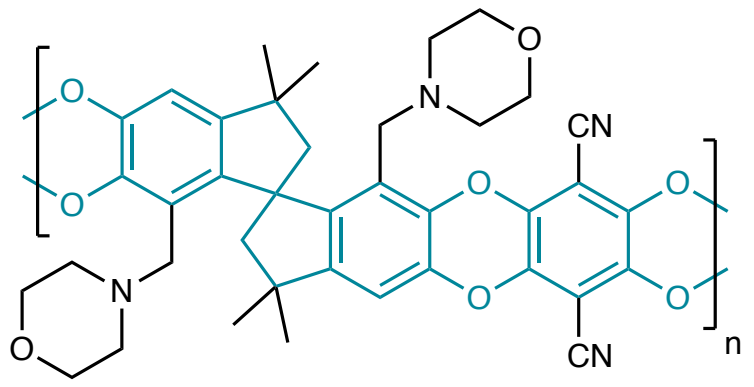

7. Number of sites of contortion along the polymer main chain per repeat unit ($N_c$).
8. Number of GIWAXS peaks ($N_G$).
9. *d*-space value of major GIWAXS peak ($L_G$).
10. Pore width of major FVE from NLDFT pore-size distributions based on $N_2$ isotherms ($L_{N2}$).

Raw data were first normalized by calculating $Z$ values based on the respective sample means and sample standard deviations for each of the seven descriptors described above. Principal component analysis (PCA) was then carried out on the normalized data ($Z$ values) using the singular value decomposition algorithm. For the PCA comparing PIMs demonstrated in this work to archetypical literature PIMs (**Fig. 1b**), descriptors 1–7 were used. The first two principal components accounted for 43% and 21% of the total variance within the data set respectively. For the PCA carried out strictly on PIMs demonstrated in this work (**Supplementary Fig. 11**), descriptors 1–6 & 8–10 were used. The first two principal components accounted for 41% and 27% of the total variance within the data set respectively. Vectors for each descriptor were also calculated, the direction and length of which indicate how each descriptor contributes to the first two principal components. A complete summary of raw data used in these calculations is included in **Supplementary Tables 3** and **4**, respectively.

**Synthetic Procedures**

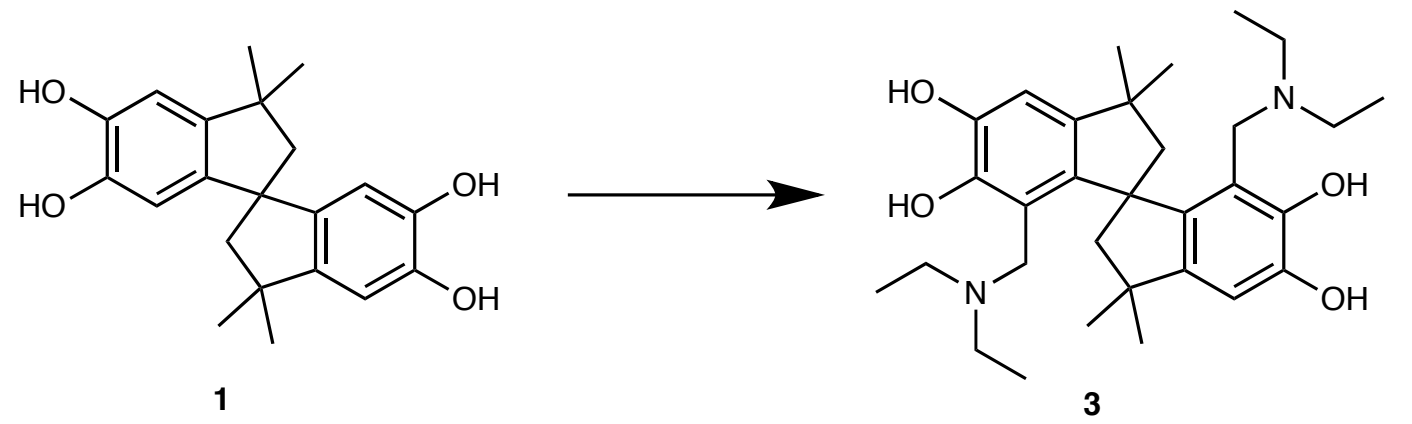

1 3

**3:** To a 500 mL round bottom flask was added toluene (250 mL), paraformaldehyde (4.45 g, 156 mmol), and diethylamine (10 mL, 156 mmol). The reaction was stirred for 10 min at 70 $^{\circ C}$ which point 3,3,3',3'-tetramethyl-1,1'-spirobisindane-5,5',6,6'-tetraol (10 g, 30.2 mmol) was added and the reaction was stirred at reflux overnight. The reaction was washed with brine, dried over $MgSO_4$, concentrated by rotary evaporator, and washed with hexanes to give 11.5 g of **3** (95%) as a yellow solid. $^1$H NMR ($CDCl_3$, 500 MHz): $\delta$ (ppm) 6.66 (s, 2H), 3.25 (d, 2H, J = 15.1 Hz), 3.14 (d, 2H, J = 15.1 Hz), 2.55 (br s, 4H), 2.38 (br s, 4H), 2.27 (d, 2H, J = 13.4 Hz), 2.17 (d, 2H, J = 13.4 Hz), 1.34 (s, 6H), 1.29 (s, 6H), 1.01 (br s, 6H), 0.93 (br s, 6H); $^{13}$C{$^1$H} NMR ($CDCl_3$, 125 MHz): δ (ppm) 145.73, 144.67, 142.02, 137.91, 116.21, 106.56, 57.89, 56.71, 53.48, 53.24, 42.73, 32.83, 28.83, 13.45; HR-MS (*m*/*z*) $[M+H]^+$: Calculated: 507.3209, Found: 507.3210; Elemental Analysis for $C_{31}H_{46}N_2O_4$ Calculated: C 72.01, H 9.08, N 5.49; Found: C 72.66, H 9.37, N 5.28.

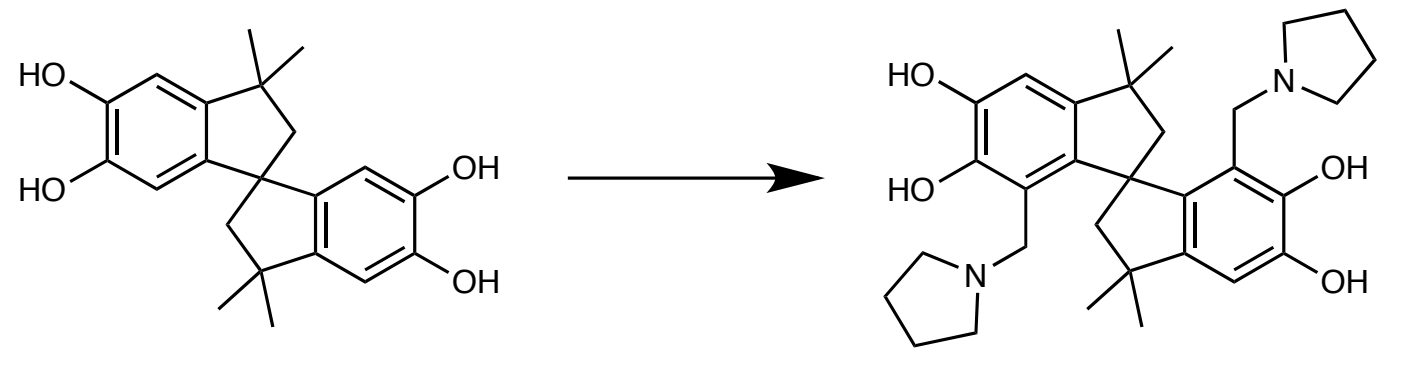

1 4

**4:** To a 500 mL round bottom flask was added ethanol (250 mL), paraformaldehyde (4.45 g, 156 mmol), and pyrrolidine (10 mL, 156 mmol). The reaction was stirred for 10 min at 70 $^{\circ C}$ which point 3,3,3',3'-tetramethyl-1,1'-spirobisindane-5,5',6,6'-tetraol 10 g (30.2 mmol) was added and the reaction was stirred at reflux overnight. The reaction was concentrated by rotary evaporator and product was recrystallized from methylene chloride/hexanes to give 11.5 g of **4** (70%) as a pale white solid. $^1$H NMR ($CDCl_3$, 500 MHz): $\delta$ (ppm) 6.60 (s, 2H), 3.27 (d, 4H, J = 5.6 Hz), 2.27 (d, 2H, J = 13.4 Hz), 2.20 (d, 2H, J = 13.4 Hz), 1.78 (br s, 16H), 1.34 (s, 6H), 1.29 (s, 6H); $^{13}$C{$^1$H} NMR ($CDCl_3$, 125 MHz): δ (ppm) 145.43, 144.28, 141.62, 137.92, 116.30, 106.76, 57.89, 56.71,

53.48, 53.24, 42.73, 32.83, 28.83, 23.45; HR-MS (*m/z*) $[M+H]^+$: Calculated: 503.2904, Found: 503.2905; Elemental Analysis for $C_{31}H_{42}N_2O_4$ Calculated: C 73.49, H 8.36, N 5.53; Found: C 73.31, H 8.33, N 5.69.

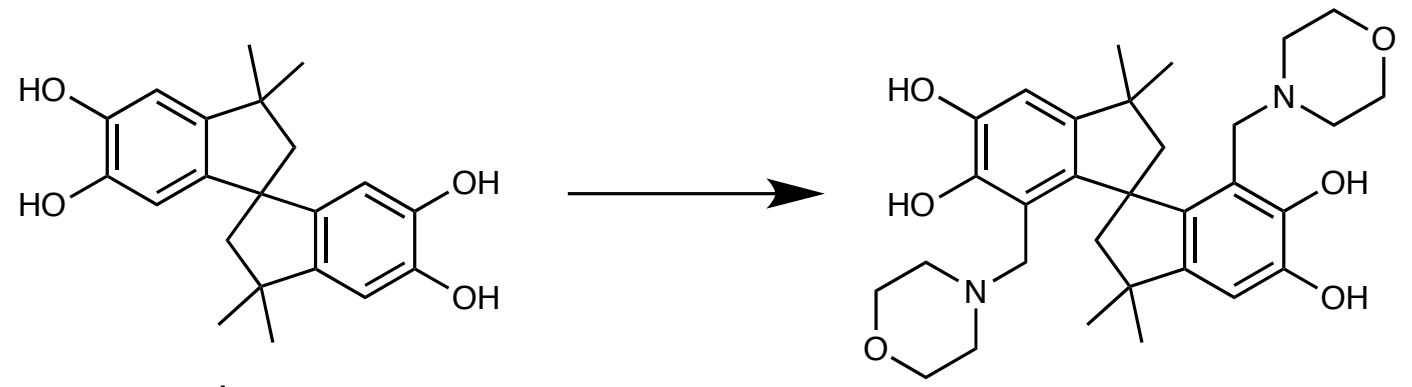

**5:** To a 500 mL round bottom flask was added ethanol (250 mL), paraformaldehyde (4.45 g, 156 mmol), and morpholine (13 mL, 156 mmol). The reaction was stirred for 10 min at 70 °C which point 3,3,3',3'-tetramethyl-1,1'-spirobisindane-5,5',6,6'-tetraol (10 g, 30 mmol) was added and the reaction was stirred at reflux overnight. 500 mL of hexanes was then added to the mixture and the it was cooled to 4 °C. The precipitate was filtered and washed with hexanes and dried to give 6.1 g of **5** (36%) as a white solid. $^1$H NMR ($CDCl_3$, 500 MHz): $\delta$ (ppm) 6.70 (s, 2H), 3.18 (d, 2H, J = 14.6 Hz), 3.08 (d, 2H, J = 14.6 Hz), 2.31 (d, 2H, J = 13.4 Hz), 2.18 (d, 2H, J = 13.4 Hz), 1.65 (br s, 16H), 1.38 (s, 6H), 1.30 (s, 6H); $^{13}C\{^1H\}$ NMR ($CDCl_3$, 125 MHz): δ (ppm) 145.91, 145.16, 141.75, 138.26, 116.00, 108.75, 66.51, 57.86, 56.73, 55.90, 52.86, 42.53, 33.21, 30.05; HR-MS (*m/z*) $[M+H]^+$: Calculated: 539.3116, Found: 539.3117; Elemental Analysis for $C_{31}H_{42}N_2O_6$ Calculated: C 69.12, H 7.86, N 5.20; Found: C 69.14, H 7.97, N 5.11.

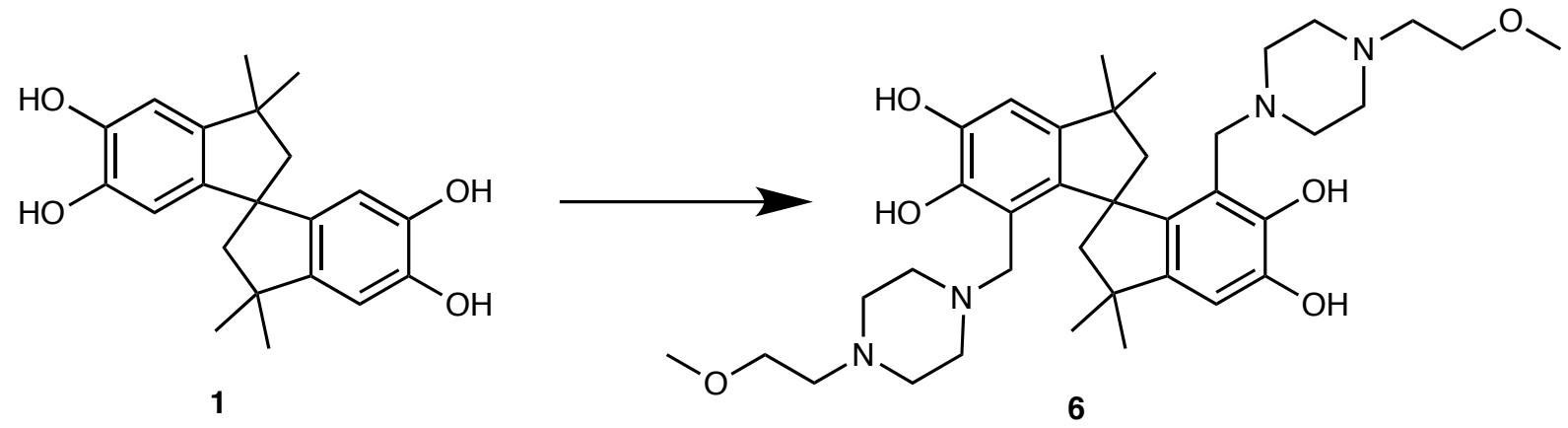

**6:** To a round bottom flask was added toluene (250 mL), paraformaldehyde (4.45 g, 156 mmol), and 2-methoxyethyl-piperazine (10 mL, 156 mmol). The reaction was stirred for 10 min at 70 °C which point 3,3,3',3'-tetramethyl-1,1'-spirobisindane-5,5',6,6'-tetraol (10 g, 30.2 mmol) was added and the reaction was stirred at reflux overnight. The reaction was concentrated by rotary evaporator and triturated with hexanes to give 11.5 g of **6** (95%) as a yellow solid. $^1$H NMR ($CDCl_3$, 500 MHz): $\delta$ (ppm) 6.65 (s, 2H), 3.48 (t, 4H, J = 5.5 Hz), 3.34 (s, 6H), 3.19 (d, 2H, J=14.8 Hz), 3.10 (d, 2H, J=14.8 Hz), 2.57 (t, 4H, J = 5.5 Hz), 2.25 (d, 2H, J = 13.4 Hz), 2.12 (d, 2H, J = 13.4 Hz), 1.34 (s, 6H), 1.29 (s, 6H); $^{13}C\{^1H\}$ NMR ($CDCl_3$, 125 MHz): δ (ppm) 146.11, 145.97, 140.35, 136.76, 116.00, 109.33, 69.94 66.51, 60.23 57.86, 56.73, 55.90, 52.86, 51.10 42.53, 33.21, 30.05; HR-MS (*m/z*) $[M+H]^+$: Calculated: 653.4273, Found: 653.4271; Elemental Analysis for $C_{37}H_{56}N_4O_6$ Calculated: C 68.07, H 8.65, N 8.58; Found: C 67.83, H 8.79, N 8.59.

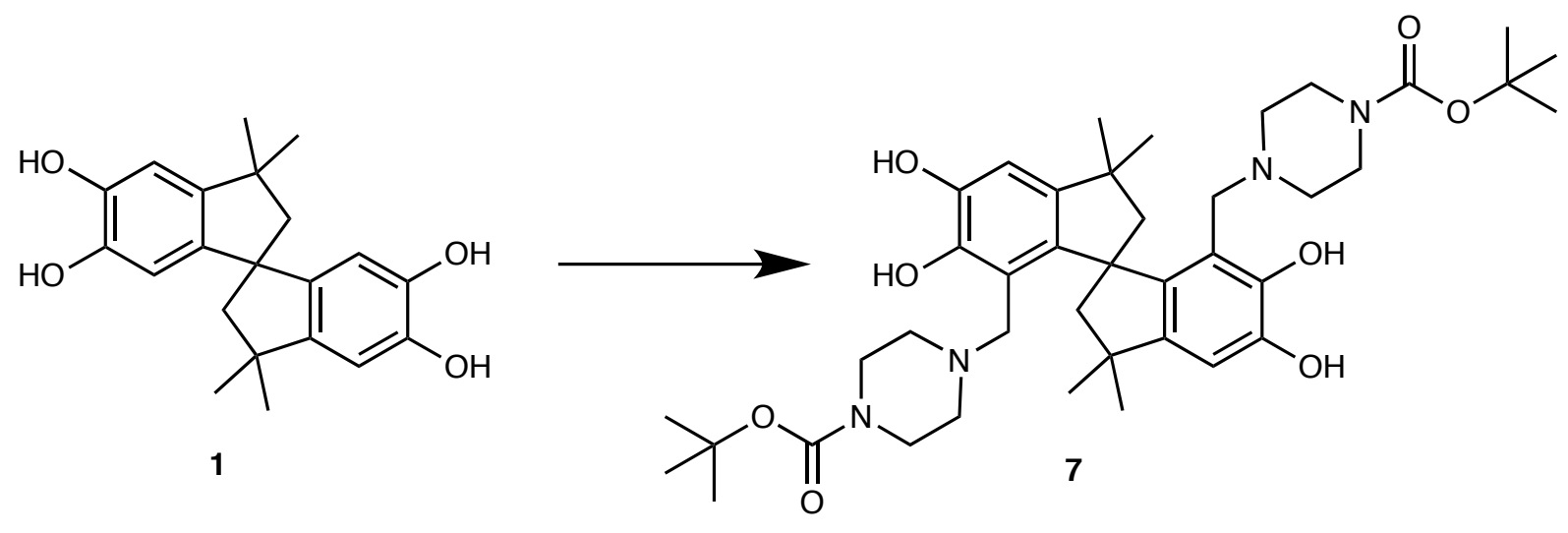

**7:** To a round bottom flask was added toluene (250 mL), paraformaldehyde (4.45 g, 156 mmol) and Boc-Piperazine (10 mL, 156 mmol). The reaction was stirred for 10 min at 70 °C, at which point 3,3,3',3'-tetramethyl-1,1'-spirobisindane-5,5',6,6'-tetraol (10 g, 30.2 mmol) was added and the reaction was stirred at reflux overnight. The reaction was washed with brine, dried over $MgSO_4$, concentrated by rotary evaporator and sonicated in hexanes to give 14.7 g of **7** (98%) as a white solid. $^1$H NMR ($CDCl_3$, 500 MHz): $\delta$ (ppm) 6.68 (s, 2H), 3.15 (d, 4H, J = 15.0 Hz), 3.11 (d, 4H, J = 15.0 Hz), 2.31 (d, 4H, J = 13.1 Hz), 2.16 (d, 2H, J = 13.1 Hz), 1.46 (s, 18H), 1.36 (s, 6H), 1.29 (s, 6H); $^{13}$C{$^1$H} NMR ($CDCl_3$, 125 MHz): δ (ppm) 146.34, 145.16, 141.75, 138.26, 116.00, 108.75, 66.51, 57.86, 57.65, 54.93, 51.43, 44.63, 32.21, 30.15 28.41; HR-MS (*m*/*z*) $[M+H]^+$: Calculated: 737.4484, Found: 737.4307; Elemental Analysis for $C_{41}H_{60}N_4O_8$ Calculated: C 66.82, H 8.21, N 7.60; Found: C 67.11, H 8.11, N 7.22.

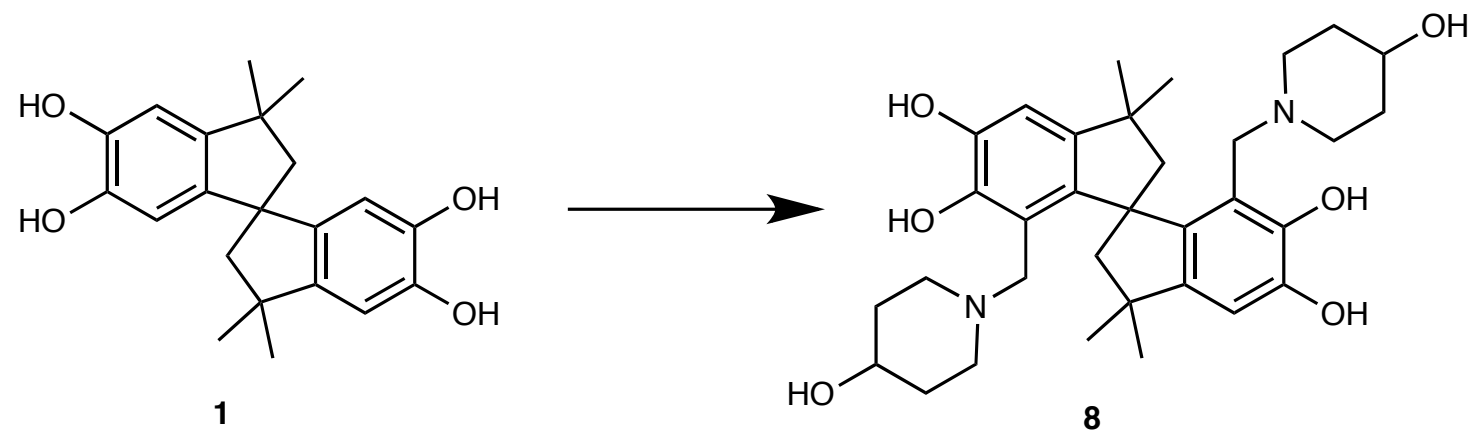

**8:** To a round bottom flask was added ethanol (250 mL), paraformaldehyde (4.4 mg, 147 mmol) and 4-hydroxypiperidine (15.1 g, 150 mmol). The reaction was stirred for 15 min at 70 °C, at which point 3,3,3',3'-tetramethyl-1,1'-spirobisindane-5,5',6,6'-tetraol (10 g, 29.4 mmol) was added and the reaction was stirred at reflux overnight. The reaction was washed with brine, dried over $MgSO_4$, concentrated by rotary evaporator and sonicated in hexanes to give 8.3 g of **8** (50%) as a white solid. $^1$H NMR (DMSO-$d_6$, 500 MHz): $\delta$ (ppm) 6.46 (br s, 2H), 4.63 (br s, 2H), 4.38 (t, 2H, J = 5.4 Hz), 3.44 (t, 2H, J = 5.4 Hz), 2.15 (d, 2H, 13.1 Hz), 2.07 (d, 2H, 13.1 Hz), 1.67 (br s, 4H), 1.28 (br s, 6H), 1.18 (br s, 6H). $^{13}$C{$^1$H} NMR (DMSO-$d_6$, 125 MHz): δ (ppm); 146.43, 145.11, 141.25, 137.99, 116.40, 108.56, 57.78, 56.69, 42.46, 33.17, 30.28; HR-MS (*m*/*z*) $[M+H]^+$: Calculated: 567.3429, Found: 567.3292; Elemental Analysis for $C_{41}H_{60}N_4O_8$ Calculated: C 66.82, H 8.21, N 7.60; Found: C 67.11, H 8.11, N 7.22.

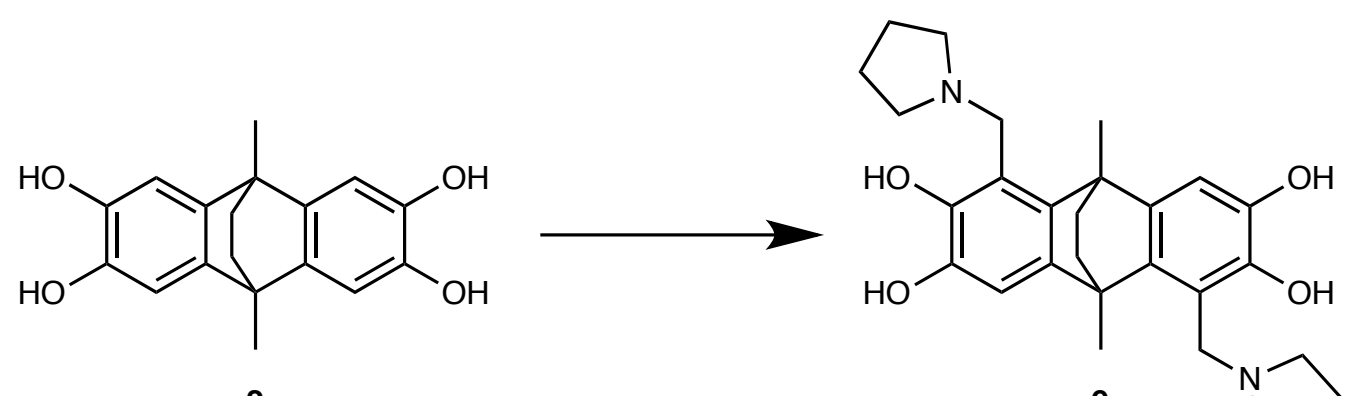

**2** **9**

**9:** To a sealable reaction vial was added toluene (12.5 mL), paraformaldehyde (500 mg, 16.65 mmol), pyrrolidine (1.40 mL, 16.76 mmol), and 9,10-dimethyl-9,10-dihydro-9,10-ethanoanthracene-2,3,6,7-tetraol (500 mg, 1.68 mmol). After sealing, the vial was heated for 8 h at 120 °C while stirring. The reaction mixture was subsequently concentrated *in vacuo*, prior to the addition of hexanes (50 mL) to precipitate the product. The product was isolated by filtration and washed with ethanol (50 mL) to give **9** as a pale pink solid (219 mg) in 28% yield after drying. $^{1}H$ NMR ($CDCl_3$, 500 MHz): $\delta$ (ppm) 6.85 (s, 2H), 4.31 (d, 2H, J = 14.1 Hz), 4.26 (d, 2H, J = 14.1 Hz), 2.67 (br s, 8H), 2.03 (s, 6H), 1.85 (br s, 12H), 1.74 (s, 2H), 1.44 (s, 2H); $^{13}C\{^{1}H\}$ NMR ($CDCl_3$, 125 MHz): δ (ppm) 143.98, 141.54, 139.71, 134.57, 117.90, 107.40, 54.77, 53.29, 43.94, 38.27, 27.04, 23.78; HR-MS (*m/z*) $[M+H]^+$: Calculated: 507.3217, Found: 507.3105.

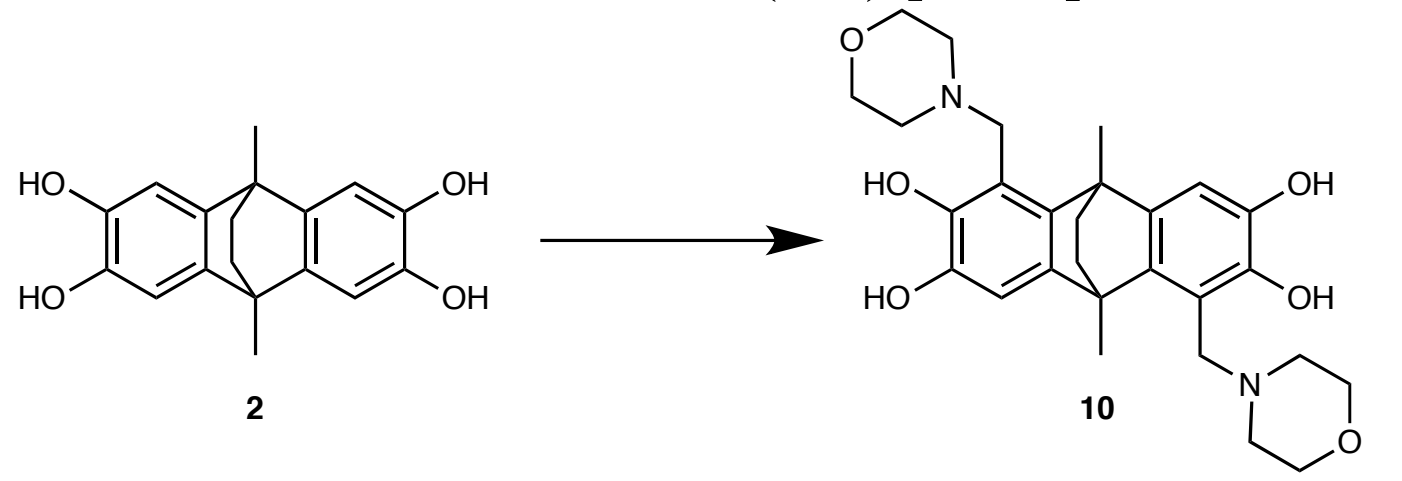

**2** **10**

**10:** To a sealable reaction vial was added toluene (12.5 mL), paraformaldehyde (500 mg, 16.65 mmol), morpholine (1.45 mL, 16.76 mmol), and 9,10-dimethyl-9,10-dihydro-9,10-ethanoanthracene-2,3,6,7-tetraol (500 mg, 1.68 mmol). After sealing, the vial was heated for 8 h at 120 °C while stirring. The reaction mixture was subsequently concentrated *in vacuo*, prior to the addition of hexanes (50 mL) to precipitate the product. The product was isolated by filtration and washed with ethanol (50 mL) to give **10** as an off-white solid (521 mg) in 63% yield after drying. $^{1}H$ NMR ($CDCl_3$, 500 MHz): $\delta$ (ppm) 6.87 (s, 2H), 4.18 (d, 2H, J = 13.8 Hz), 4.12 (d, 2H, J = 13.8 Hz), 3.72 (br s, 8H), 2.61 (br s, 8H), 2.05 (s, 6H), 1.74 (s, 2H), 1.46 (s, 2H); $^{13}C\{^{1}H\}$ NMR ($CDCl_3$, 125 MHz): δ (ppm) 143.05, 141.72, 140.44, 135.41, 116.80, 108.07, 66.89, 57.08, 52.72, 43.94, 38.07; HR-MS (*m/z*) $[M+H]^+$: Calculated: 497.2646, Found: 497.2527.

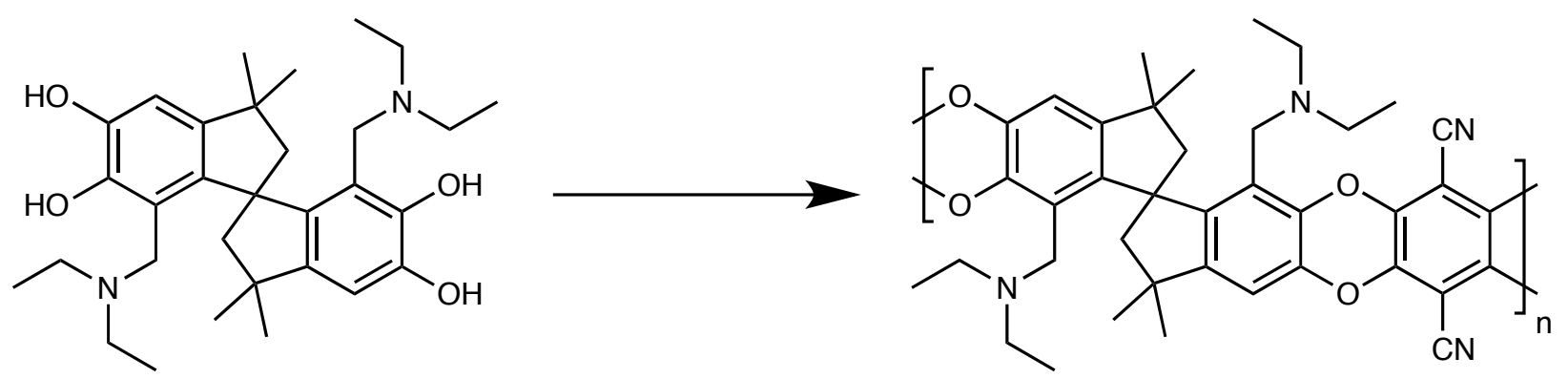

**3** **11**

**11: 3** (11 g, 22 mmol) and tetrafluoroterephthalonitrile (4.4 g, 22 mmol) were dissolved in anhydrous DMF (100 mL) at 65 °C. Freshly ground potassium carbonate (12 g, 87 mmol) was slowly added and the mixture stirred for 48 h. The reaction mixture was then poured into water (500 mL) to precipitate the crude polymer as a yellow solid, which was then filtered and washed with an additional portion of water (250 mL). The crude polymer was dissolved in hot chloroform (200 mL) and subsequently precipitated into methanol (1.0 L), filtered, and washed with an additional portion of methanol (200 mL). This process was repeated. After drying *in vacuo* at 65 °C, **11** was isolated as a yellow solid (12 g) in 85% yield. $^{1}$H NMR ($CDCl_3$, 500 MHz): $\delta$ (ppm) 6.80 (br s, 2H), 3.53 (br s, 4H), 3.16 (br s, 2H), 2.99 (br s, 2H), 2.24 (br s, 8H), 1.39 (br s, 6H), 1.31 (br s, 6H), 0.69 (br s, 12H); Elemental Analysis for $C_{39}H_{42}N_4O_4$ Calculated: C 74.26, H 6.71, N 8.88; Found: C 69.91 H 5.90, N 8.16; SEC (THF): $M_n$ = 63.7 kg mol$^{-1}$, $M_w$ = 80.3 kg mol$^{-1}$, *PDI* = 1.53; BET surface area: 446 m$^2$ g$^{-1}$.

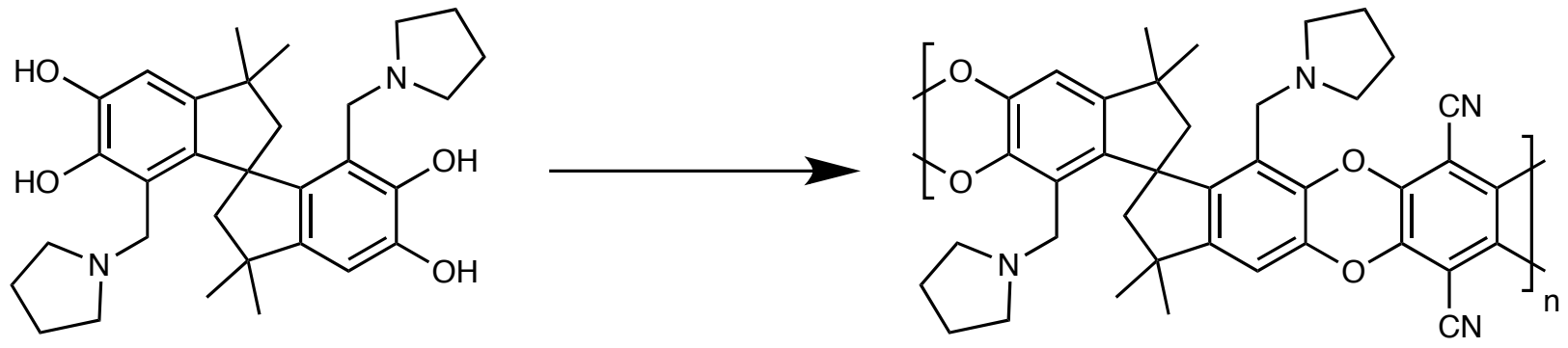

**4** **12**

**12: 4** (8.8 g, 17.5 mmol) and tetrafluoroterephthalonitrile (3.5 g, 17.4 mmol) were dissolved in anhydrous DMF (50 mL) at 65 °C. Freshly ground potassium carbonate (9.7 g, 70 mmol) was slowly added and the mixture stirred for 48 h. The reaction mixture was poured into water (800 mL) to precipitate the crude polymer as a yellow solid, which was then filtered and washed with an additional portion of water (200 mL). The crude polymer was dissolved in hot THF (200 mL) and subsequently precipitated into acetone (1.0 L), filtered, and washed with an additional portion of methanol (200 mL). An additional precipitation of the polymer into methanol (1.0 L) was conducted from a solution **12** in hot chloroform (200 mL). After drying *in vacuo* at 65 °C, **12** was isolated as a yellow solid (9.6 g) in 92% yield. $^{1}$H NMR ($CDCl_3$, 500 MHz): $\delta$ (ppm) 7.01 (s, 2H), 2.33 (br s, 4H), 1.99 (br s, 4H), 1.63 (br s, 16H), 1.42 (br s, 12H); Elemental Analysis for $C_{39}H_{38}N_4O_4$ Calculated: C 74.74, H 6.11, N 8.94; Found: C 73.91 H 5.90, N 8.16; SEC (THF): $M_n$ = 126 kg mol$^{-1}$, $M_w$ = 146 kg mol$^{-1}$, *PDI* = 1.16; BET surface area: 434 m$^2$ g$^{-1}$.

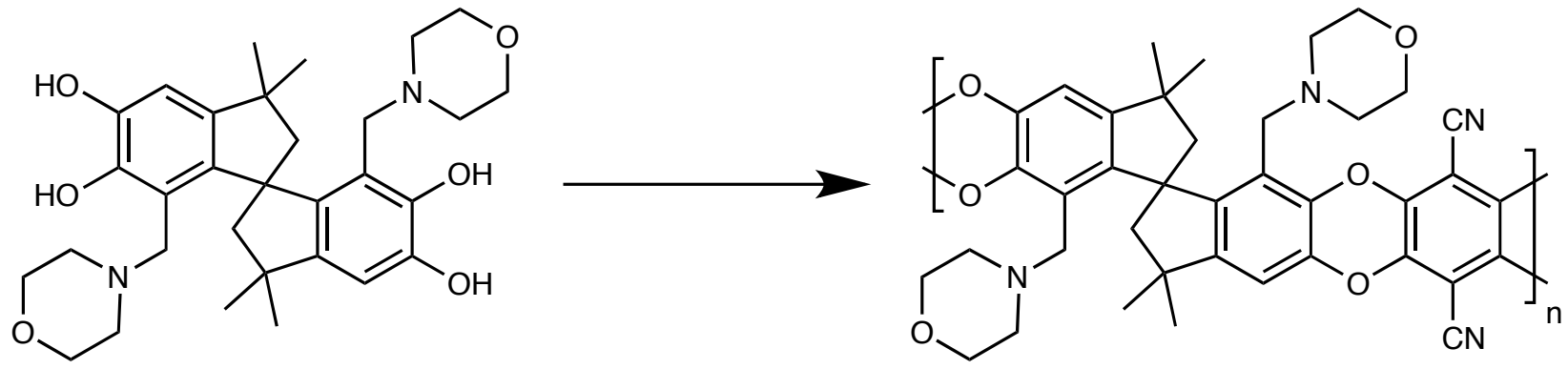

5 13

**13:** **5** (2.03 g, 3.77 mmol) and tetrafluoroterephthalonitrile (0.75 g, 3.77 mmol) were dissolved in anhydrous DMF (38 mL) at 65 °C. Freshly ground potassium carbonate (2.08 g, 15.07 mmol) was slowly added and the mixture stirred for 48 h. The reaction mixture was poured into water (400 mL) to precipitate the crude polymer as a yellow solid, which was then filtered and washed with an additional portion of water (200 mL). The crude polymer was dissolved in hot THF (50 mL) and subsequently precipitated into methanol (500 mL), filtered, and washed with an additional portion of methanol (200 mL). This process was repeated in hot chloroform (50 mL) and methanol (500 mL). After drying *in vacuo* at 65 °C, **13** was isolated as a yellow-brown solid (1.94 g) in 78% yield. $^1$H NMR ($CDCl_3$, 500 MHz): $\delta$ (ppm) 6.80 (br s, 2H), 3.53 (br s, 4H), 3.16 (br s, 2H), 2.99 (br s, 2H), 2.24 (br s, 16H), 1.39 (br s, 6H), 1.31 (br s, 6H); Elemental Analysis for $C_{39}H_{38}N_4O_6$ Calculated: C 71.11, H 5.81, N 8.51; Found: C 69.91 H 5.90, N 8.16; SEC (THF): $M_n$ = 66.7 kg mol$^{-1}$, $M_w$ = 121 kg mol$^{-1}$, *PDI* = 1.81; BET surface area: 505 m$^2$ g$^{-1}$.

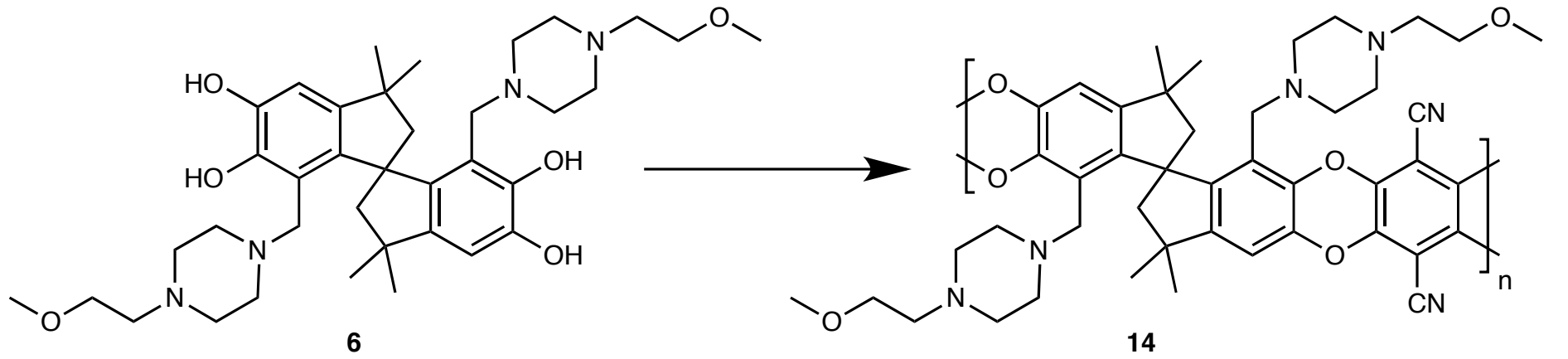

6 14

**14:** **6** (5.0 g, 7.7 mmol) and tetrafluoroterephthalonitrile (1.5 g, 7.7 mmol) were dissolved in anhydrous DMF (50 mL) at 65 °C. Freshly ground potassium carbonate (4.2 g, 31 mmol) was then slowly added and the mixture stirred for 3 h. The reaction mixture was poured into water (200 mL) to precipitate the crude polymer as a yellow solid, which was then filtered and washed with an additional portion of water (200 mL). The crude polymer was dissolved in hot chloroform (75 mL) and subsequently precipitated into methanol (750 mL), filtered, and washed with an additional portion of methanol (200 mL). This process was repeated. After drying *in vacuo* at 65 °C, **14** was isolated as a yellow solid (3.4 g) in 54% yield. $^1$H NMR (DMSO-$d_6$, 500 MHz): $\delta$ (ppm) 7.00 (br s, 2H), 3.36 (br s, 6H), 3.19 (br s, 12H), 2.13–2.57 (br m, 20H), 1.48 (br s, 6H), 1.38 (br s, 6H); Elemental Analysis for $C_{45}H_{52}N_6O_6$ Calculated: C 69.93, H 6.78, N 10.87; Found: C 69.77 H 6.75, N 10.53; SEC (THF): $M_n$ = 26.1 kg mol$^{-1}$, $M_w$ = 29.9 kg mol$^{-1}$, *PDI* = 1.15; BET surface area: 63 m$^2$ g$^{-1}$.

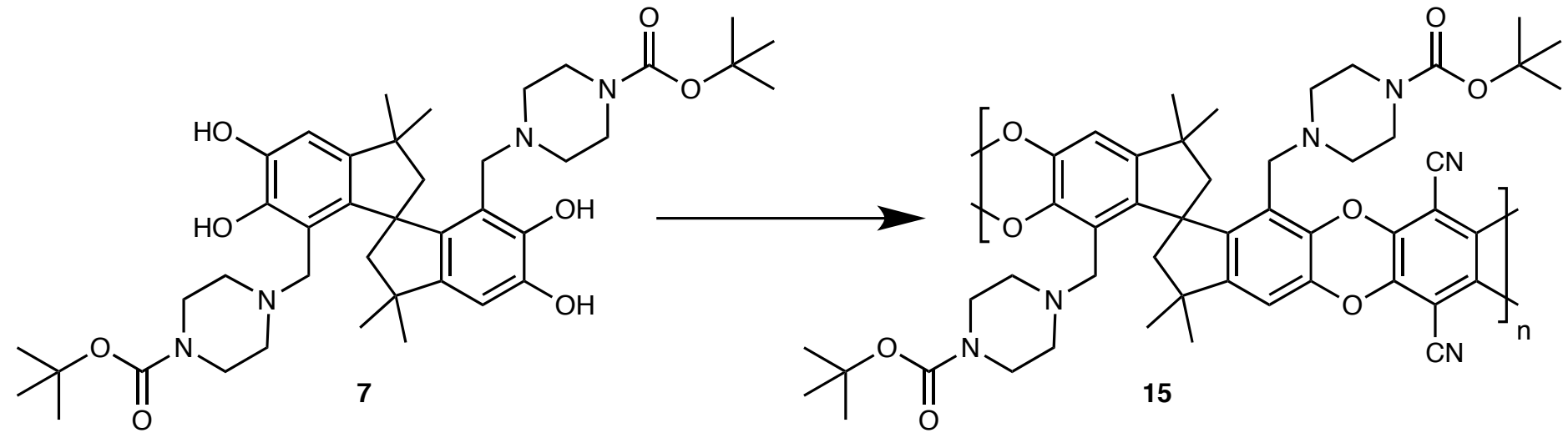

**15: 7** (5.0 g, 6.8 mmol) and tetrafluoroterephthalonitrile (1.4 g, 6.8 mmol) were dissolved in anhydrous DMF (60 mL) at 65 °C. Freshly ground potassium carbonate (3.8 g, 27 mmol) was then slowly added and the mixture stirred for 3 h. The reaction mixture was poured into water (200 mL) to precipitate the crude polymer as a yellow solid, which was then filtered and washed with an additional portion of water (200 mL). The crude polymer was dissolved in hot THF (100 mL) and subsequently precipitated into methanol (500 mL), filtered, and washed with an additional portion of methanol (200 mL). This process was repeated in hot chloroform (100 mL) and methanol (500 mL). After drying *in vacuo* at 65 °C, **15** was isolated as a yellow solid (4.7 g) in 82% yield. $^{1}$H NMR ($CDCl_3$, 500 MHz): $\delta$ (ppm) 6.80 (br s, 2H), 3.53 (br s, 4H), 3.16 (br s, 2H), 2.99 (br s, 2H), 2.24 (br s, 16H), 1.31–1.45 (br m, 30H); Elemental Analysis for $C_{49}H_{56}N_6O_8$ Calculated: C 68.67, H 6.59, N 9.81; Found: C 71.10 H 6.54, N 8.25; SEC (THF): $M_n$ = 46.9 kg mol$^{-1}$, $M_w$ = 51.5 kg mol$^{-1}$, *PDI* = 1.11; BET surface area: 234 m$^2$ g$^{-1}$.

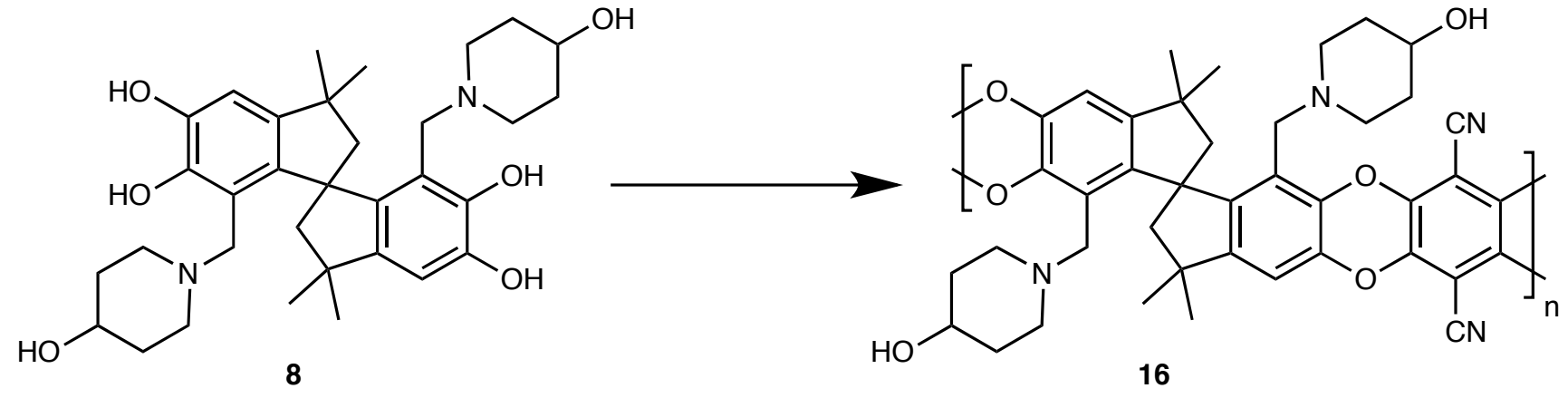

**16: 8** (1.0 g, 1.8 mmol) and tetrafluoroterephthalonitrile (352 mg, 1.8 mmol) were dissolved in anhydrous DMF (10 mL) at 65 °C. Freshly ground potassium carbonate 971 mg (7.0 mmol) was then slowly added and the mixture stirred for 48 h. The reaction mixture was poured into water (100 mL) to precipitate the crude polymer as a yellow solid, which was then filtered and washed with an additional portion of water (100 mL). The crude polymer was dissolved in hot chloroform (20 mL) and subsequently precipitated into methanol (200 mL), filtered, and washed with an additional portion of methanol (200 mL). This process was repeated. After drying *in vacuo* at 65 °C, **16** was isolated as a yellow solid (600 mg) in 45% yield. $^{1}$H NMR (DMSO-$d_6$, 500 MHz): $\delta$ (ppm) 7.09 (br s, 2H), 4.53 (br s, 2H), 2.92 (br s, 4H), 2.03-2.45 (br m, 8H), 1.85 (br s, 4H), 1.56 (br s, 10H), 1.39 (br s, 6H), 1.29 (br s, 6H); SEC (DMF): $M_n$ = 11.9 kg mol$^{-1}$, $M_w$ = 19.2 kg mol$^{-1}$, *PDI* = 1.61; BET surface area: 11 m$^2$ g$^{-1}$.

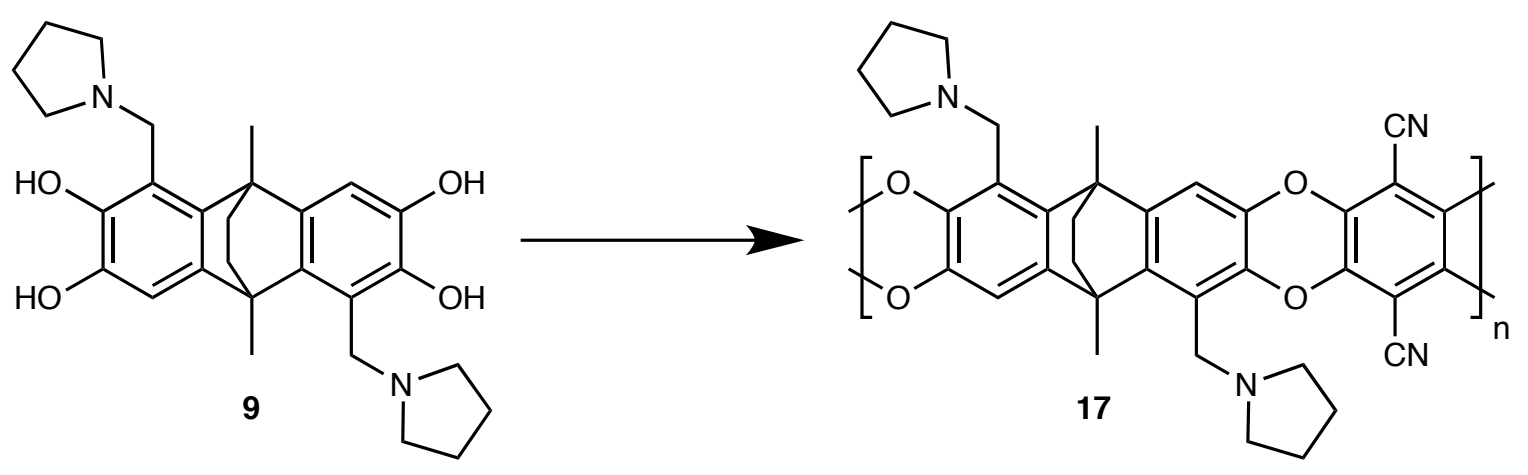

**17: 9** (3.0 g, 5.6 mmol) and tetrafluoroterephthalonitrile (1.2 g, 5.6 mmol) were dissolved in anhydrous DMF (20 mL) at 65 °C. Freshly ground potassium carbonate (3.1 g, 22.4 mmol) was then slowly added and the mixture stirred for 48 h. The reaction mixture was poured into water (50 mL) to precipitate the crude polymer as an orange solid, which was then filtered and washed with an additional portion of water (50 mL). The crude polymer was dissolved in hot THF (5 mL) and subsequently precipitated into methanol (50 mL), filtered, and washed with an additional portion of methanol (50 mL). This process was repeated from hot chloroform (5 mL) using methanol (50 mL). After drying *in vacuo* at 65 °C, **17** was isolated as an orange solid (214 mg) in 90% yield. $^1$H NMR ($CDCl_3$, 500 MHz): $\delta$ (ppm) 7.00 (br s, 2H), 3.89 (br d, 4H), 2.55 (br s, 8H), 2.32 (br s, 4H), 1.75 (br s, 6H), 1.58 (br s, 8H); Elemental Analysis for $C_{36}H_{32}N_4O_4$ Calculated: C 73.95, H 5.52, N 9.58; BET surface area: 444 m$^2$ g$^{-1}$.

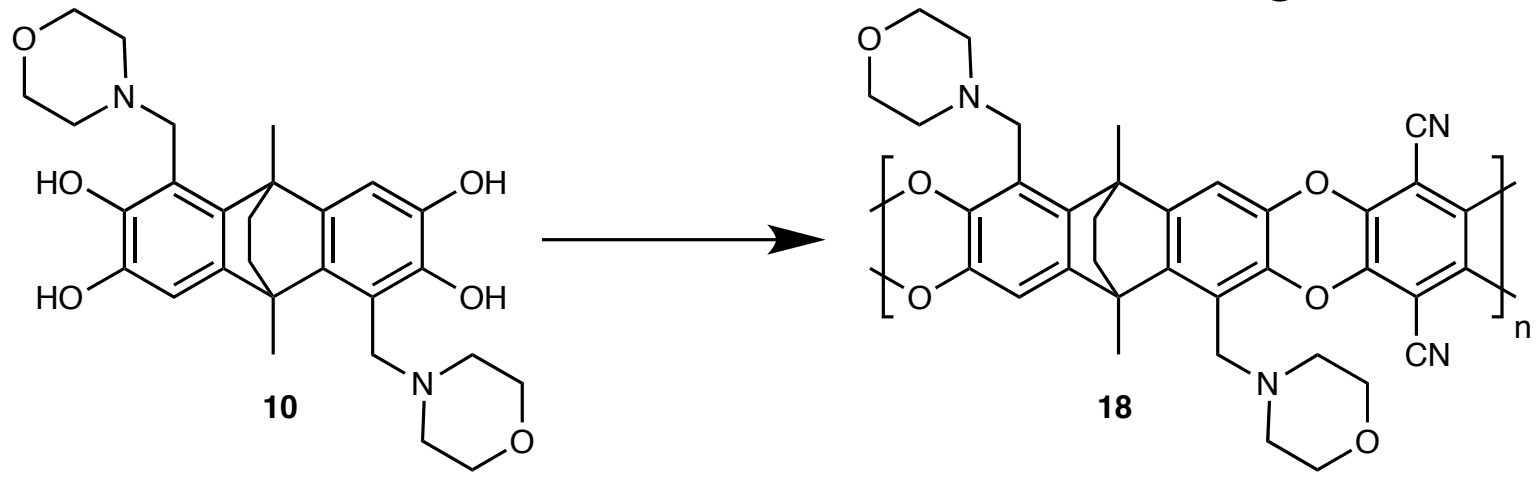

**18: 10** (200 mg, 402 μmol) and tetrafluoroterephthalonitrile (82.9 mg, 414 μmol) were dissolved in anhydrous DMF (3.4 mL) at 65 °C. Freshly ground potassium carbonate (229 mg, 1.66 mmol) was then slowly added and the mixture was stirred for 96 h. The reaction mixture was poured into water (50 mL) to precipitate the crude polymer as a yellow solid, which was then filtered and washed with an additional portion of water (50 mL). The crude polymer was dissolved in hot THF (5 mL) and subsequently precipitated into methanol (50 mL), filtered, and washed with an additional portion of methanol (50 mL). This process was repeated in hot chloroform (5 mL) and methanol (50 mL). After drying *in vacuo* at 65 °C, **18** was isolated as a yellow solid (211 mg) in 80% yield. $^1$H NMR ($CDCl_3$, 500 MHz): $\delta$ (ppm) 7.00 (br s, 2H), 3.66 (br s, 12H), 2.48 (br s, 8H), 2.29 (br s, 6H), 1.84 (br s, 2H), 1.47 (br s, 2H); Elemental Analysis for $C_{36}H_{32}N_4O_6$ Calculated: C 70.12, H 5.23, N 9.09; BET surface area: 19 m$^2$ g$^{-1}$.

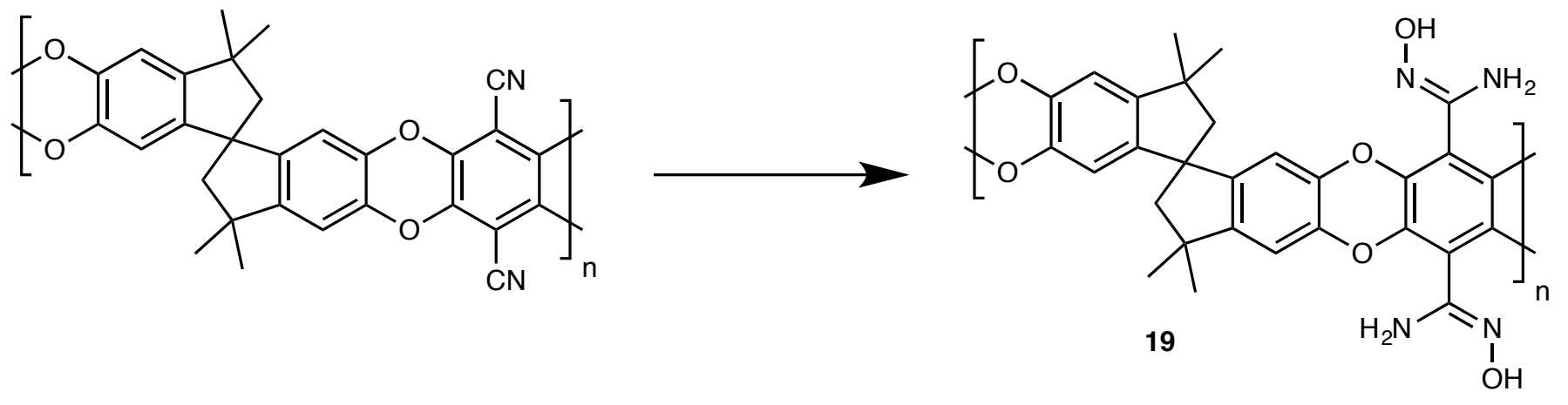

**19:** To a round bottom flask was added PIM-1 (5.0 g, 9.6 mmol) and tetrahydrofuran (380 mL). The mixture was heated to at 60 °C to dissolve the polymer. A solution of 50 wt% aqueous hydroxylamine (50 mL, 760 mmol) was then added dropwise. The reaction mixture was stirred overnight and then cooled before precipitating the polymer into methanol (1320 mL). The solid was filtered and then rinsed twice with methanol (500 mL) before drying *in vacuo* at 50 °C to give **19** as a pale yellow powder (3.93 g) in 69% yield. $^{1}$H NMR (DMSO-$d_6$, 500 MHz): $\delta$ (ppm) 9.47 (br t, 2H), 6.82 (br s, 2H), 6.16 (br s, 2H), 5.82 (br s, 4H), 2.17 (br d, 4H), 1.32 (br s, 6H), 1.25 (br s, 6H); Elemental Analysis for $C_{29}H_{26}N_4O_6$ Calculated: C 66.15, H 4.98, N 10.64; Found: C 64.62 H 5.09, N 10.17; SEC (DMF): $M_n$ = 42.7 kg mol$^{-1}$, $M_w$ = 119 kg mol$^{-1}$, *PDI* = 2.78; BET surface area: 454 m$^2$ g$^{-1}$.

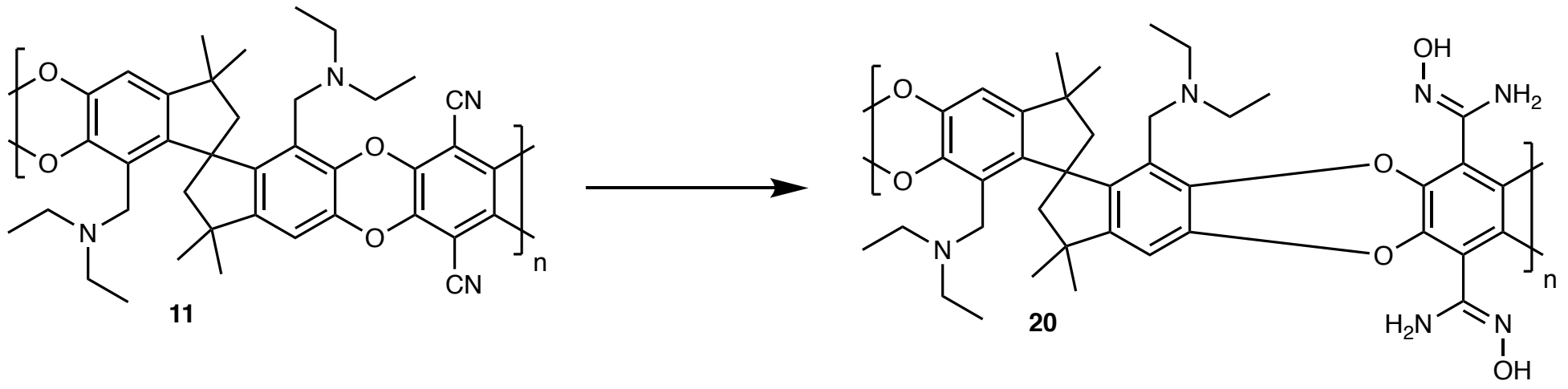

**20:** To a round bottom flask was added **11** (7.5 g, 12.5 mmol) and tetrahydrofuran (375 mL). The mixture was heated to at 60 °C to dissolve the polymer. A solution of 50 wt% aqueous hydroxylamine (60 mL, 875 mmol) was then added dropwise. The reaction mixture was stirred overnight and then cooled before precipitating the polymer into a 1:1 (v/v) solution of water and methanol (1875 mL). The solid was filtered and then rinsed twice with methanol (500 mL) before drying *in vacuo* at 50 °C to give **20** as a white powder (1.45 g) in 18% yield. $^{1}$H NMR (DMSO-$d_6$, 500 MHz): $\delta$ (ppm) 9.48 (br s, 2H), 6.74 (br s, 2H), 5.81 (br s, 4H), 3.63 (br s, 4H), 2.91 (br s, 4H), 2.18 (br s, 8H), 1.31 (br s, 6H), 1.26 (br s, 6H), 0.69 (br s, 12H); SEC (DMF): $M_n$ = 5.0 kg mol$^{-1}$, $M_w$ = 14.2 kg mol$^{-1}$, *PDI* = 2.83; BET surface area: 79 m$^2$ g$^{-1}$.

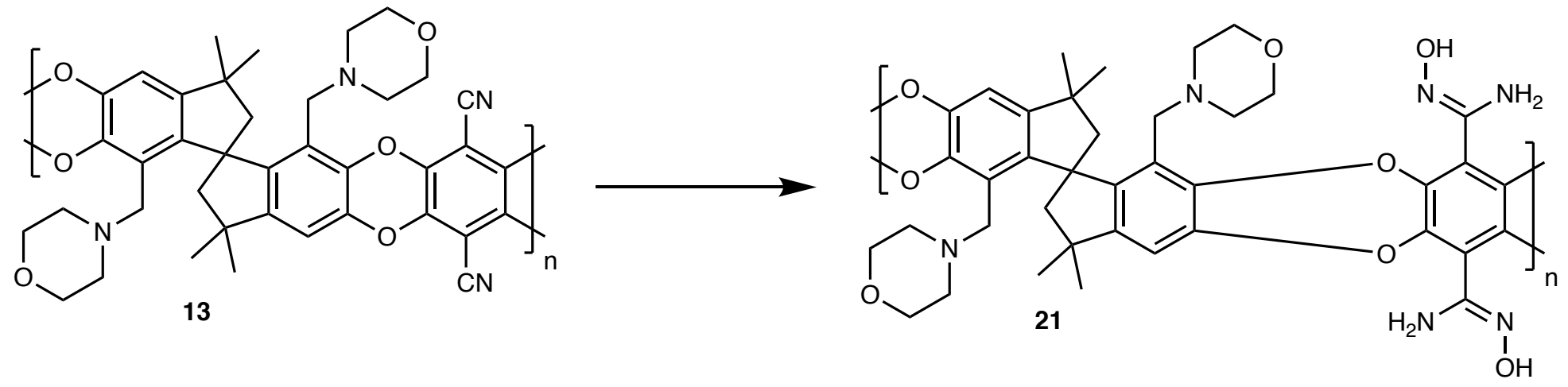

**21:** To a round bottom flask was added **13** (2.7 g, 3.8 mmol) and tetrahydrofuran (120 mL). The mixture was heated to at 60 °C to dissolve the polymer. A solution of 50 wt% aqueous hydroxylamine (20 mL, 280 mmol) was then added dropwise. The reaction mixture was stirred overnight and then cooled before precipitating the polymer into methanol (400 mL). The solid was filtered and then rinsed twice with methanol (500 mL) before drying *in vacuo* at 50 °C to give **21** as a white powder (2.0 g) in 68% yield. $^1$H NMR (DMSO-$d_6$, 500 MHz): $\delta$ (ppm) 9.50 (br s, 2H), 6.79 (br s, 2H), 5.86 (br s, 4H), 3.43 (br s, 4H), 3.02 (br s, 2H), 2.84 (br s, 2H) 2.11 (br s, 16H), 1.36 (br s, 6H), 1.26 (br s, 6H); SEC (DMF): $M_n$ = 29.7 kg mol$^{-1}$, $M_w$ = 62.8 kg mol$^{-1}$, *PDI* = 2.12; BET Surface area: 152 m$^2$ g$^{-1}$.

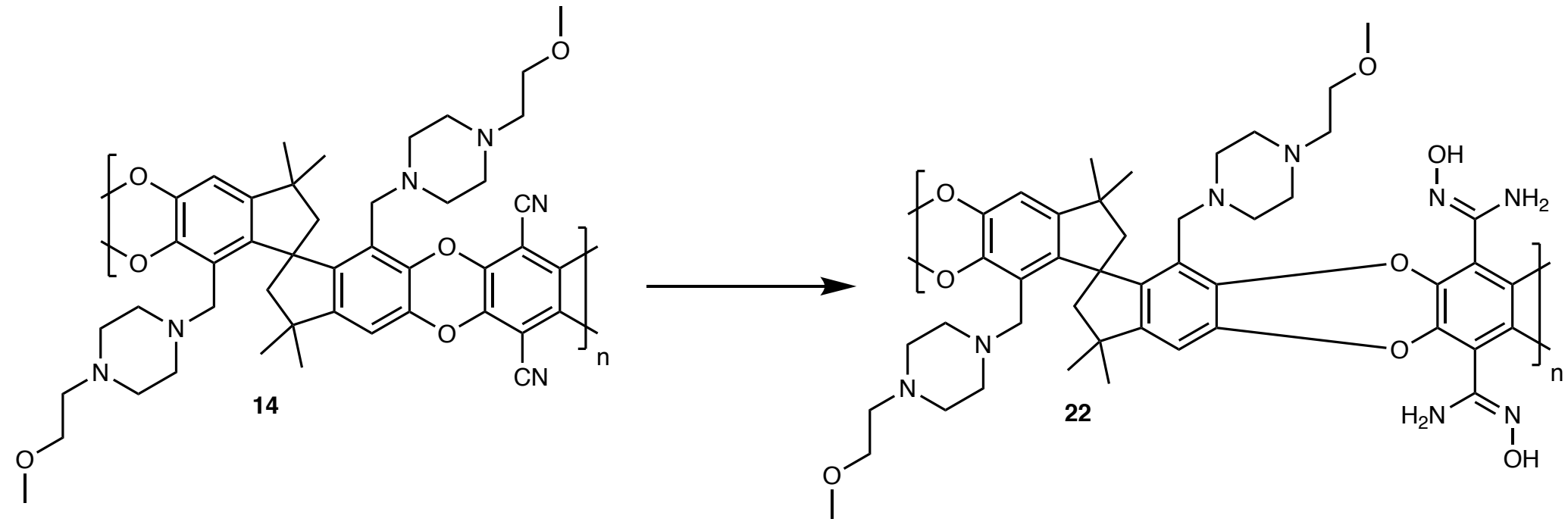

**22:** To a round bottom flask was added **14** (1.5 g, 1.4 mmol) and tetrahydrofuran (75 mL). The mixture was heated to at 60 °C to dissolve the polymer. A solution of 50 wt% aqueous hydroxylamine (12 mL, 95 mmol) was then added dropwise. The reaction mixture was stirred overnight and then cooled before precipitating the polymer into a 1:1 (v/v) solution of water and methanol (450 mL). The solid was filtered and then rinsed twice with methanol (500 mL) before drying *in vacuo* at 50 °C to give **22** as a pale yellow powder (1.2 g) in 74% yield. $^1$H NMR (DMSO-$d_6$, 500 MHz): $\delta$ (ppm) 9.47 (br s, 2H), 6.77 (br s, 2H), 5.85 (br s, 4H), 3.17 (br s, 14H), 1.86–2.54 (br m, 24H), 1.37 (br s, 6H), 1.25 (br s, 6H); SEC (DMF): $M_n$ = 11.5 kg mol$^{-1}$, $M_w$ = 22.6 kg mol$^{-1}$, *PDI* = 1.96; BET Surface area: 10 m$^2$ g$^{-1}$.

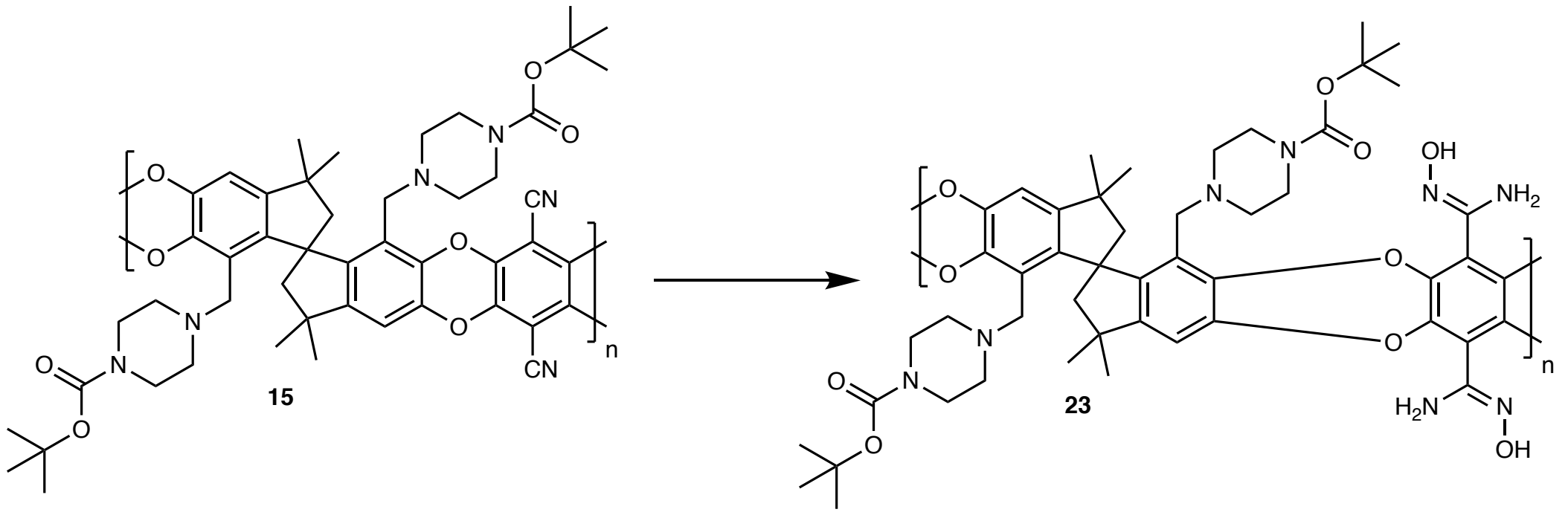

**23:** To a round bottom flask was added **15** (1.0 g, 1.2 mmol) and tetrahydrofuran (60 mL). The mixture was heated to at 60 °C to dissolve the polymer. A solution of 50 wt% aqueous hydroxylamine (11 mL, 163 mmol) was then added dropwise. The reaction mixture was stirred overnight and then cooled before precipitating the polymer into water (600 mL). The solid was filtered and then rinsed twice with methanol (500 mL) before drying *in vacuo* at 50 °C to give **23** as a pale yellow powder (950 mg) in 89% yield. $^1$H NMR (DMSO-$d_6$, 500 MHz): $\delta$ (ppm) 7.80 (br s, 2H), 6.81 (br s, 2H), 5.75 (br s, 4H), 3.12 (br s, 8H), 2.03 (br s, 16H), 1.18–1.46 (br m, 30H)); SEC (DMF): $M_n$ = 8.0 kg mol$^{-1}$, $M_w$ = 12.4 kg mol$^{-1}$, $PDI$ = 1.56; BET Surface area: 12 m$^2$ g$^{-1}$.

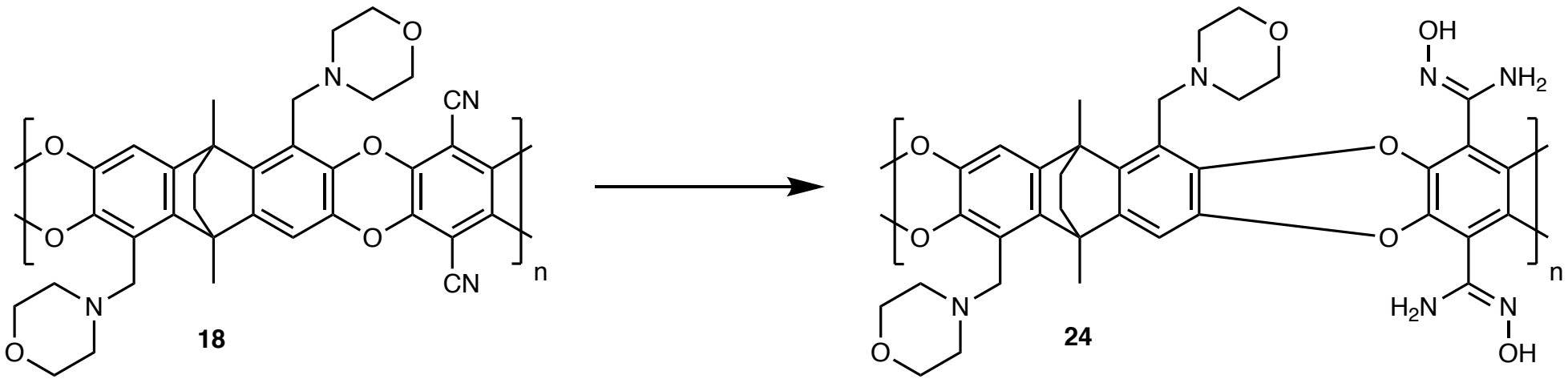

**24:** To a round bottom flask was added **18** (350 mg, 0.57 mmol) and tetrahydrofuran (22 mL). The mixture was heated to 70 °C to dissolve the polymer. A solution of 50 wt% aqueous hydroxylamine (2.623 mL, 39.7 mmol) was then added dropwise. The reaction was stirred for 72 h and then cooled before precipitating the polymer in methanol (120 mL). The solid was filtered and rinsed twice with methanol (50 mL) before drying *in vacuo* 50 °C to give **24** as a white powder (322.9 mg) in 82% yield. $^1$H NMR (DMSO-$d_6$, 500 MHz): $\delta$ (ppm) 9.57 (br s, 2H), 6.77 (br s, 4H), 5.83 (br s, 4H), 3.49 (br t, 8H), 2.23–1.33 (broad m, 22H); SEC (DMF): $M_n$ = 22.7 kg mol$^{-1}$, $M_w$ = 36.7 kg mol$^{-1}$, $PDI$ = 1.52; BET Surface area: 10 m$^2$ g$^{-1}$.

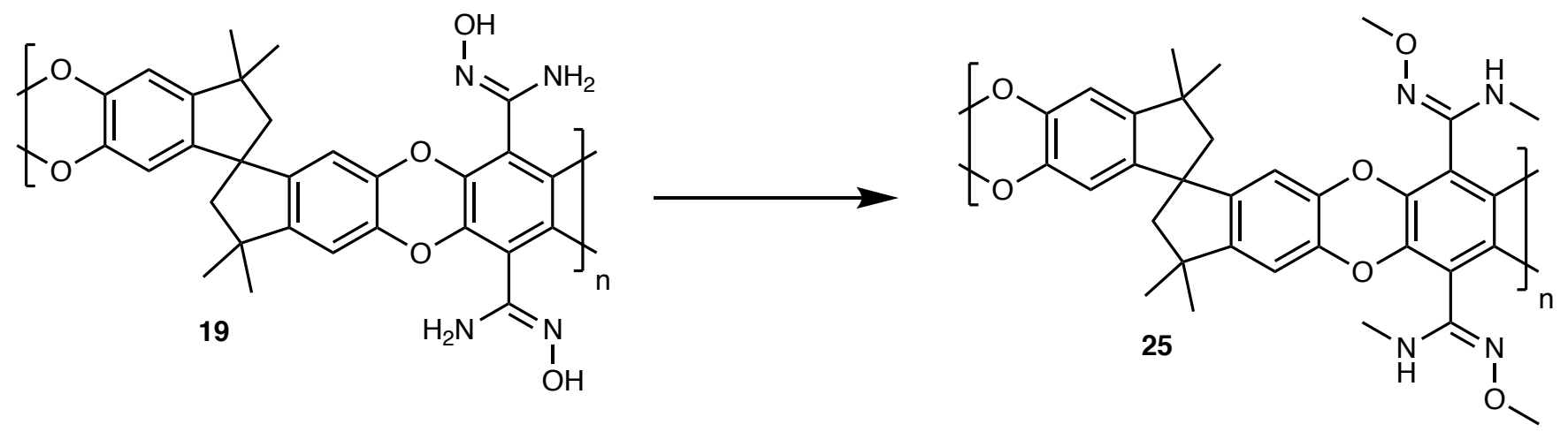

**25:** To a solution of **19** (1.50 g, 2.6 mmol) in dimethyl sulfoxide (57 mL) was added a solution of lithium hydroxide monohydrate (0.478 g, 11.40 mmol) in MilliQ water (5.75 mL) dropwise over 1 min. After stirring for 1 h under nitrogen, the reaction mixture was placed on ice, and dimethyl sulfate (1.08 mL, 11.40 mmol) was added in two portions. The reaction was allowed to reach room temperature and stirred for 72 h, before quenching with sufficient sodium hydroxide to increase the pH above 7. The product was precipitated in MilliQ water (400 mL), filtered and rinsed with two portions of MilliQ water (100 mL each), and then dried *in vacuo* at 60 °C to give **25** as a dark yellow solid (1.28 g) in 81% yield. $^{1}$H NMR (DMSO-$d_6$, 500 MHz): $\delta$ (ppm) 6.67 (br s, 2H), 6.28 (br s, 2H), 5.83 (br s, 2H), 3.42–4.08 (br m, 6H), 2.79–3.38 (br m, 6H), 2.20 (br d, 4H), 1.28 (br s, 6H), 1.33 (br s, 6H); Elemental Analysis for $C_{33}H_{34}N_4O_6$ Calculated: C 68.03, H 5.88, N 9.62; Found: C 65.21 H 5.76, N 8.93; SEC (DMF): $M_n$ = 17.5 kg mol$^{-1}$, $M_w$ = 102 kg mol$^{-1}$, *PDI* = 5.85; BET Surface area: 7 m$^2$ g$^{-1}$.

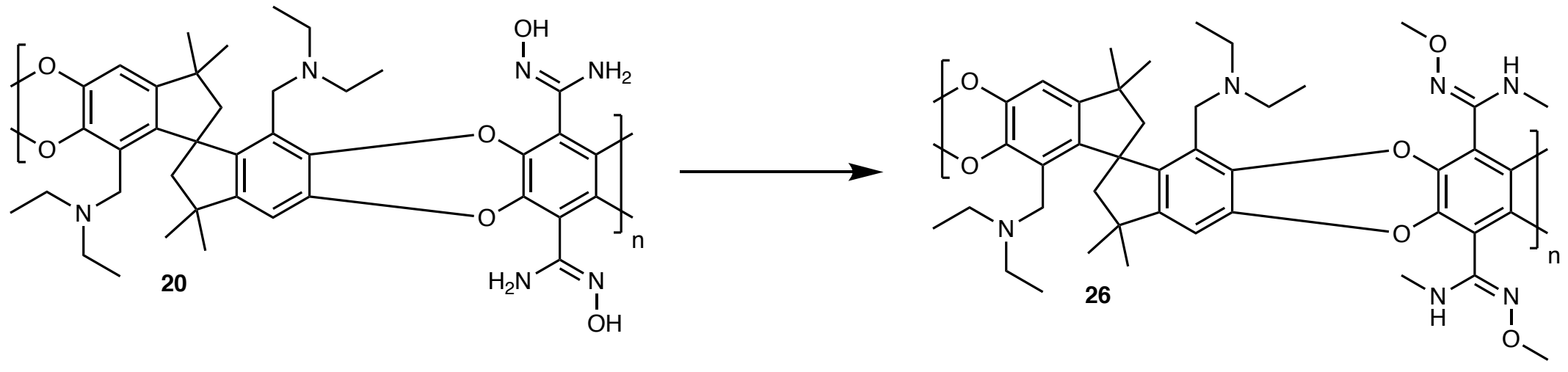

**26:** To a solution of **20** (1.50 g, 2.2 mmol) in dimethyl sulfoxide (43 mL) was added a solution of lithium hydroxide monohydrate (361 mg, 8.6 mmol) in MilliQ water (4 mL) dropwise over 1 min. After stirring for 1 h under nitrogen, the reaction mixture was placed on ice, and dimethyl sulfate (820 µL, 8.6 mmol) was added in two portions. The reaction was allowed to reach room temperature and stirred for 72 h, before quenching with 5 N sodium hydroxide (2 mL). The product was precipitated in MilliQ water (500 mL), filtered and rinsed with two portions of MilliQ water (100 mL each), followed by an additional precipitation from dimethyl sulfoxide into MilliQ water, and then dried *in vacuo* at 60 °C to give **26** as a dark yellow solid (940 mg) in 46% yield. $^{1}$H NMR (DMSO-$d_6$, 500 MHz): $\delta$ (ppm) 6.76 (br s, 2H), 5.82 (br s, 2H), 3.64 (br s, 4H), 2.91 (br s, 4H), 2.18 (br s, 20H), 1.32 (br s, 6H), 1.27 (br s, 6H), 0.69 (br s, 12H); SEC (DMF): $M_n$ = 6.9 kg mol$^{-1}$, $M_w$ = 13.8 kg mol$^{-1}$, *PDI* = 1.99; BET Surface area: 8 m$^2$ g$^{-1}$.

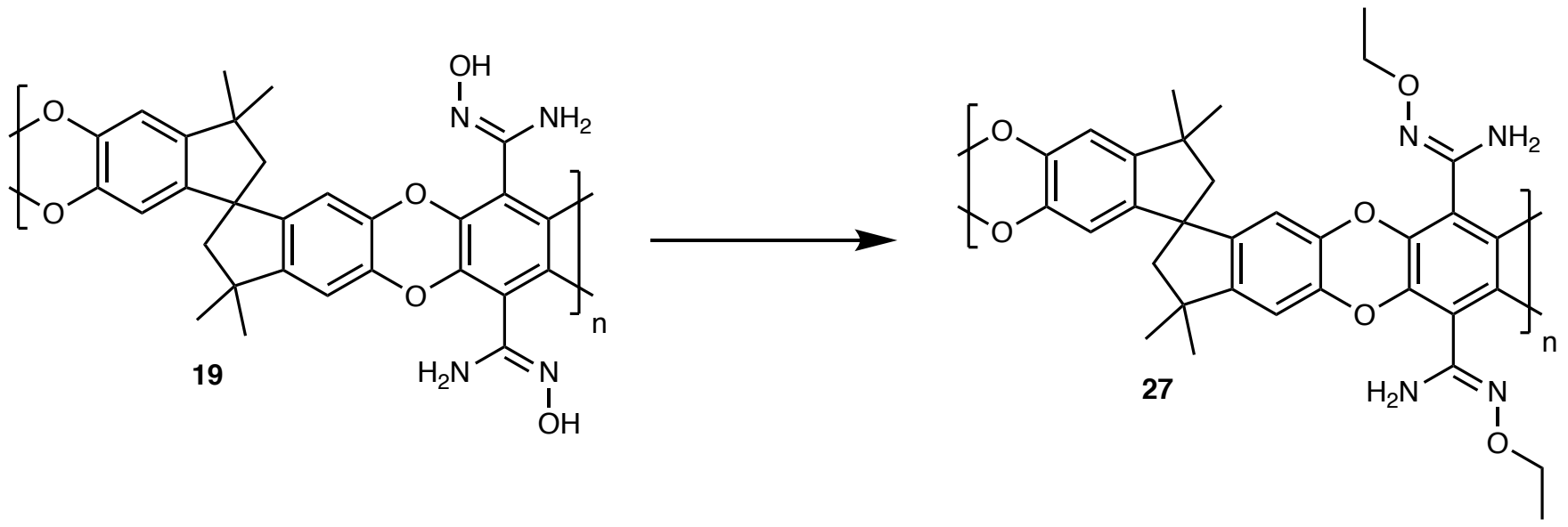

**27:** To a solution of **19** (1.50 g, 2.6 mmol) in dimethyl sulfoxide (57 mL) was added a solution of lithium hydroxide monohydrate (0.478 g, 11.40 mmol) in MilliQ water (5.75 mL) dropwise over 1 min. After stirring for 1 h under nitrogen, the reaction mixture was placed on ice, and diethyl sulfate (1.757 g, 11.40 mmol) was added in two portions. The reaction was allowed to reach room temperature and stirred for 72 h, before quenching with sufficient sodium hydroxide to increase the pH above 7. The product was precipitated in MilliQ water (400 mL), then filtered and rinsed with two portions of MilliQ water (100 mL each), and then dried *in vacuo* at 60 °C to give **27** as a dark yellow solid (1.34 g) in 84.8% yield. $^1$H NMR (DMSO-$d_6$, 500 MHz): $\delta$ (ppm) 6.82 (br s, 2H), 6.18 (br s, 2H), 5.82 (br s, 4H), 3.88 (br t, 4H), 2.17 (br d, 4H), 0.76–1.72 (br m, 18H); Elemental Analysis for $C_{33}H_{34}N_4O_6$ Calculated: C 68.03, H 5.88, N 9.62; Found: C 66.17 H 6.08, N 8.85; SEC (DMF): $M_n$ = 45.4 kg mol$^{-1}$, $M_w$ = 222 kg mol$^{-1}$, *PDI* = 4.87; BET Surface area: 10 m$^2$ g$^{-1}$.

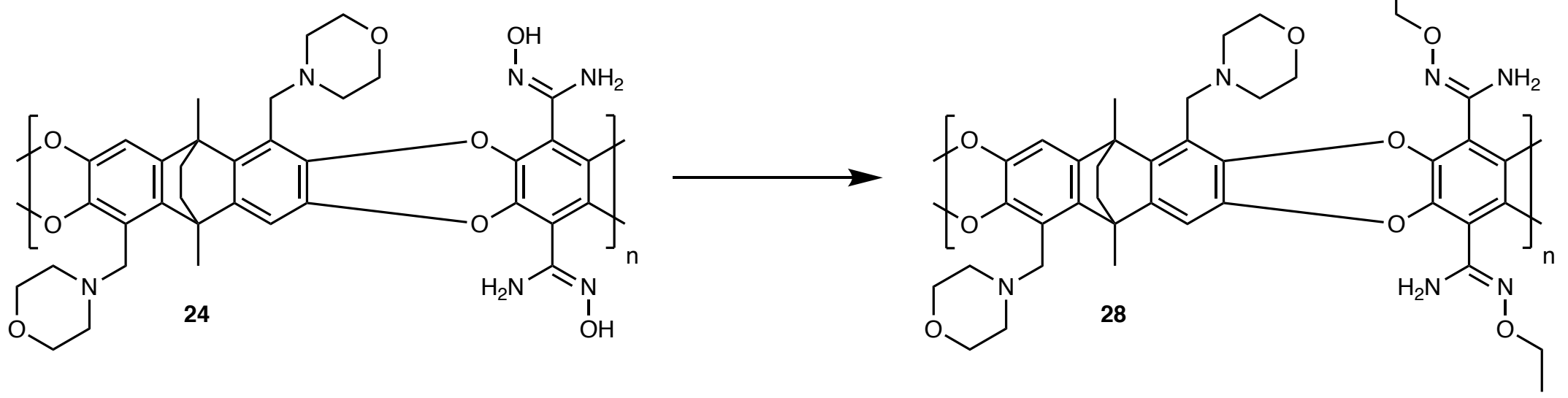

**28:** To a solution of **24** (200 mg, 0.32 mmol) in dimethyl sulfoxide was added lithium hydroxide monohydrate (53.4 mg, 1.27 mmol) in water (0.6 mL). The mixture was stirred for 1 h on ice before adding diethyl sulfate (0.164 mL, 1.27 mmol) and warming to room temperature. Reaction was stirred for two days, and then quenched by adding sufficient 5 N sodium hydroxide to increase the pH of the solution above 7. The polymer was precipitated in MilliQ water (100 mL) and then filtered, rinsed with an additional two portions of water (20 mL), and dried *in vacuo* at 50 °C to give **28** as a (231.2 mg) in 98% yield. $^1$H NMR (DMSO-$d_6$, 500 MHz): $\delta$ (ppm), 6.77 (br s, 4H), 6.05 (br s, 4H), 3.94 (br d, 4H), 3.48 (br s, 8H) 2.23–1.25 (broad m, 32H); SEC (DMF): $M_n$ = 16.6 kg mol$^{-1}$, $M_w$ = 41.9 kg mol$^{-1}$, PDI = 2.54; BET Surface area: 24 m$^2$ g$^{-1}$.

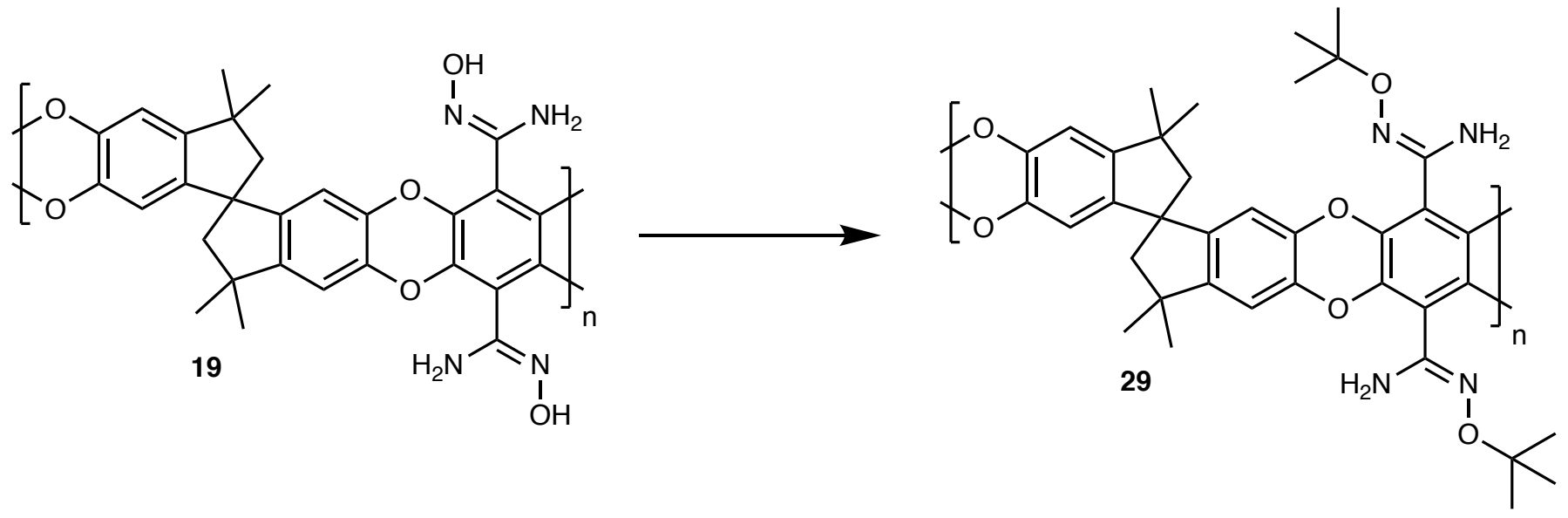

**29:** To a solution of **19** (1.00 g, 1.7 mmol) in dimethyl sulfoxide (38 mL) was added magnesium perchlorate (0.08g, 0.38 mmol) and di-*tert*-butyl dicarbonate (2.14 g, 7.60 mmol). The reaction was stirred for 24 h at 40 °C, then cooled and precipitated in MilliQ water (400 mL) with a few drops of sodium hydroxide and then filtered. The solid was rinsed with three portions of MilliQ water (50 mL each), and then two portions of chloroform (50 mL each) to remove excess di-*tert*-butyl dicarbonate, and then dried *in vacuo* at 60 °C to give **29** as a dark yellow solid (0.90 g) in 74% yield. $^1$H NMR (DMSO-$d_6$, 500 MHz): $\delta$ (ppm) 6.87 (br s, 4H), 6.25 (br s, 2H), 2.18 (br d, 4H), 0.93–1.73 (br m, 34H); Elemental Analysis for $C_{37}H_{42}N_4O_6$ Calculated: C 69.57, H 6.63, N 8.77; Found: C 61.29 H 5.47, N 8.09; SEC (DMF): $M_n$ = 43.2 kg mol$^{-1}$, $M_w$ = 154 kg mol$^{-1}$, *PDI* = 3.56; BET Surface area: 11 m$^2$ g$^{-1}$.

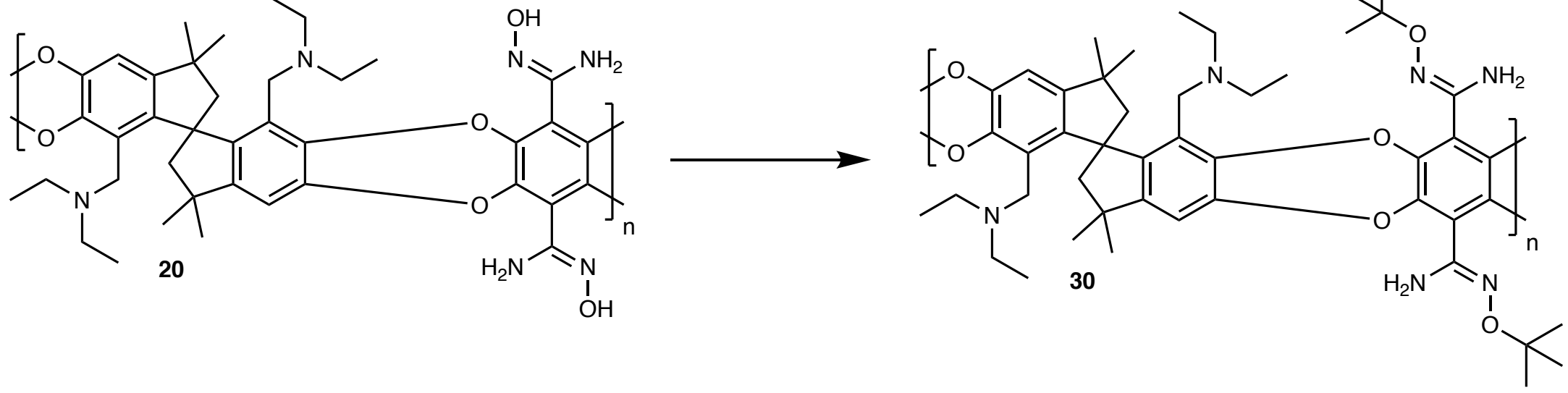

**30:** To a solution of **20** (1.5 g, 2.2 mmol) in dimethyl sulfoxide (43 mL) was added magnesium perchlorate (95 mg, 430 μmol) and di-*tert*-butyl dicarbonate (1.9 g, 8.6 mmol). The reaction was stirred for 24 h at 40 °C, then cooled and precipitated in MilliQ water (430 mL) with a few drops of 5N sodium hydroxide and then filtered. The solid was rinsed with three portions of MilliQ water (50 mL each), and then two portions of chloroform (50 mL each) to remove excess di-*tert*-butyl dicarbonate, and then dried *in vacuo* at 60 °C to give **30** as a dark yellow solid (980 mg) in 49% yield. $^1$H NMR (DMSO-$d_6$, 500 MHz): $\delta$ (ppm) 6.80 (br s, 6H), 3.27 (br s, 4H), 2.93 (br s, 4H), 2.14 (br s, 8H), 1.37 (br m, 30H), 0.67 (br s, 12H); SEC (DMF): $M_n$ = 10.9 kg mol$^{-1}$, $M_w$ = 15.2 kg mol$^{-1}$, *PDI* = 1.39; BET Surface area: 48 m$^2$ g$^{-1}$.

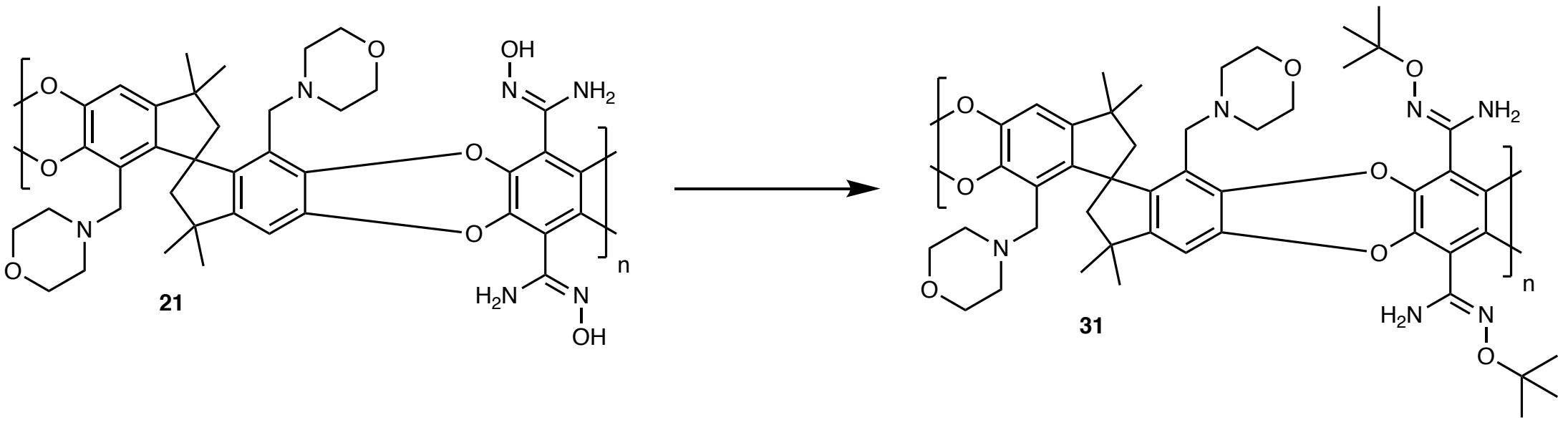

**31:** To a solution of **21** (1.5 g, 2.2 mmol) in dimethyl sulfoxide (43 mL) was added magnesium perchlorate (95 mg, 430 µmol) and di-*tert*-butyl dicarbonate (1.9 g, 8.6 mmol). The reaction was stirred for 24 h at 40 °C, then cooled and precipitated in MilliQ water (400 mL) with a few drops of sodium hydroxide and then filtered. The solid was rinsed with three portions of MilliQ water (50 mL each), and then two portions of chloroform (50 mL each) to remove excess di-*tert*-butyl dicarbonate, and then dried *in vacuo* at 60 °C to give **31** as a dark yellow solid (1.2 g) in 56% yield. $^1$H NMR (DMSO-$d_6$, 500 MHz): $\delta$ (ppm) 6.84 (br s, 6H), 2.11 (br s, 24H), 1.14-1.55 (br m, 30H); SEC (DMF): $M_n$ = 43.0 kg mol$^{-1}$, $M_w$ = 137 kg mol$^{-1}$, *PDI* = 3.19; BET Surface area: 31 m$^2$ g$^{-1}$.

**Polymer solubility**. Qualitative solubility tests were determined by stirring 5 mg of polymer and 2 mL of solvent in a test tube for 12 h. Complete solubility is indicated with "+++", partial solubility with "+", and poor solubility with "–" as indicated in **Supplementary Table 1**.

**Electrochemical Methods**

**Electrochemical cell assembly.** All electrochemical studies were performed using CR2032 coin cells with a single spacer and spring pressed at 1000 psi, Celgard 2325 separators, 70 µL of liquid electrolyte (1.0 M $LiPF_6$ in EC:DEC (1:1 v/v) with fluoroethylene carbonate (FEC, 10% w/w) and vinylene carbonate (VC, 1% w/w) and 750-µm thick Li electrodes, onto which PIMs were drop cast as inks in THF and dried overnight before cell assembly.

**Electrochemical cells for conductivity, activation energy, and transference number measurements.** Symmetric cells were assembled with PIM interlayers cast on both Li electrodes (loading and interlayer thickness given below), with coating facing toward the separator. Cells were allowed to equilibrate at ambient temperature at least 24 h before testing.

| Polymers | Total polymer thickness in cell (µm) |
|---|---|
| PIM-1, **11–15** | 15 |
| **17** | 2 |
| **18, 21, 23–26, 28, 30, 31** | 6 |

Electrochemical impedance spectroscopy (EIS) was measured at open-circuit voltage (OCV) with a Biologic VMP3 potentiostat from 1 MHz to 1 Hz with AC voltage amplitude of 10 mV. Conductivity measurements were taken in a temperature control oven, which was allowed to equilibrate to 25, 40, 55, and 70 °C for 2.5 h before acquiring impedance spectra at each temperature. Spectra were modelled using the equivalent circuit below.

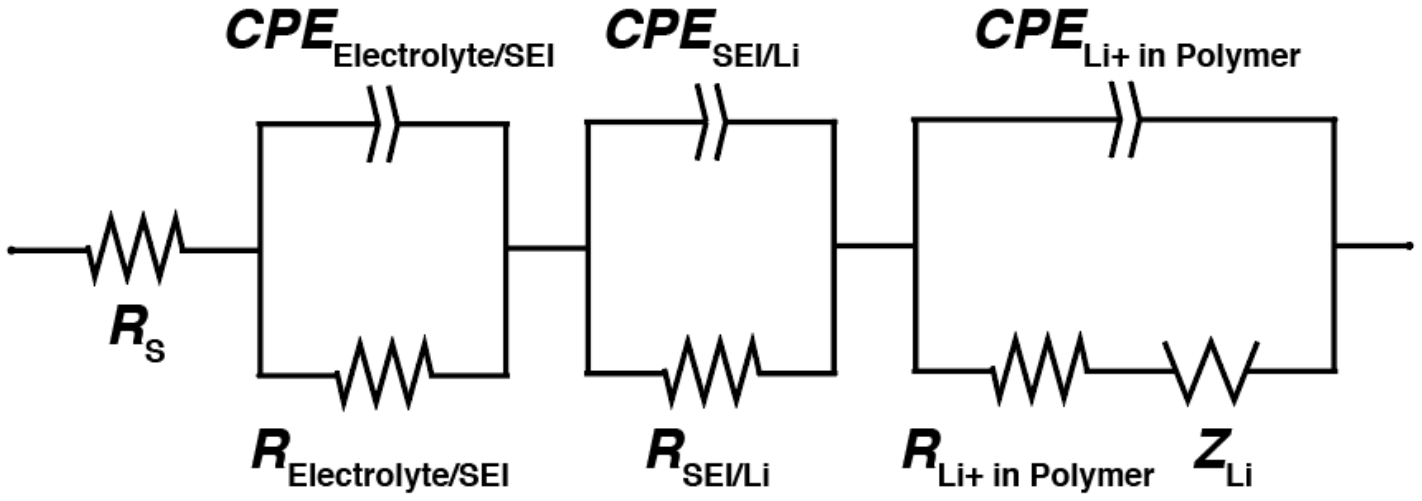

The resistance for a given polymer was determine by subtracting the solution resistance ($R_S$) of a cell with no interlayer (Celgard only) at the same temperature from the solution resistance of the given cell. Conductivity was then determined from the interlayer thickness ($l$), resistance ($R_{Polymer}$), and electrode area ($A$, 1.267 $cm^2$).

$$R_{Polymer} = R_{S\ Polymer} - R_{S\ Celgard}$$

$$C_{Polymer} = \frac{l}{R_{Polymer}A}$$

Conductivity values at 25 °C are reported in **Fig. 2**. Activation energy values were determined from the slope of a plot of $ln(CT)$ vs. $1/T$, shown in **Supplementary Figs. 12–28**.

Transference number was determined by EIS before and after potentiostatic polarization at 10 mV for 2 h, shown in **Supplementary Figs. 29–44**. The applied voltage ($\Delta V$) initial current ($I_0$), steady state current ($I_{ss}$), and solution resistance values before ($R_0$), and after polarization ($R_{ss}$) were used to calculate cation transference number ($t_+$).

$$t_+ = \frac{I_{ss}\,(\Delta V - I_0 R_0)}{I_0\,(\Delta V - I_{ss} R_{ss})}$$

**Li–NMC622 Cells.** Lithium metal anodes left bare, or optionally coated with either PIM-1 or **13** (~3 µm-thick overlayers), were paired with $LiNi_{0.6}Mn_{0.2}Co_{0.2}O_2$ (NMC622) composite cathodes with an areal capacity of 1.44 mAh $cm^{-2}$ (supplied by the CAMP facility at Argonne National Lab); the cathodes were 12 mm in diameter. Celgard 2325, 19 mm in diameter, and used as separator. The electrolyte was 1.0 M $LiPF_6$ in EC:DEC (1:1 v/v) with FEC (10% w/w) and VC (1% w/w); 70 μL was used in each cell. Cycling was carried out at 25 °C on an Arbin BT-2043 System with three initialization cycles at 0.1 mA $cm^{-2}$ before cycling at a current density of 1 mA $cm^{-2}$.

## Computational Methods

**General Methodology.** To explore the functional role of $Li^+$ solid solvation cages we use a multiscale computational approach with the main focus on the study of $Li^+$ ion transport between cages when in contact with the liquid electrolyte. Specifically, we analyze the structural properties and thermodynamics of $Li^+$ in solid solvation cages of PIM **13** in dimethyl carbonate (DMC). The processes of dissociation of the cation from a cage and its solvation/desolvation was performed in three steps.

In the first step, we used quantum mechanical (density functional theory, DFT) calculations (ignoring finite temperature effects) of small molecular clusters, involving the PIM 13 unit, the solvated $Li^+$ ion and DMC molecules. We calculated the equilibrium configurations of polymer units with and without the $Li^+$ ion. Long-range solvent effects are included in these quantum chemical calculations using the polarizable continuum model (PCM). These calculations not only provide structural and electronic details (the equilibrium configurations and atomic partial charges, based on natural bond orbitals) but also their sensitivity to dielectric screening from the electrolyte solvent. Using thermodynamic cycles we evaluated the cation solvation energies and relative stability of different configurations (e.g., contrasting $Li^+$ in the cage versus in the solvent) by reference to the corresponding condensed phases.

In the second step, sampling of the ensemble of molecular configurations accessible at finite temperature, with inclusion of explicit solvent molecules, was provided by classical molecular dynamics (MD) simulations. We tested a various force fields and different charge schemes. Special attention was paid to partial charges of the lithium ion, as well as the coordinating oxygen and nitrogen moieties in the cages.

In the third step, we employed metadynamics to explore the free-energy landscape associated with $Li^+$ transport: the kinetic barriers and the relative free energy differences of various states of the $Li^+$ cation, on its path from solid solvation cage to bulk solvent or to another cage. The metadynamics approach also incorporates the entropy of all the thermodynamically accessible configurations of the polymer, cation and solvent molecules. We used 1D and 2D free energy analyses with particular choices of collective variables that reflect the transport processes of interest (see below).

**Quantum chemistry calculations.** Optimized geometries, relative energies, and molecular orbitals were calculated using DFT, with the GPU-based TeraChem package.[12] As suggested in previous extensive computational studies of aprotic ionic liquids, we used the B3LYP5-D3 functional with the 6-311++G** basis set,[13] employing the third version of Grimme's empirical dispersion correction.[14] We used the L-BFGS geometry optimization method[15] with the termination criterion for the maximum energy gradient component of $4.5 \times 10^{-4}$ au. The wave function convergence threshold was set as $3.0 \times 10^{-5}$. The two-electron integral threshold was set as $1.0 \times 10^{-12}$, and the basis set linear dependency threshold was $1.0 \times 10^{-4}$. The numerical accuracy was set to double precision. Partial charges were computed using the full natural bond orbital[16] (NBO) and Mulliken analysis. The structure of an isolated DMC molecule is shown in **Supplementary Fig. 46**. The analysis of the stability of the $Li^+$-ion in DMC revealed that the most stable configuration is the 4-fold coordinated cation, shown in **Supplementary Fig. 47**, as previously found.[17–20] The equilibrium configuration of the PIM **13** fragment is shown in **Supplementary Fig. 48**. Note the "chair" configuration of morpholine. When $Li^+$ is added to the polymer, it finds the most stable configuration in the cage indicated by indices in **Supplementary Fig. 49**; the "boat" configuration of morpholine has been adopted, which provides both nitrogen and oxygen atoms to coordinate the cation. NBO charges for nitrogen and oxygen atoms of the cages are shown in table **Supplementary Table 6**.

Next we used nudged elastic band[21] (NEB) calculations to evaluate the kinetic barrier for transition of the $Li^+$ ion between neighboring cages (the PIM **13** unit contains two adjacent though not fully identical cages). No continuum dielectric media or explicit solvent molecules were used. NEB energies are shown in **Supplementary Fig. 50**. Since the cages are not fully identical, the minima have slightly different energies, and the left minimum corresponds to the cage shown in **Supplementary Fig. 48**. As **Supplementary Fig. 51** suggests, the kinetic barrier of 4.51 eV for motion of the $Li^+$ ion is unsurmountable for the transport between cages. Note that these are energies at $T = 0$ K, and no entropy effects are considered. Moreover, since the ends of the polymer unit are not fixed, one may expect higher energies as the molecular configuration could be more compact because of the finite size effect.

**Hybrid cluster/continuum quantum mechanical calculations: Ion solvation free energies.** Quantum mechanical calculations (B3LYP5-D3/6-311++G**) with polarizable continuum media (PCM) calculations with COSMO[22] model were used to account for the effects of the solvent and to calculate the solvation energies of species of interest. The dielectric constant of DMC was set as 3.087, and atomic radii as suggested by Bondi[23]. To evaluate the free energy of dissociation of the $Li^+$ ion from the cage to the bulk solvent we use the thermodynamic cycle shown in **Supplementary Fig. 51**.

From this cycle, $\Delta G^*_{diss} = \Delta G_1 + \Delta G_2 + \Delta G_3 + \Delta G^*_{sol}(Li^+)$, with $\Delta G_1$ – free energy of desolvation of the polymer (PIM **13**) unit with the $Li^+$ ion inside the cage, $\Delta G_2$ – dissociation energy into a free unit and an isolated cation in vacuum, $\Delta G_3$ – free energy of solvation of the polymer unit, and finally, $\Delta G^*_{sol}(Li^+)$ – $Li^+$ ion absolute solvation free energy. Since we ignore the concentration of PIM, no thermodynamic correction terms for the standard 1.0 M state were applied. Moreover, due to the large size of the polymer unit and low dielectric constant of DMC, we assume $\Delta G_3 = 0$. To evaluate $\Delta G^*_{sol}(Li^+)$ we employed the "monomer" thermodynamic cycle[24] shown in **Supplementary Fig. 52**.

Using this cycle, the ion solvation energy can be calculated as follows:

$$\Delta G^*_{solv}(Li^+) = \Delta G^o_{g,bind}(I) + \Delta G^*_{solv}([Li(DMC)_n]^+) - n\Delta G^*_{solv}(DMC) - nRT\ln[DMC] - n\Delta G^{o-*}$$

Where,

$\Delta G^{o}_{g,bind}(I)$ is the cluster $Li^+(DMC)_n$ formation energy referenced to the dissociated state in vacuum;

$\Delta G^{*}_{solv}([Li(DMC)_n]^+) = G([Li(DMC)_n]^+)_{DMC} - G([Li(DMC)_n]^+)_{gas}$ is the solvation energy of the cluster (calculated with the COSMO model);

$\Delta G^{*}_{solv}(DMC)$ is the free energy of self-solvation of DMC in DMC (i.e., the negative vaporization energy), which could be obtained either from the pressure of the saturated vaper at $T$ = 298.15 K in equilibrium with the liquid DMC: $\Delta G^{*}_{solv}(DMC) = RTln\left(\frac{[DMC_g]}{[DMC_{liq}]}\right)$ or from PCM calculations: $\Delta G^{*}_{solv}(DMC) = G(DMC)_{THF} - G(DMC)_{gas}$;

$RT\ln[DMC] = 1.4654\ kcal/mol$ is the thermodynamic correction for the standard state (1 M) of the solvent, i.e., the free energy cost of transferring of a compound from a hypothetical ideal gas at 1.0 M into a hypothetical ideal 1.0 M solution. Molar concentration of DMC [DMC(l)] = 11.879 M;

$\Delta G^{o-*} = 1.89\ kcal/mol$ is the thermodynamic correction of conversion of an ideal gas standard state of 1.0 atm to a standard state of 1.0 M.

Note that the formation of a cluster of a solvated ion in liquid DMC does not involve any change in free energy, so the lower cycle leg has only $RT\ln[DMC]$ correction. As suggested from our previous calculations, the first solvation sphere contains four DMC molecules (**Supplementary Fig. 47**), so we put $n = 4$. The results of quantum mechanical calculations are summarized in **Supplementary Tables 7** and **8**.

As follows from these calculations, solvation energy of the $Li^+$ ion in DMC (–6.86 eV) is on par with that in water (–5.49 eV)[25] despite the difference in dielectric constants. It is also close to the values of solvation energies of other monovalent cations (e.g., $Na^+$) in methanol and acetonitrile[26] (–7.043 eV and –6.90 eV, respectively). These calculations suggest that the dissociation of $Li^+$ from the cage in the bulk solvent is significantly renormalized (+0.457 eV) as compared to that in vacuum (+5.827 eV), and remains thermodynamically unfavorable: the equilibrium $Li^+_{sol} + cage \leftrightarrow Li^+_{cage}$ is then expected to be strongly shifted to the right. The explicit free energy sampling (see below) shows even further renormalization of the dissociation energy, up to 0.14–0.15 eV (see main text).

**Classical molecular dynamics calculations.** Classical molecular dynamics (MD) simulations were conducted using the LAMMPS simulation package.[27] Long-range electrostatic interactions were treated within the particle-mesh Ewald (PME) method with a cutoff distance 1.0 nm with grid spacing in *k*-space of $10^{-5}$. A cut-off of 1.0 nm with a spline from 0.9 to 1.0 nm was used for Lennard-Jones interactions. The relaxation of the initial structures was performed in two steps, first using steepest descent with a convergence criterion of $10^{-4}$ kcal mol$^{-1}$ for energies and $10^{-4}$ kcal mol$^{-1}$ Å$^{-1}$ for forces. The systems were first heated to 298 K in the canonical ensemble (NVT). To remove any "memory" effects, the systems were first melted at 400 K and then annealed back to 298 K three times (evolving the trajectory 2 ns for annealing each step). Then, isothermal-isobaric (NPT, isotropic $P$ = 1 atm, $T$ = 298 K) simulations were performed for 2 ns (2 fs time step) to obtain the correct density using a Nose/Hoover thermostat[28] and Nose/Hoover barostat.[29] The temperature coupling constant was 0.1 ps, the pressure piston constant was 2.0 ps. Afterwards, NVT simulations were performed ($T$ = 298 K) for 1 ns (1 fs time step) to equilibrate. Structural properties were obtained from 5 ns MD simulation runs with an integration time step 1 fs in NVT ensemble with coordinates and velocities saved every 5 ps for post-trajectory analysis.

Throughout all classical MD calculations (regular and metadynamics protocols) the Generalized Amber force field[30] (GAFF) was used for ions and solvent molecules. The diagonal vdW parameters of the Lennard-Jones potential for the $Li^+$ ion are: 2.050 and 0.028 for $\sigma$ and $\varepsilon$, respectively, as suggested by Aqvist.[31] The charge of the $Li^+$ ion is fixed as +1 (NBO charge of Li in the $[Li(DMC)_4]^+$ complex is ca. 0.91). For the anion $PF_6$ we used the harmonic potential for the P – F bond ($k$ = 507.6 kcal mol$^{-1}$, $r_0$ = 1.579 Å) and the periodic cos potential for F – P – F angle ($k$ = 190 kcal mol$^{-1}$, $b$ = 1, $n$ = 4) to ensure the octahedral configuration.

We used two strands of PIM **13**, eight units long each (712 atoms), per a simulation box, as shown in **Supplementary Fig. 53**. For the oxygens and nitrogen atoms of the cage we used NBO charges listed in **Supplementary Table 6**. We performed simulation of pure DMC (to test the proper choice of the force field capable to reproduce the bulk density of the solvent), and two different concentrations of $LiPF_6$ (0.2 M and 0.7 M) in the presence of the polymers to see the population of the occupied cages. A typical simulation box is shown in **Supplementary Fig. 54**. Parameters of simulation boxes are summarized in **Supplementary Table 9**.

We performed classical free energy sampling to evaluate the free energy profiles along the collective variables (CV) of interest. For one-dimensional free energy sampling we used $\boldsymbol{r_1}$ – the distance between the $Li^+$-ion and the center of the masses of a cage, for two-dimensional free energy analysis we used linear combinations $\boldsymbol{d} = \boldsymbol{r_1} - \boldsymbol{r_2}$ and $\boldsymbol{s} = \boldsymbol{r_1} + \boldsymbol{r_2}$ of distances $\boldsymbol{r_1}$ and $\boldsymbol{r_2}$ between the $Li^+$-ion and the centers of the masses of two cages (see further details below). Specifically, we use the metadynamics[32,33] protocol to evaluate the potential of mean force (PMF) along the collective variables. For a system of N particles in a canonical ensemble at temperature T, the probability density associated with a generalized coordinate $\lambda$ having a value of $\lambda_0$ can be expressed in terms of a generalized partition function:

$$\Omega(\lambda_0) = \int dq^{3N} dp^{3N} \delta(\lambda - \lambda_0) e^{-\beta H(q^N, p^N)},$$

where $q$ and $p$ are conjugated generalized coordinates and momenta of particles and we assume that additional constraints associated with general integrals of motion are already subtracted.

Accordingly, the configurational part of the partition function can be written as $Z(\lambda_0) = \int dq^{3N}\, \delta(\lambda - \lambda_0) e^{-\beta H(q^N;\lambda)}$. The PMF is then defined as $W(\lambda) = -kT\ln\Omega(\lambda) + const.$ The integration constant is chosen so min $\{W(\lambda)\} = W(\lambda_{min}) = 0$, where $\lambda_{min}$ – is the value of CV at which $W(\lambda)$ reaches its global minima.

In **Supplementary Figure 55** we show the definitions of collective variables used in our free energy analysis. Since we want to explore the free energy profile along the pathway from one cage to another as close as possible to the backbone of the polymer, just $\mathbf{r}_1$ and $\mathbf{r}_2$ will not be an appropriate choice for collective variables for 2D free energy sampling. To solve this problem we used $\mathbf{d} = \mathbf{r}_1 - \mathbf{r}_2$ and $\mathbf{s} = \mathbf{r}_1 + \mathbf{r}_2$ and apply a constraint that limits the range for **d** and **s** in such a way that we explore efficiently the phase space available for a system when a Li+ ion moves along the plausible way that connects adjacent cages. Therefore, we limit the phase space: $|d| \leq D$ and $s \geq D + \Delta$, with $D$ – the distance between centers of masses of cages and $\Delta$ – limits how far from the straight line connecting the centers of cages the $Li^+$ ion can deviate. **Supplementary Fig. 56** illustrates the meaning of the constraints.

For a motion with $d = const$ or $s = const$ the $Li^+$ ion will move along hyperbolae and ellipses (in 2D) described by equations:

$$\frac{x^2}{d^2} - \frac{y^2}{D^2 - d^2} = \frac{1}{4}, d = const, |d| \leq D,$$
$$\frac{x^2}{s^2} + \frac{y^2}{s^2 - D^2} = \frac{1}{4}, s = const, s \geq D.$$

$d = -D$ and $s = D$ corresponds to the $Li^+$ ion in the cage A, and $d = D$ and $s = D$ are when $Li^+$ ion in the cage B (in centers of their masses, as defined). The phase space in coordinates ***d*** and ***s*** is shown in **Supplementary Fig. 57**. The left and right bottom corners of the rectangles correspond to centers of cages A and B. In simulations we do not control the positions and the mutual distance between centers of adjacent cages, so the actual distance *D* fluctuates, so do the minima of the free energies corresponding to local minima of the $Li^+$ ion residing inside the cages. These fluctuations explain the tilt of the regions around the minima.

**Classical MD metadynamics protocol.** An equilibrated simulation cell with solvent molecules, two polymer strands and ions (see regular classical MD simulations section) was used as initial configuration for free energy sampling with the metadynamics protocol. We employed dilute limit (one $Li^+$ and one $PF_6^-$ per box). In **Supplementary Table 6**, we summarize the parameters of the metadynamics free energy sampling: height of the Gaussian hills (*H*, kcal $mol^{-1}$), frequency of hill creation (*freq*, steps), and width of hills ($W$, Å) along with the ranges for the collective variables, and the overall simulation times (*T*, ns) for each system. The free energy profile as a function of $r_1$ is shown in **Supplementary Fig. 58**.

To ensure the convergence of the sampling, we ran metadynamics until the diffusive motion along the CV is reached, as previously suggested.[33] In **Supplementary Fig. 59**, we show the time evolution of the collective variable.

**Supplementary Tables**

**Supplementary Table 1.** Polymer solubility in a variety of organic solvents, where +++ indicates highly soluble, + indicates slightly soluble, and – indicates insoluble.

| **Polymer** | **Solubility** | | |
|---|---|---|---|
| | **THF** | **$CHCl_3$** | **DMSO** |
| **11** | +++ | +++ | + |
| **12** | +++ | +++ | + |
| **13** | +++ | +++ | +++ |
| **14** | +++ | +++ | – |
| **15** | +++ | +++ | +++ |
| **16** | +++ | + | +++ |
| **17** | + | + | – |
| **18** | + | + | +++ |
| **19** | – | + | +++ |
| **20** | +++ | + | +++ |
| **21** | +++ | + | +++ |
| **22** | + | + | +++ |
| **23** | + | + | +++ |
| **24** | + | – | +++ |
| **25** | +++ | +++ | +++ |
| **26** | +++ | – | + |
| **27** | – | – | +++ |
| **28** | +++ | +++ | +++ |
| **29** | + | – | + |
| **30** | +++ | +++ | + |
| **31** | +++ | +++ | + |

**Supplementary Table 2**. GIWAXS 2D and 1D plots.

| Compound Number | GIWAXS 2-D Plot | GIWAXS 1-D Plot | Peak $q$ value (Å$^{-1}$) | Peak $d$-space value (Å) |
|---|---|---|---|---|
| **11** | 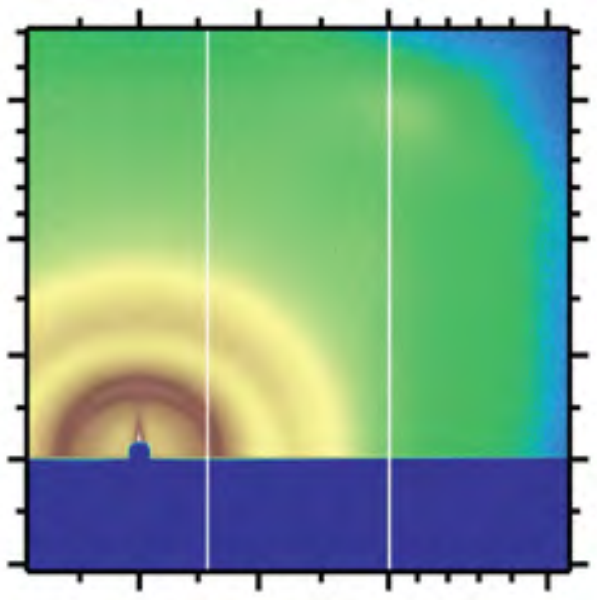 | 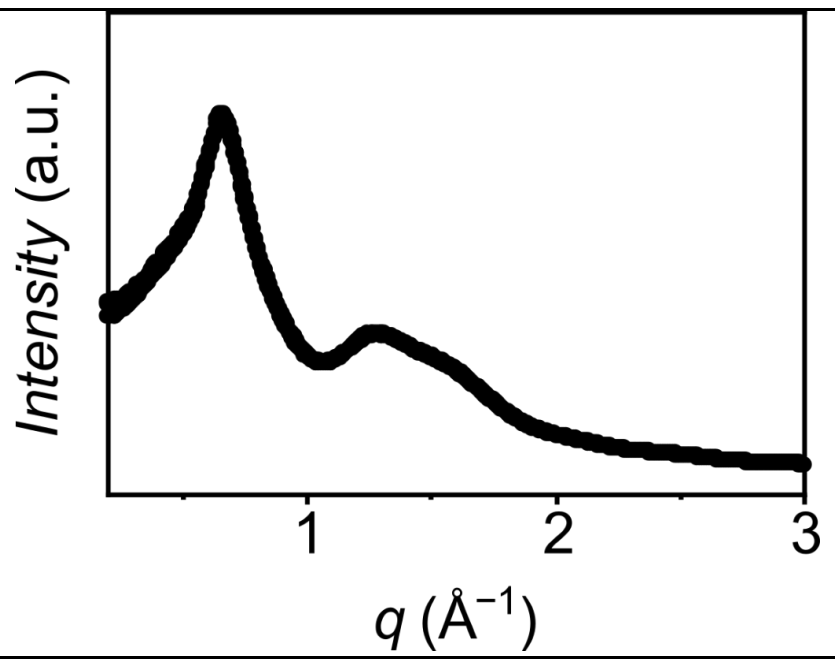  | 0.66<br>1.28 | 9.6<br>4.9 |
| **12** | 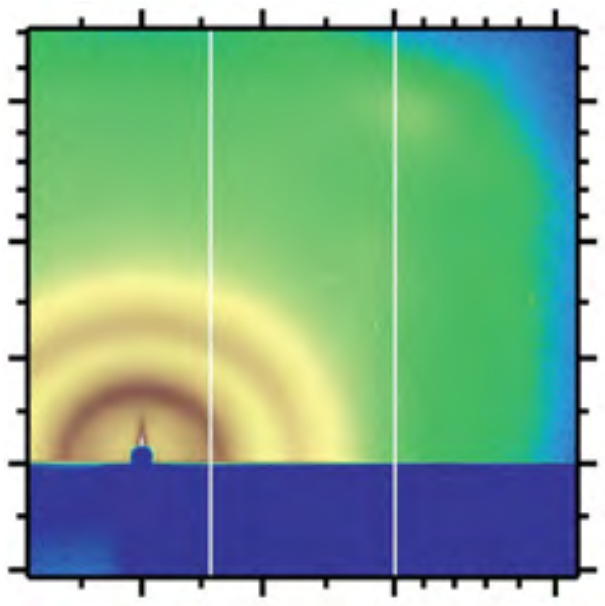 | 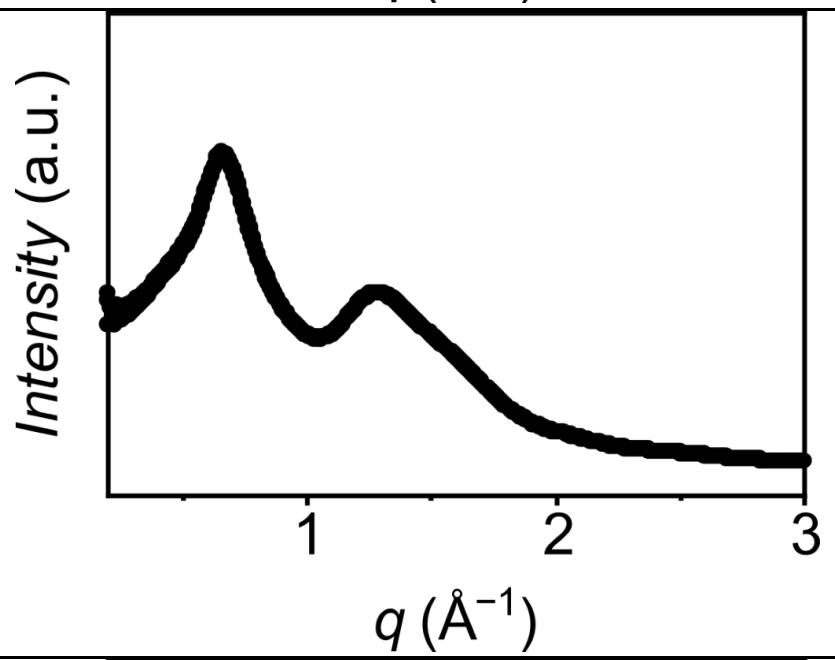  | 0.66<br>1.28 | 9.5<br>4.9 |
| **13** | 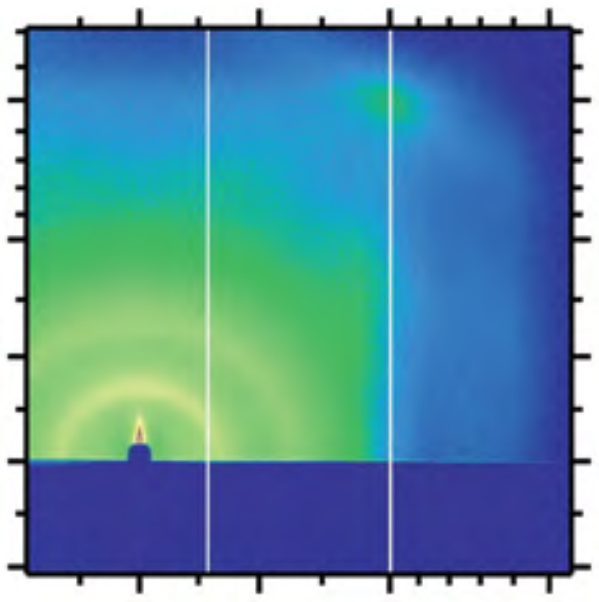 | 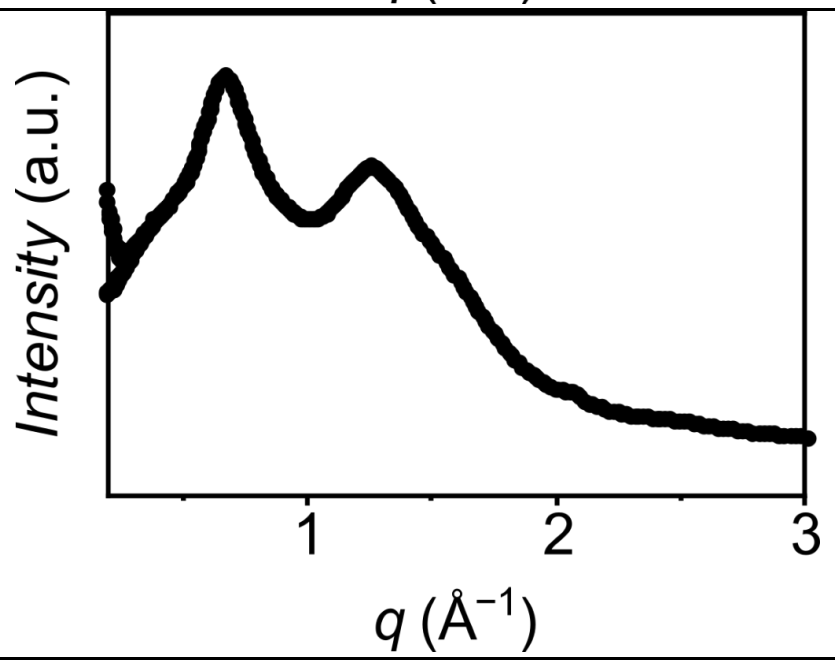  | 0.67<br>1.27 | 9.3<br>5.0 |
| **14** | 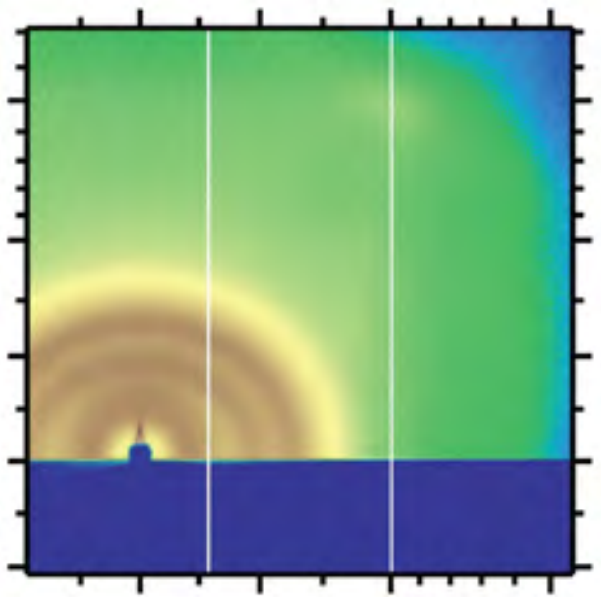 | 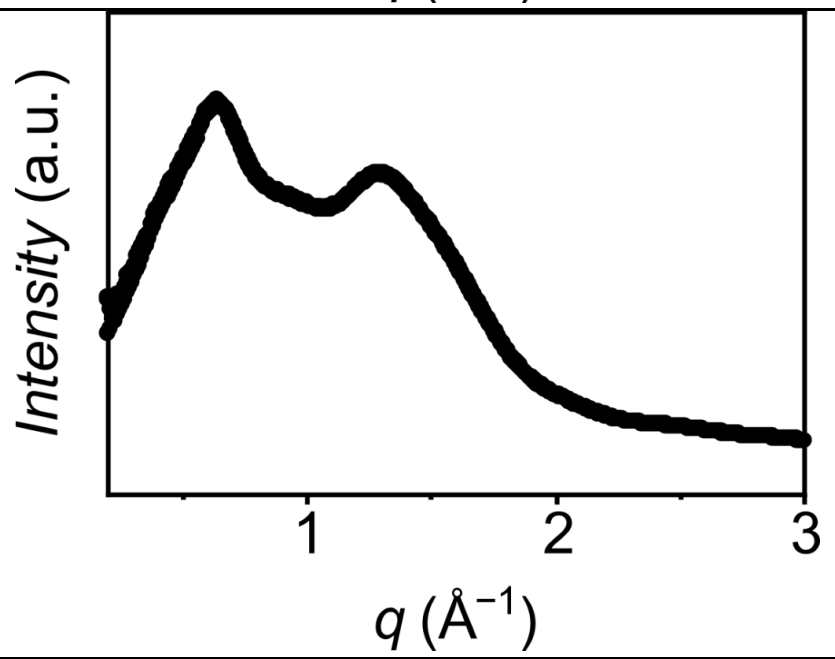  | 0.64<br>1.29 | 9.8<br>4.9 |
| **15** | 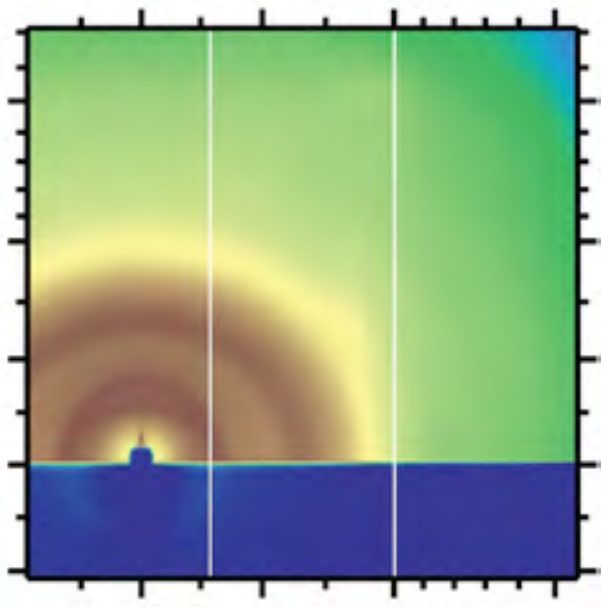 | 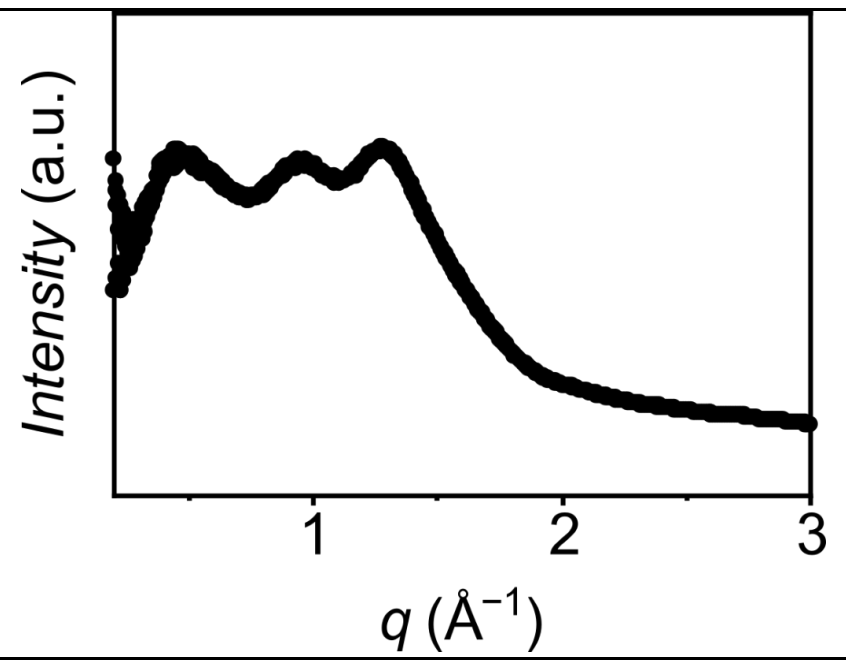 | 0.47<br>0.95<br>1.28 | 13.3<br>6.6<br>4.9 |
| **16** | 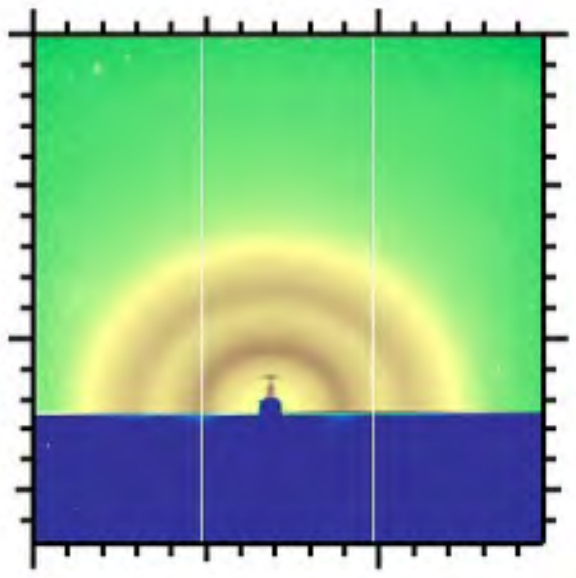 | 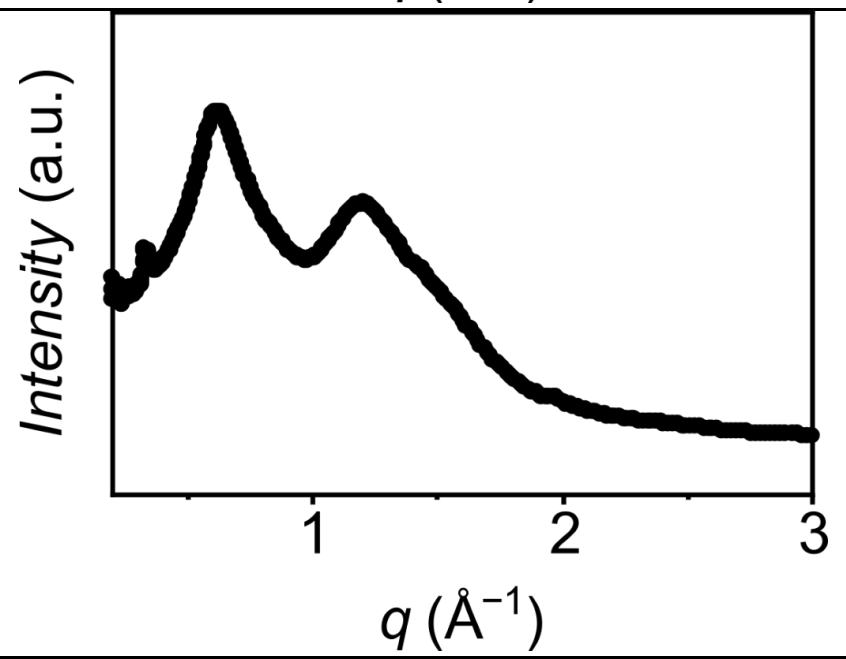 | 0.63<br>1.20 | 10.0<br>5.2 |
| **17** | 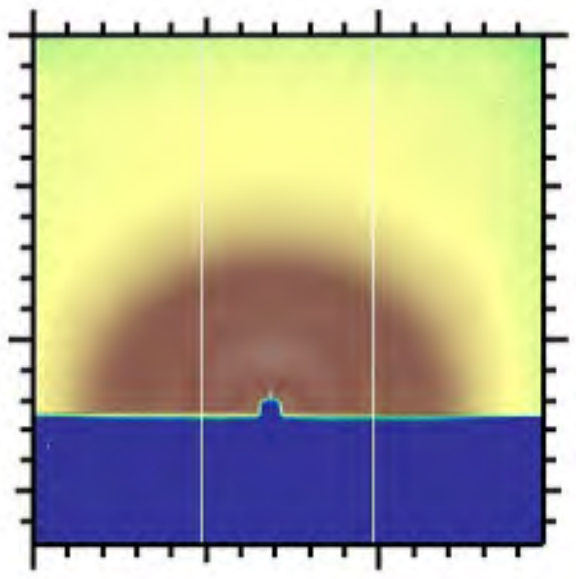 | 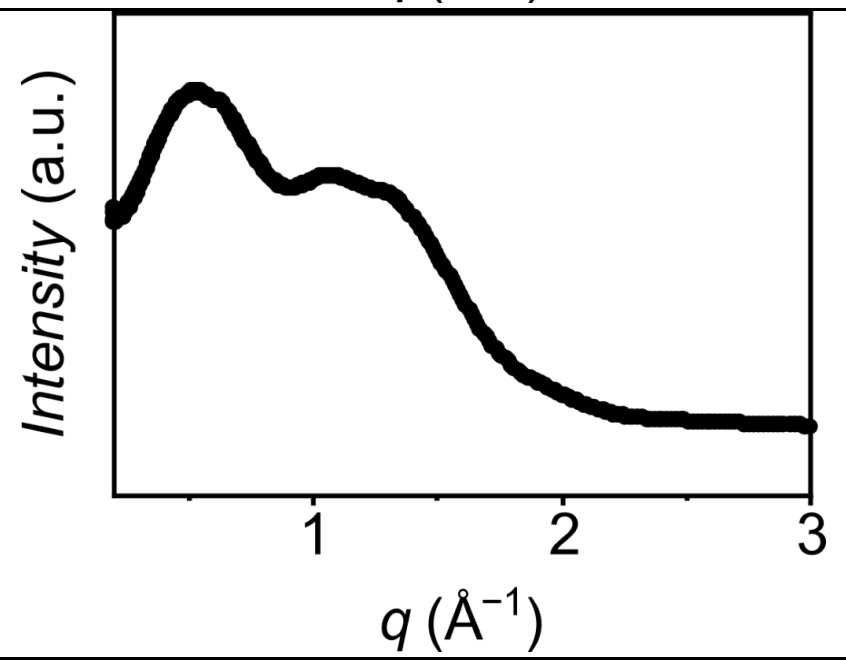 | 0.53<br>1.07<br>1.32 | 11.9<br>5.9<br>4.8 |
| **18** | 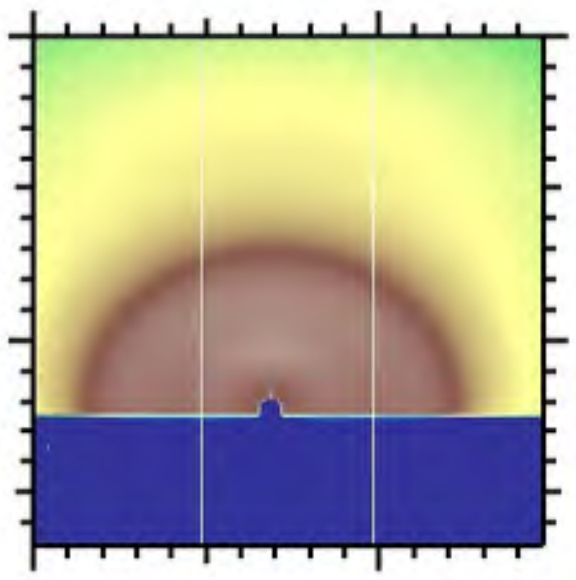 | 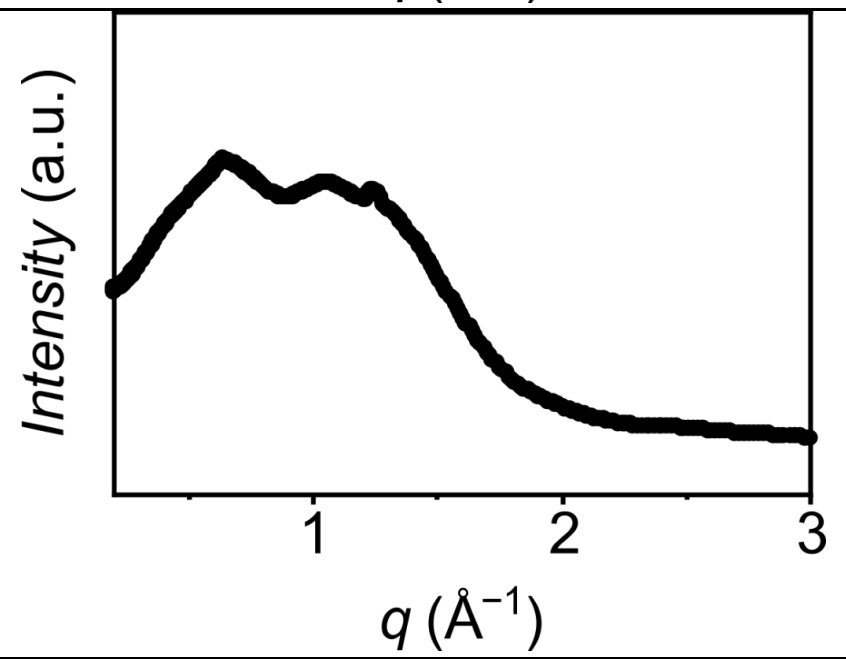 | 0.65<br>1.05 | 9.6<br>6.0 |
| **19** | 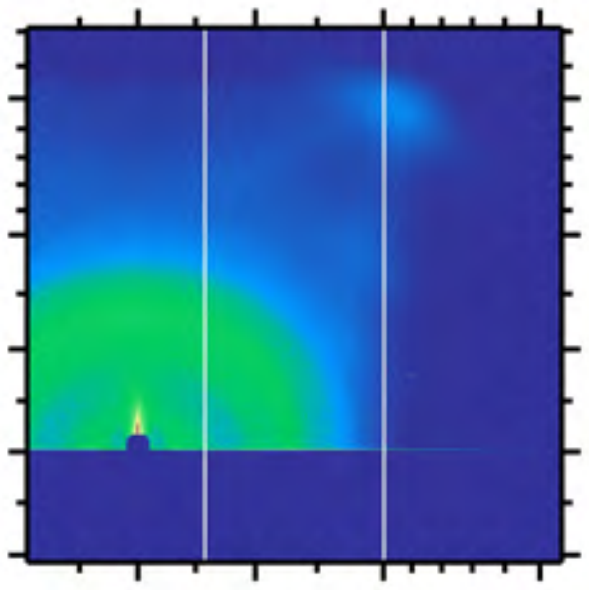 | 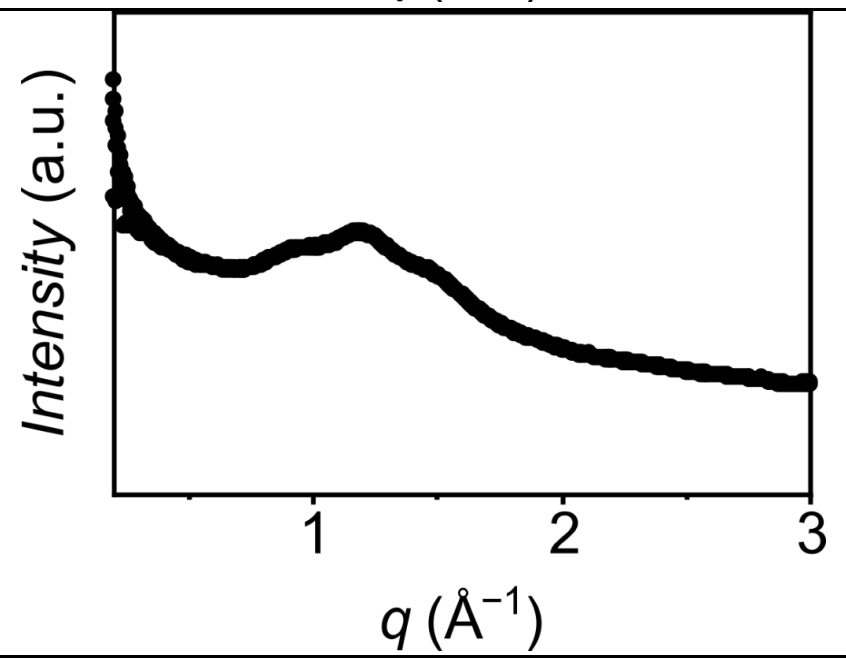 | 0.93<br>1.19<br>1.51 | 6.8<br>5.3<br>4.2 |
| **20** | 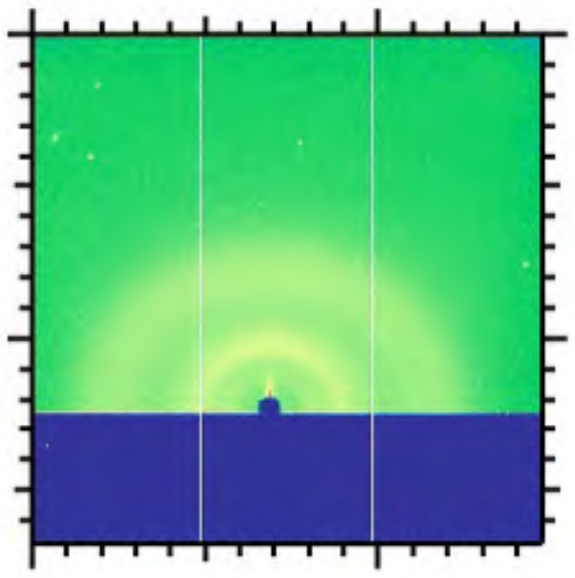 | 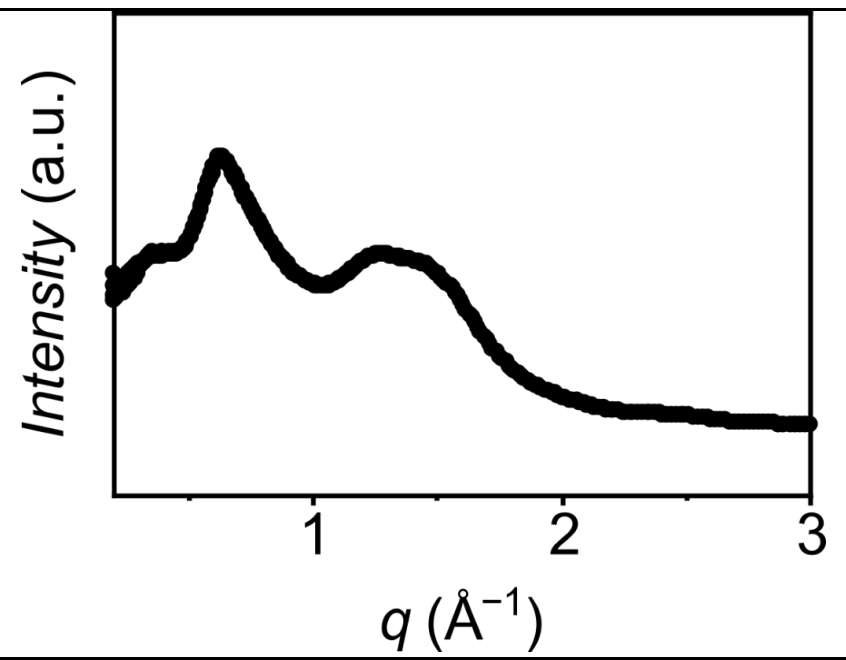 | 0.63<br>1.40 | 9.9<br>4.5 |
| **21** | 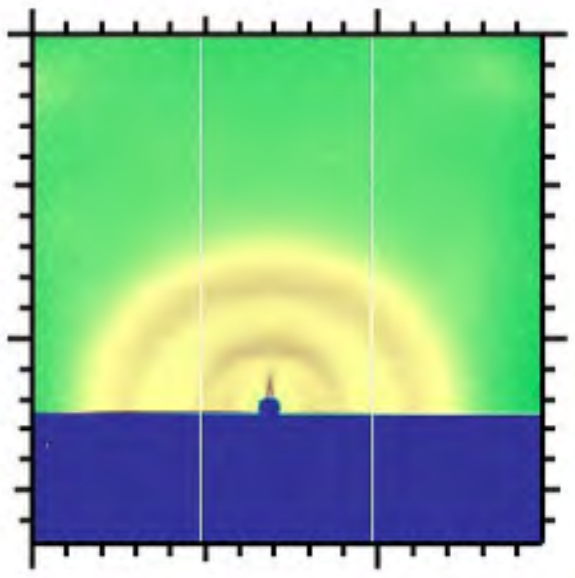 | 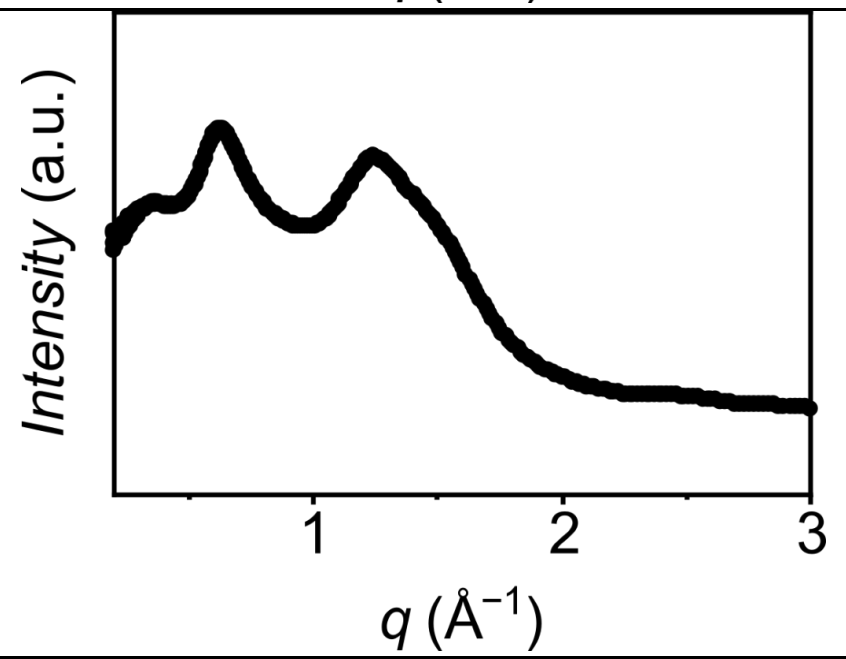 | 0.63<br>1.25 | 10.0<br>5.0 |
| **22** | 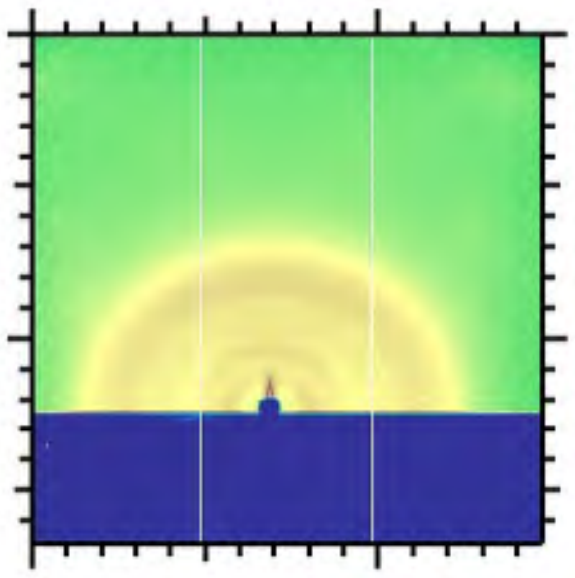 | 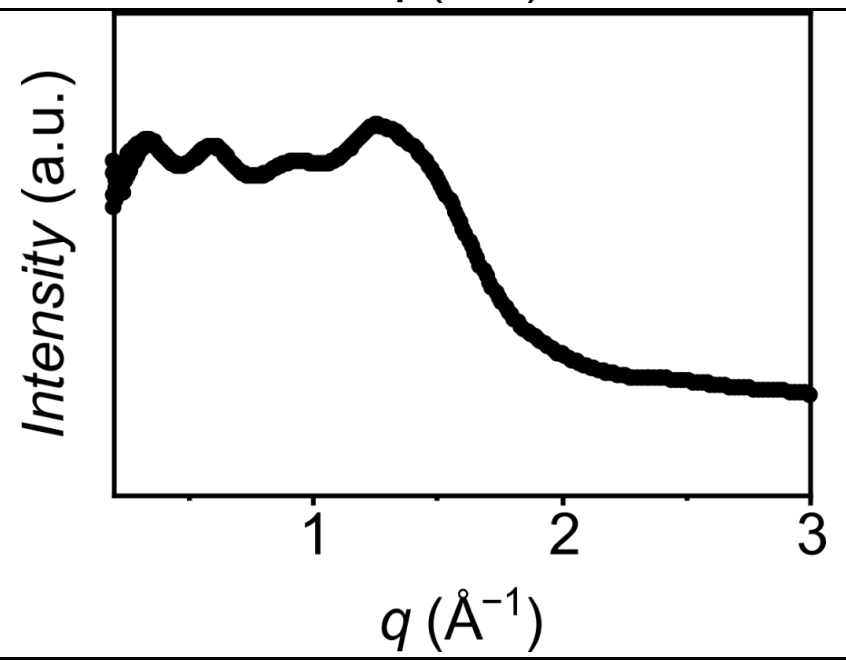 | 0.34<br>0.60<br>0.94<br>1.26 | 18.4<br>10.5<br>6.7<br>5.0 |
| **23** | 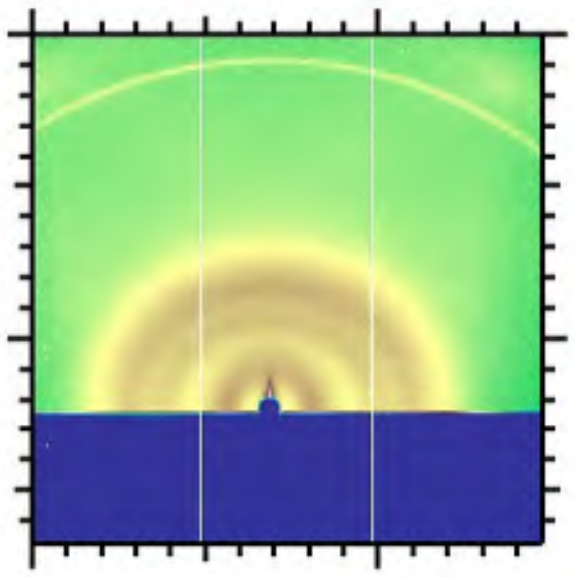 | 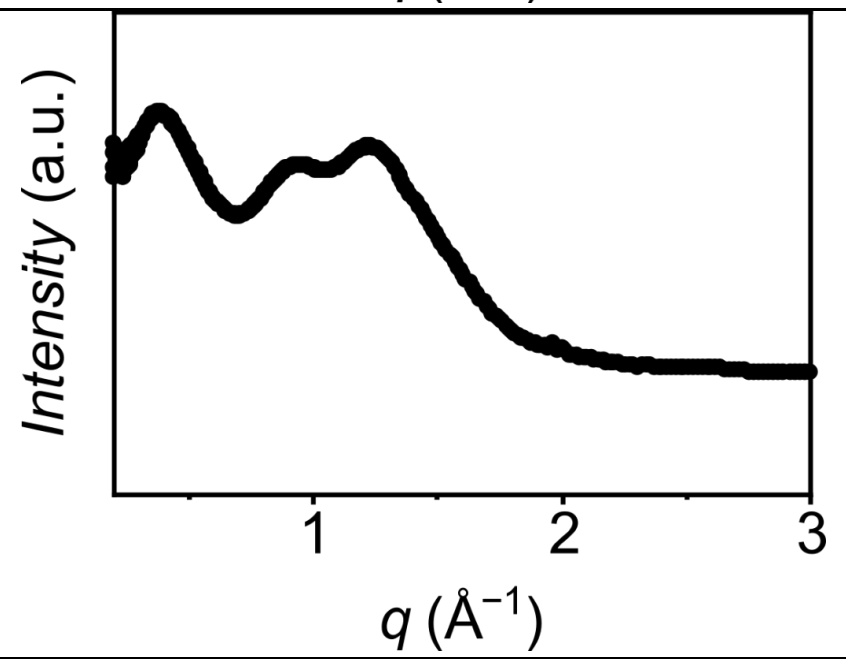 | 0.39<br>0.94<br>1.23 | 16.0<br>6.7<br>5.1 |
| **24** | N.D. | N.D. | N.D. | N.D. |
| **25** | 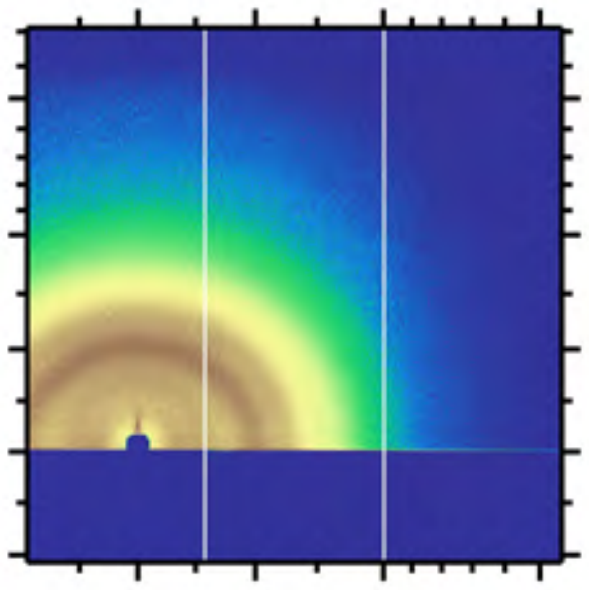 | 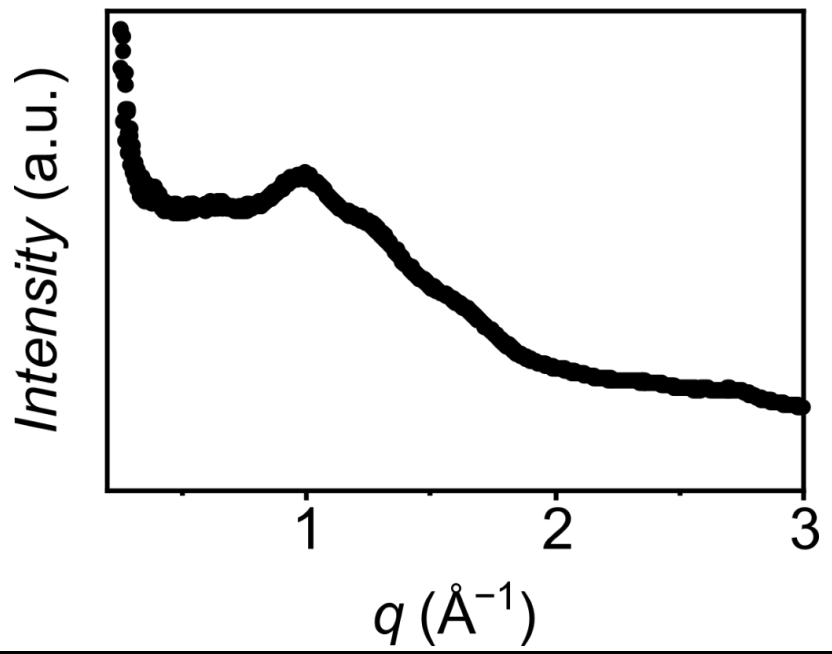 | 0.99<br>1.31<br>1.64 | 6.3<br>4.8<br>3.8 |
| **26** | 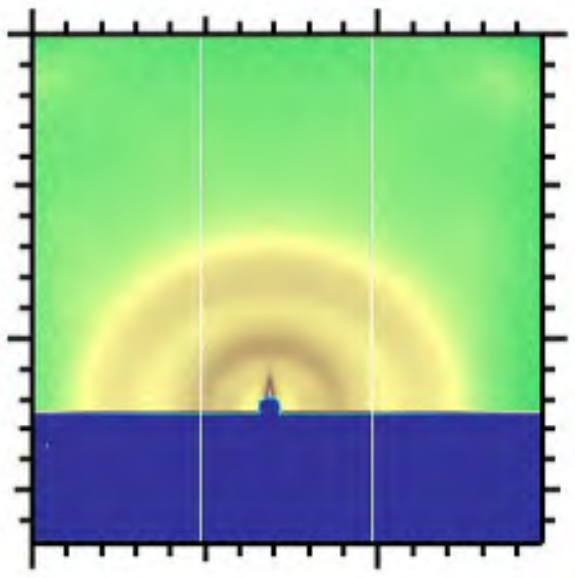 | 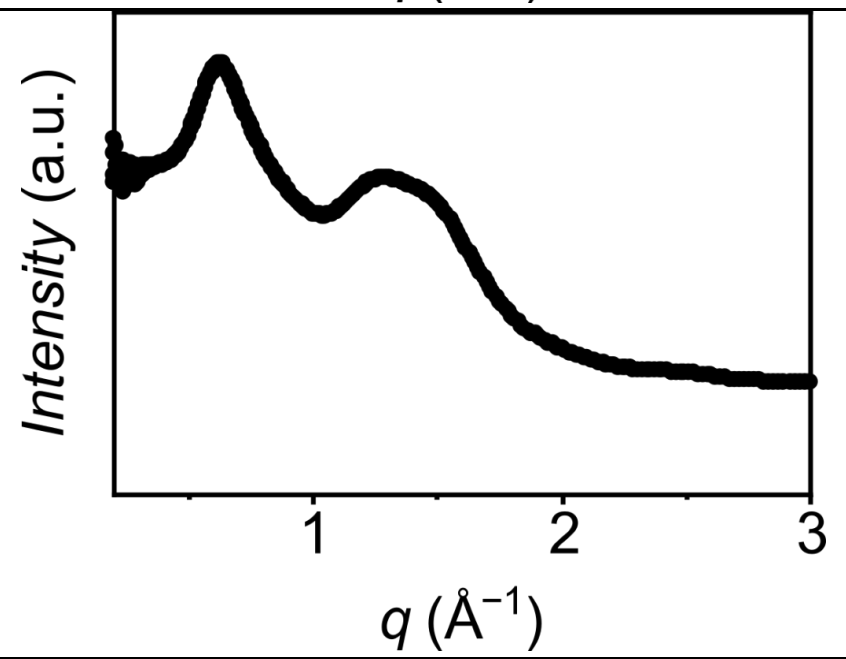 | 0.63<br>1.37 | 10.0<br>4.6 |
| **27** | 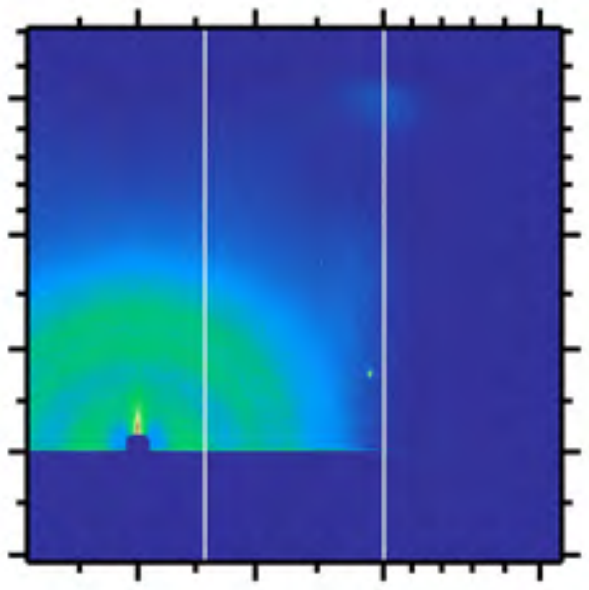 | 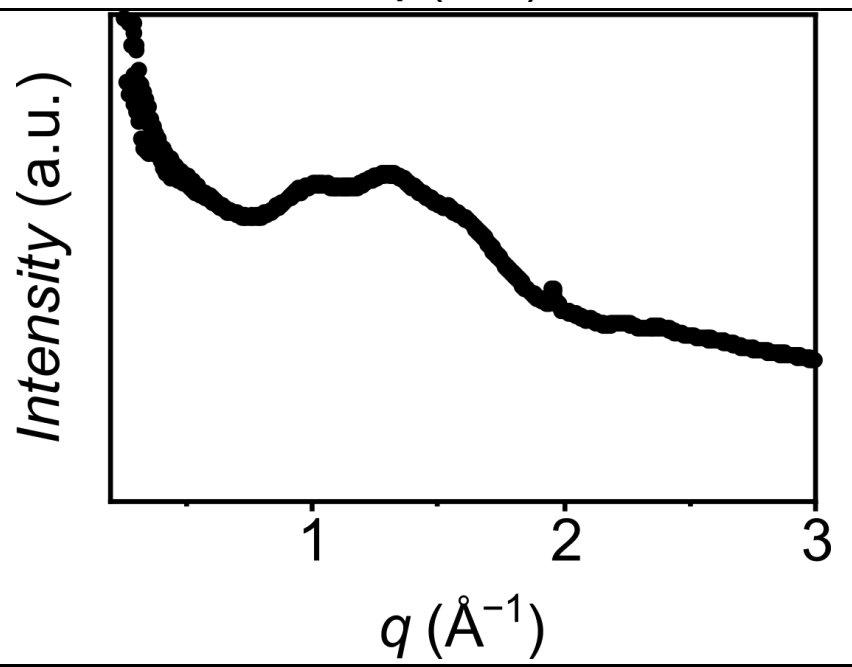 | 1.01<br>1.31<br>1.64 | 6.2<br>4.8<br>3.8 |
| **28** | N.D. | N.D. | N.D. | N.D. |
| **29** | 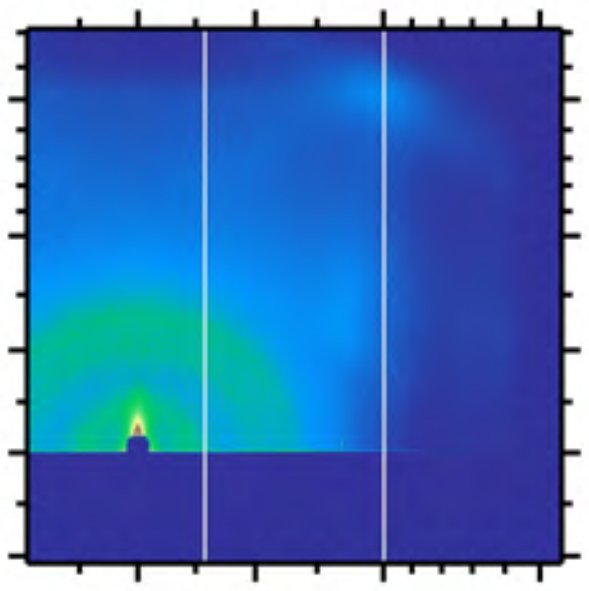 | 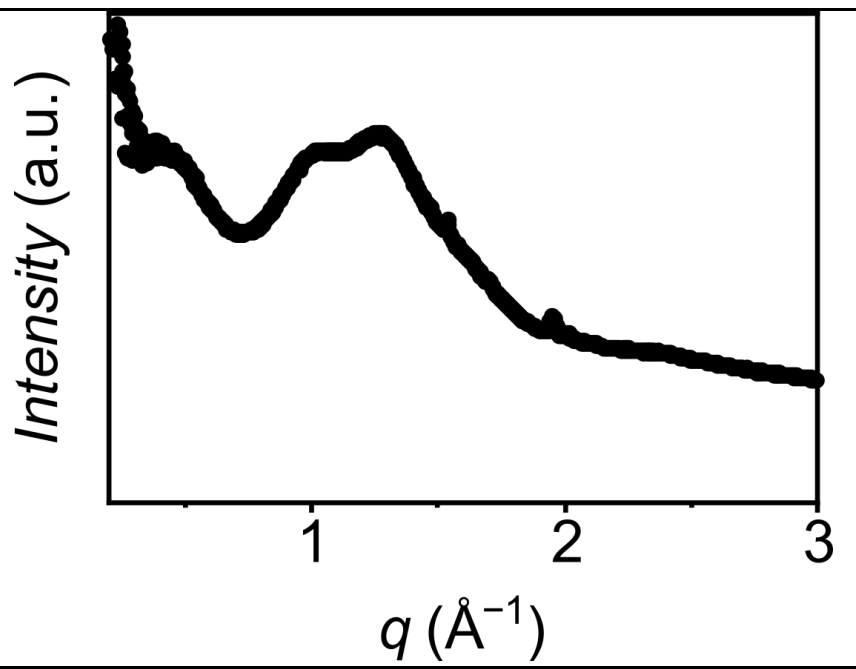 | 1.04<br>1.27 | 6.1<br>4.9 |
| **30** | 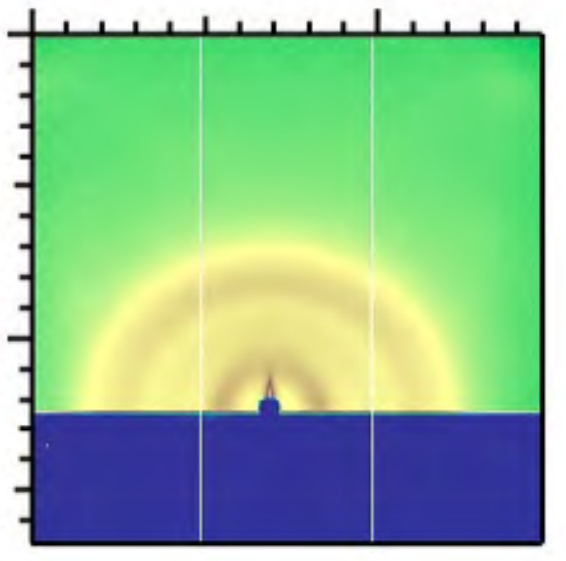 | 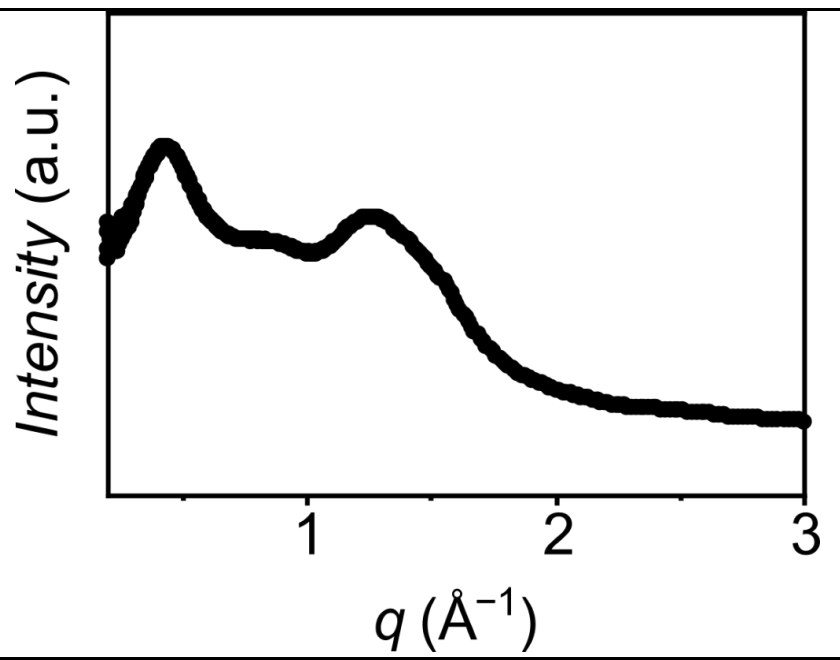 | 0.43<br>0.86<br>1.26 | 14.5<br>7.3<br>5.0 |
| **31** | 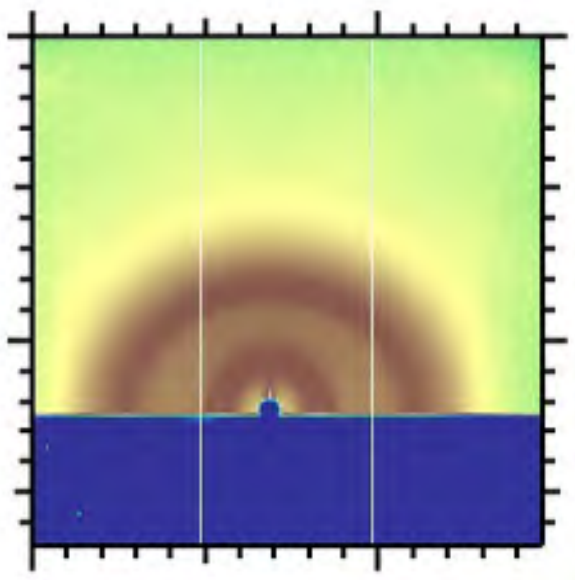 | 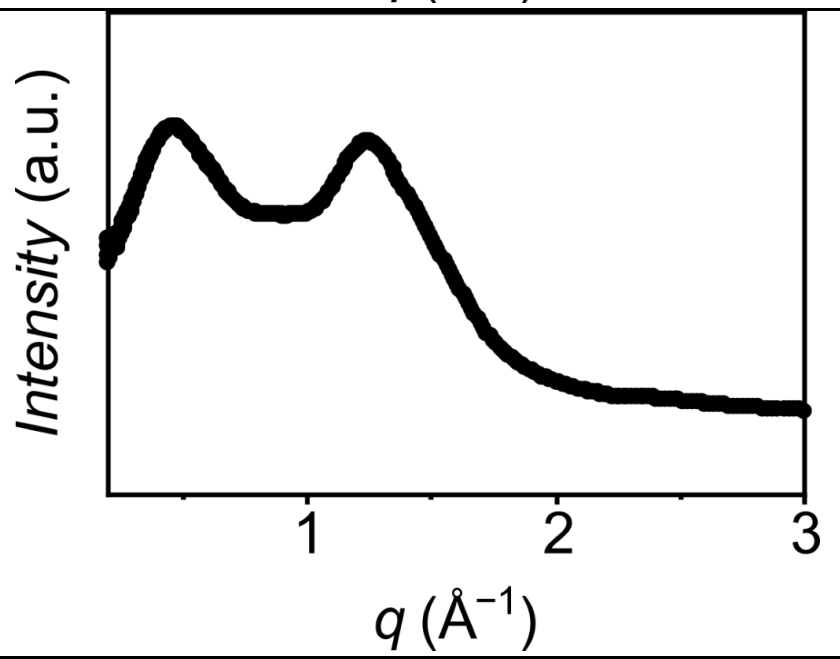 | 0.46<br>1.25 | 13.5<br>5.0 |

**Supplementary Table 3.** Raw data for the principle component analysis of polymers presented in this paper, used to generate **Fig. 1b**; where $M_s$ is the molecular weight of the side chains per repeat, $A_{BET}$ is the BET surface area, $N_O$, $N_N$, and $N_s$ are the numbers of oxygen atoms, nitrogen atoms, and side groups per unit respectively, $M_b$ is the molecular weight of the main chain repeat unit, and $N_C$ is the number of sites of contortion along the polymer main chain repeat unit.

| **Polymer** | $M_s$ (g mol$^{-1}$) | $A_{BET}$ (m$^2$ g$^{-1}$) | $N_O$ | $N_N$ | $N_s$ | $M_b$ (g mol$^{-1}$) | $N_c$ |
|---|---|---|---|---|---|---|---|
| **11** | 284 | 446 | 4 | 4 | 3 | 346 | 1 |
| **12** | 280 | 434 | 4 | 4 | 3 | 346 | 1 |
| **13** | 312 | 505 | 6 | 4 | 3 | 346 | 1 |
| **14** | 427 | 63 | 6 | 6 | 3 | 346 | 1 |
| **15** | 511 | 234 | 8 | 6 | 3 | 346 | 1 |
| **16** | 340 | 11 | 6 | 4 | 3 | 346 | 1 |
| **17** | 278 | 444 | 4 | 4 | 4 | 306 | 1 |
| **18** | 310 | 19 | 6 | 4 | 4 | 306 | 1 |
| **20** | 351 | 79 | 6 | 6 | 3 | 346 | 1 |
| **21** | 379 | 152 | 8 | 6 | 3 | 346 | 1 |
| **22** | 493 | 10 | 8 | 8 | 3 | 346 | 1 |
| **23** | 577 | 12 | 10 | 8 | 3 | 346 | 1 |
| **24** | 376 | 10 | 8 | 6 | 4 | 306 | 1 |
| **25** | 234 | 7 | 6 | 4 | 2 | 348 | 1 |
| **26** | 407 | 8 | 6 | 6 | 3 | 346 | 1 |
| **27** | 234 | 10 | 6 | 4 | 2 | 348 | 1 |
| **28** | 433 | 24 | 8 | 6 | 4 | 306 | 1 |
| **29** | 290 | 11 | 6 | 4 | 2 | 348 | 1 |
| **30** | 463 | 48 | 6 | 6 | 3 | 346 | 1 |
| **31** | 491 | 31 | 8 | 6 | 3 | 346 | 1 |
| Thioamide-PIM-1 | 180 | 263 | 4 | 2 | 2 | 348 | 1 |
| Amidoxime-PIM-1 | 178 | 490 | 6 | 4 | 2 | 348 | 1 |
| Amine-PIM-1 | 120 | 0 | 4 | 2 | 2 | 348 | 1 |
| Carboxylated-PIM-1-5H | 150 | 378 | 8 | 0 | 2 | 348 | 1 |
| TZPIM-3 | 198 | 30 | 4 | 8 | 2 | 348 | 1 |
| PIM-1 | 112 | 800 | 4 | 2 | 2 | 348 | 1 |
| PIM-EA-TB | 72 | 1030 | 0 | 2 | 3 | 228 | 2 |
| PIM-SBI-TB | 74 | 745 | 0 | 2 | 2 | 268 | 2 |
| PIM-EA | 110 | 630 | 4 | 2 | 3 | 308 | 1 |
| PIM-Trip-TB | 90 | 900 | 0 | 2 | 2 | 257 | 2 |
| PIM-TMN-SBI | 376 | 1015 | 4 | 2 | 2 | 344 | 1 |
| TPIM-1 | 162 | 860 | 2 | 2 | 2 | 306 | 1 |
| Tp-Ad-Me | 166 | 615 | 0 | 2 | 3 | 214 | 2 |
| PIM-TMN-Trip | 318 | 1050 | 4 | 2 | 3 | 308 | 1 |
| PIM-SBF-1 | 156 | 803 | 4 | 2 | 2 | 344 | 1 |
| PIM-HPB-2 | 360 | 537 | 4 | 2 | 2 | 358 | 1 |
| PIM-5 | 218 | 540 | 4 | 0 | 2 | 354 | 2 |
| Cardo-PIM-1 | 224 | 621 | 4 | 4 | 2 | 580 | 2 |
| Pf5 | 84 | 501 | 6 | 2 | 2 | 348 | 1 |
| Pf6 | 206 | 560 | 4 | 2 | 2 | 350 | 1 |
| Pf7 | 356 | 895 | 4 | 2 | 2 | 348 | 1 |
| PIM-4bIII | 381 | 660 | 6 | 3 | 3 | 308 | 1 |
| PIM-c1 | 156 | 818 | 5 | 2 | 3 | 346 | 1 |
| PIM-PI-1 | 184 | 682 | 8 | 2 | 2 | 572 | 1 |
| PIM-PI-3 | 262 | 471 | 8 | 2 | 3 | 664 | 2 |
| PIM-PI-10 | 258 | 699 | 4 | 2 | 3 | 434 | 1 |
| PIM-PI-11 | 184 | 220 | 4 | 2 | 2 | 502 | 2 |
| PIM-PI-EA | 256 | 620 | 4 | 2 | 4 | 458 | 1 |
| PIM-6FDA-OH | 296 | 225 | 6 | 2 | 4 | 450 | 2 |
| PIM-PMDA-OH | 158 | 190 | 6 | 2 | 3 | 362 | 1 |
| SPFDA-DMN | 302 | 686 | 4 | 2 | 3 | 430 | 1 |
| 6FDA-SBF | 306 | 240 | 4 | 2 | 3 | 448 | 2 |
| PI-TB-2 | 140 | 580 | 5 | 4 | 3 | 454 | 2 |
| 4MTBDA-6FDA | 262 | 584 | 4 | 4 | 4 | 454 | 2 |
| 4MTBDA-SBFDA | 228 | 739 | 4 | 4 | 4 | 500 | 2 |
| TBDA1-SBI-PI | 154 | 560 | 8 | 4 | 3 | 718 | 2 |
| KAUST-PI-4 | 226 | 420 | 9 | 2 | 3 | 628 | 2 |
| spiroTR-PBO-6F | 198 | 368 | 2 | 2 | 2 | 460 | 2 |
| PIM-PBO-3 | 60 | 360 | 6 | 2 | 1 | 584 | 1 |
| SPDA-SBF-PBO | 164 | 480 | 6 | 2 | 2 | 718 | 2 |

**Supplementary Table 4.** Raw data for the principle component analysis of polymers presented in this paper, used to generate **Supplementary Fig. 11**; where $M_s$ is the molecular weight of the side chains per repeat, $A_{BET}$ is the BET surface area, $N_O$, $N_N$, and $N_s$ are the numbers of oxygen atoms, nitrogen atoms, and side groups per unit respectively, $M_b$ is the molecular weight of the main chain repeat unit, $N_G$ is the number of GIWAXS reflections, and $L_G$, $L_{N2}$, and $L_{CO2}$ are the pore diameters based to the real value of the major GIWAXS reflection, the pore size distributions from the $N_2$ and $CO_2$ adsorption isotherms respectively.

| Polymer | $M_s$ (g mol$^{-1}$) | $A_{BET}$ (m$^2$ g$^{-1}$) | $N_O$ | $N_N$ | $N_s$ | $M_b$ (g mol$^{-1}$) | $N_G$ | $L_G$ (Å) | $L_{N2}$ (Å) | $L_{CO2}$ (Å) |
|---|---|---|---|---|---|---|---|---|---|---|
| **11** | 284 | 446 | 4 | 4 | 3 | 346 | 2 | 9.6 | 10.2 | |
| **12** | 280 | 434 | 4 | 4 | 3 | 346 | 2 | 9.5 | 5.9 | |
| **13** | 312 | 505 | 6 | 4 | 3 | 346 | 2 | 9.3 | 9.5 | |
| **14** | 427 | 63 | 6 | 6 | 3 | 346 | 2 | 9.8 | 14.1 | |
| **15** | 511 | 234 | 8 | 6 | 3 | 346 | 3 | 13.3 | 14.4 | |
| **16** | 340 | 11 | 6 | 4 | 3 | 346 | 2 | 10 | 22.5 | 6.1 |
| **17** | 278 | 444 | 4 | 4 | 4 | 306 | 3 | 11.9 | 10.5 | |
| **18** | 310 | 19 | 6 | 4 | 4 | 306 | 2 | 9.6 | 25.5 | 5.6 |
| **20** | 351 | 79 | 6 | 6 | 3 | 346 | 2 | 9.9 | 12.1 | |
| **21** | 379 | 152 | 8 | 6 | 3 | 346 | 2 | 10 | 15.2 | |
| **22** | 493 | 10 | 8 | 8 | 3 | 346 | 4 | 5 | 24.3 | |
| **23** | 577 | 12 | 10 | 8 | 3 | 346 | 3 | 16 | 19.9 | 8.4 |
| **24** | 376 | 10 | 8 | 6 | 4 | 306 | N.D. | N.D. | 24.3 | 5.2 |
| **25** | 234 | 7 | 6 | 4 | 2 | 348 | 3 | 6.3 | 25.5 | 5.4 |
| **26** | 407 | 8 | 6 | 6 | 3 | 346 | 2 | 10 | 19.9 | 5.4 |
| **27** | 234 | 10 | 6 | 4 | 2 | 348 | 3 | 4.8 | 20.4 | 5.8 |
| **28** | 433 | 24 | 8 | 6 | 4 | 306 | N.D. | N.D. | 18.5 | |
| **29** | 290 | 11 | 6 | 4 | 2 | 348 | 2 | 4.9 | 26.8 | 5.2 |
| **30** | 463 | 48 | 6 | 6 | 3 | 346 | 3 | 14.5 | 14.8 | |
| **31** | 491 | 31 | 8 | 6 | 3 | 346 | 2 | 13.5 | 17.6 | |

**Supplementary Table 5.** Summary of properties presented in **Fig. 2**

| Polymer | BET Surface Area ($m^2 g^{-1}$) | Conductivity at 25 °C ($mS cm^{-1}$) | Activation Energy (eV) | Transference Number |
|---|---|---|---|---|
| Celgard | | 0.413 | 0.092 | 0.302 |
| PIM-1 | 800 | 0.0377 | 0.109 | 0.672 |
| **11** | 446 | 0.0296 | 0.156 | 0.664 |
| **12** | 434 | 0.0280 | 0.161 | 0.446 |
| **13** | 505 | 0.206 | 0.168 | 0.696 |
| **14** | 63 | 0.195 | 0.165 | 0.842 |
| **15** | 234 | 0.0418 | 0.189 | 0.760 |
| **17** | 444 | 0.0279 | 0.224 | 0.241 |
| **18** | 19 | 0.0602 | 0.222 | 0.250 |
| **21** | 152 | 0.0243 | 0.236 | 0.390 |
| **23** | 12 | 0.0634 | 0.252 | 0.317 |
| **24** | 10 | 0.147 | 0.288 | 0.718 |
| **25** | 7 | 0.107 | 0.310 | 0.638 |
| **26** | 8 | 0.100 | 0.106 | 0.224 |
| **28** | 24 | 0.0882 | 0.225 | 0.426 |
| **30** | 48 | 0.122 | 0.270 | 0.515 |
| **31** | 31 | 0.192 | 0.262 | 0.258 |

**Supplementary Table 6.** NBO charges of the cage atoms

| Atom | NBO charges (b3lyp, 6-311++G**) |
|---|---|
| O1 | –0.61560 |
| N1 | –0.81660 |
| O2 | –0.59112 |
| N2 | –0.55780 |

**Supplementary Table 7.** Energies of ions, molecules and clusters in a.u. obtained with B3LYP5-D3/6-311++G** and COSMO model.

| | $Li^+$ in cage of the polymer unit | Polymer unit | $[Li(DMC)_4]^+$ | $Li^+$ | $DMC$ |
|---|---|---|---|---|---|
| Vacuum | –2345.6303715028 | –2338.1387910 | –1381.7020501288 | –7.2774402290 | –343.5552004527 |
| PCM DMC ε = 3.087 | –2345.6851565112 | – | –1381.7566919966 | –7.3597988781 | –343.5621266985 |

**Supplementary Table 8.** Binding and solvation energies of molecules and clusters in kcal mol$^{-1}$ obtained with B3LYP5-D3/6-311++G** and COSMO model.

| $\Delta G_1$ | $\Delta G_2$ | $\Delta G^o_{g,bind} Li(DMC)_4$ | $\Delta G^*_{solv}([Li(DMC)_4]^+)$ | $\Delta G^*_{solv}(DMC)$ | $\Delta G^*_{sol}(Li^+)$ | $\Delta G^*_{diss}$ |
|---|---|---|---|---|---|---|
| 34.377 | 134.371 | –127.889 | –34.287 | –4.346 | **–158.214** | **+10.535** |

**Supplementary Table 9.** Parameters of classical MD simulations.

| System | # of atoms in the box | Equilibrated box size dimension (Å) |
|---|---|---|
| DMC | 21600 | 62.2525 (density 1.09 vs. 1.07 exp) |
| PIM **13**/$LiPF_6$/DMC (0.7M) | 19832 | 60.2553 |
| PIM **13**/$LiPF_6$/DMC (0.2M) | 19360 | 60.0254 |
| PIM **13**/single $LiPF_6$/DMC | 19167 | 59.4678 |

**Supplementary Table 10.** Parameters of classical MD metadynamics free energy sampling.

| | PIM **13**/single $LiPF_6$/DMC 1D, $\mathbf{r_1}$, [0; 12] Å | PIM **13**/single $LiPF_6$/DMC 2D, **d**, [–8; 8] Å | PIM **13**/single $LiPF_6$/DMC 2D, **s**, [7; 10] Å |
|---|---|---|---|
| *H* | 0.02 | 0.02 | 0.02 |
| *W* | 0.05 | 0.1 | 0.05 |
| *freq* | 500 | 200 | 200 |
| *T* | 300 | 700 | |

**Supplementary Figures**

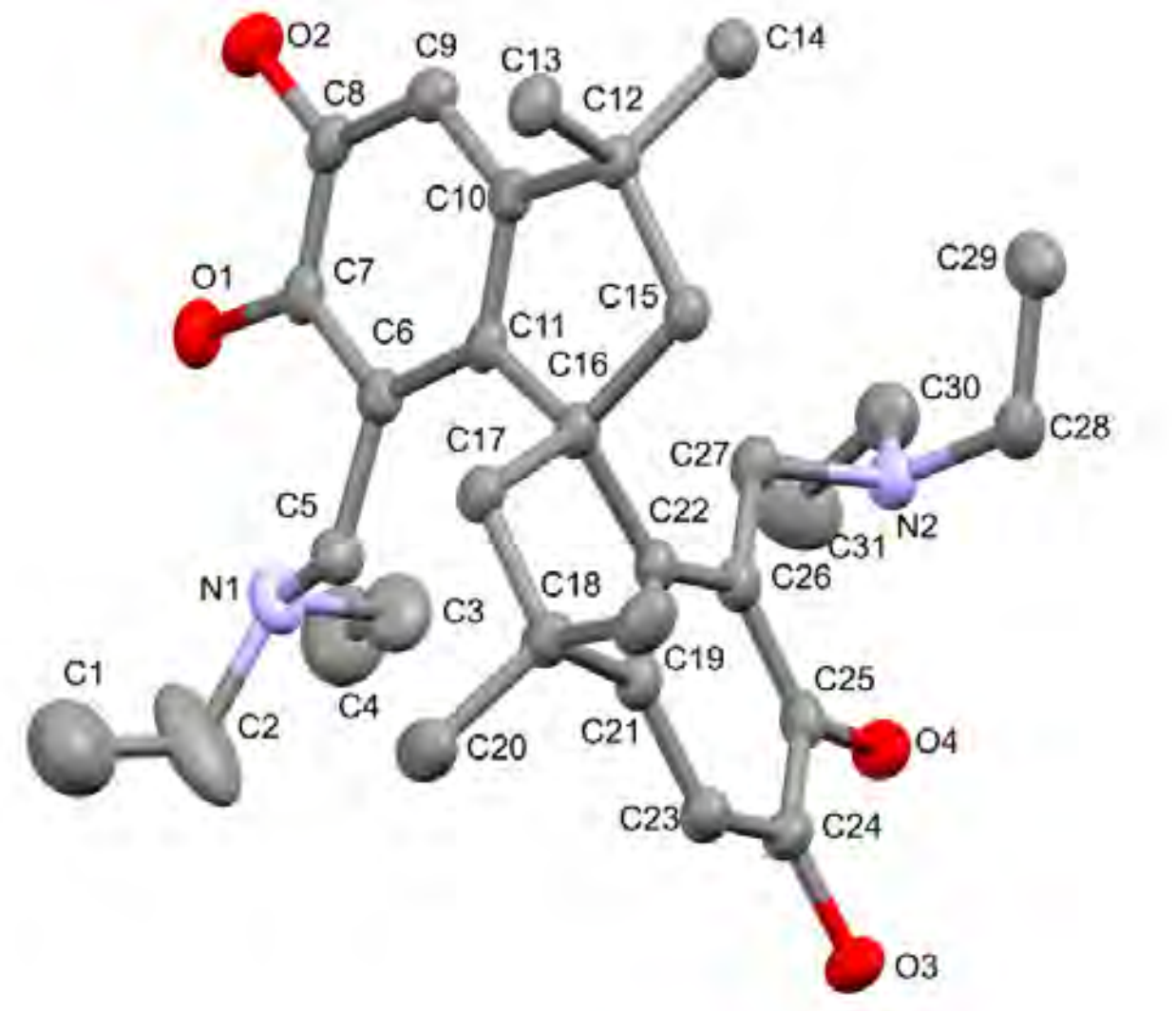

**Supplementary Figure 1 | Structure of compound 3 determined by single crystal X-ray diffraction.** Hydrogen atoms have been omitted for clarity, 50% probability ellipsoids, red oxygen, blue nitrogen and grey carbon.

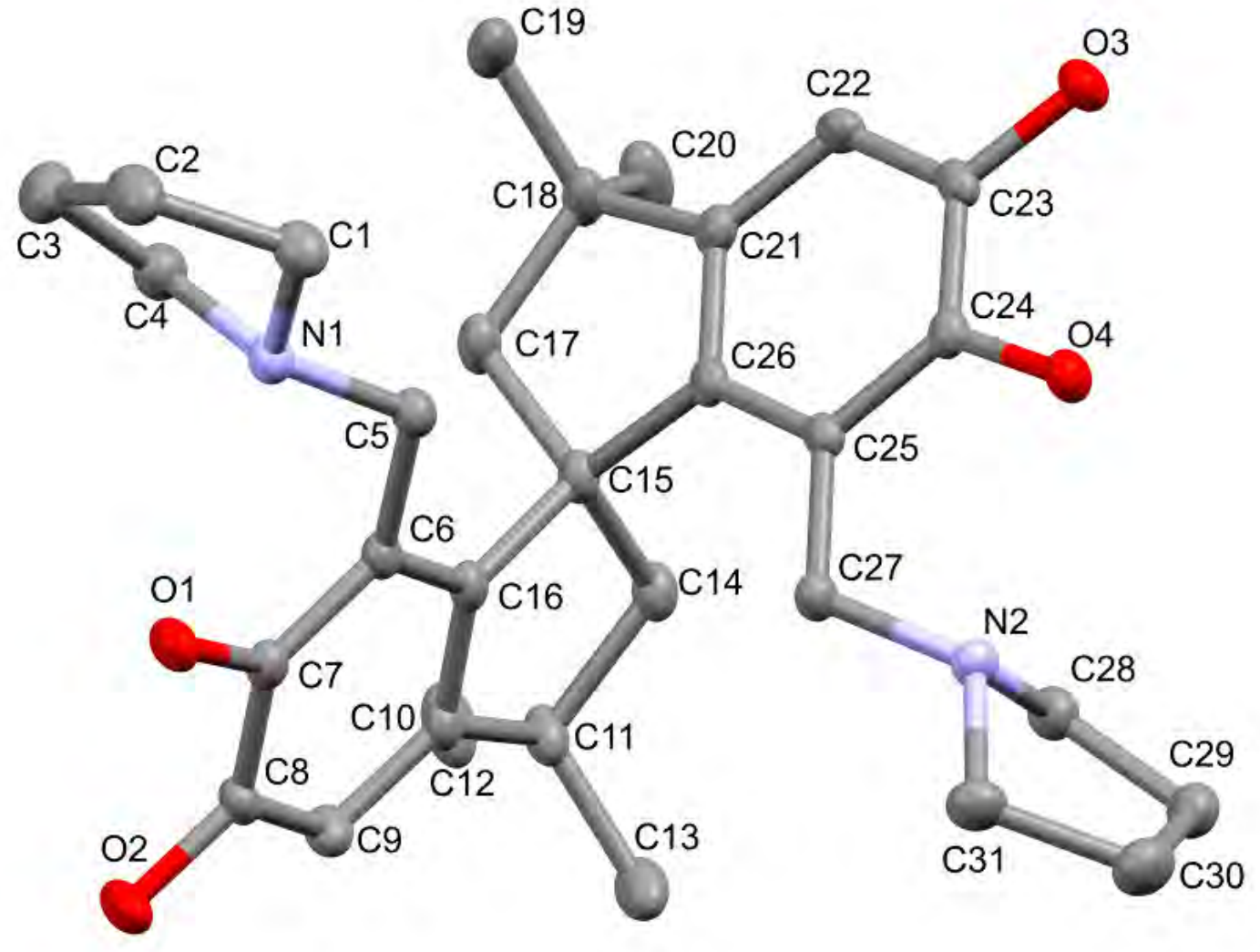

**Supplementary Figure 2 | Structure of compound 4 determined by single crystal X-ray diffraction.** Hydrogen atoms have been omitted for clarity, 50% probability ellipsoids, red oxygen, blue nitrogen and grey carbon.

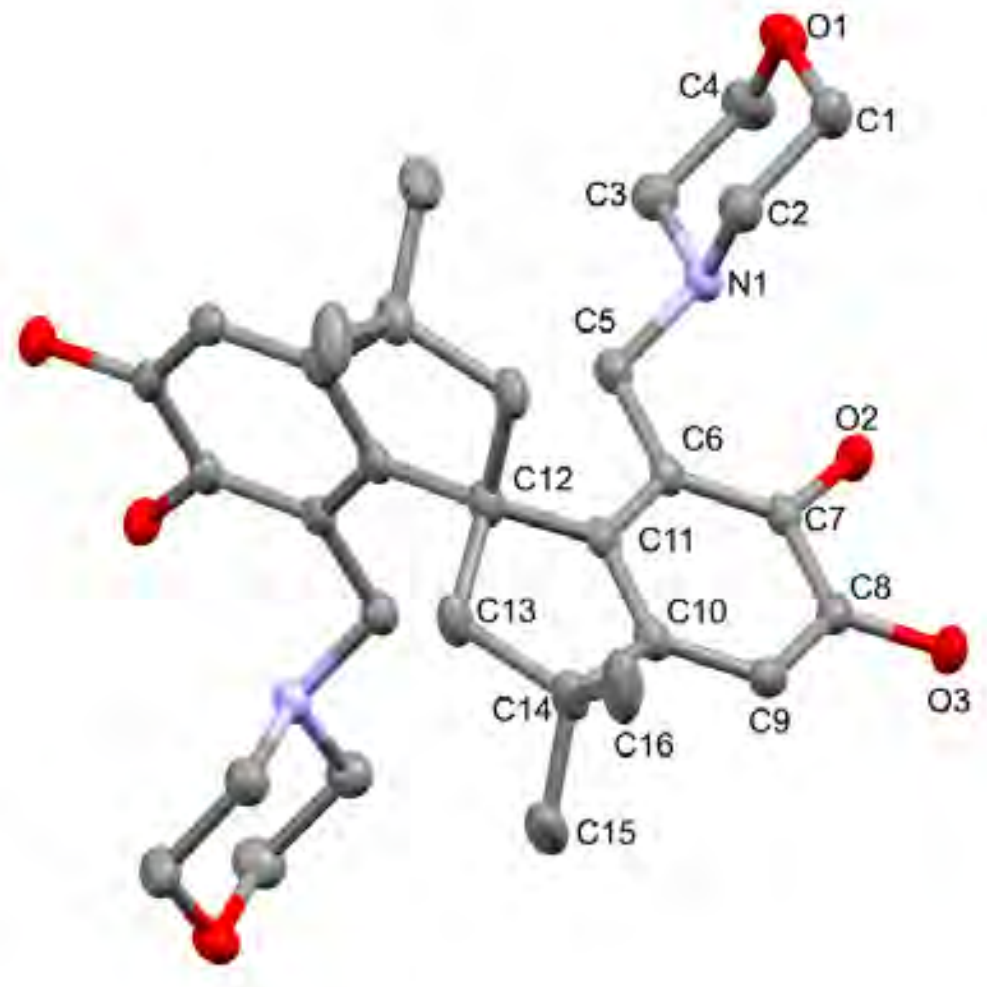

**Supplementary Figure 3 | Structure of compound 5 determined by single crystal X-ray diffraction.** Hydrogen atoms have been omitted for clarity, 50% probability ellipsoids, red oxygen, blue nitrogen and grey carbon. As half the molecule is generated by symmetry only the asymmetry unit has been labelled.

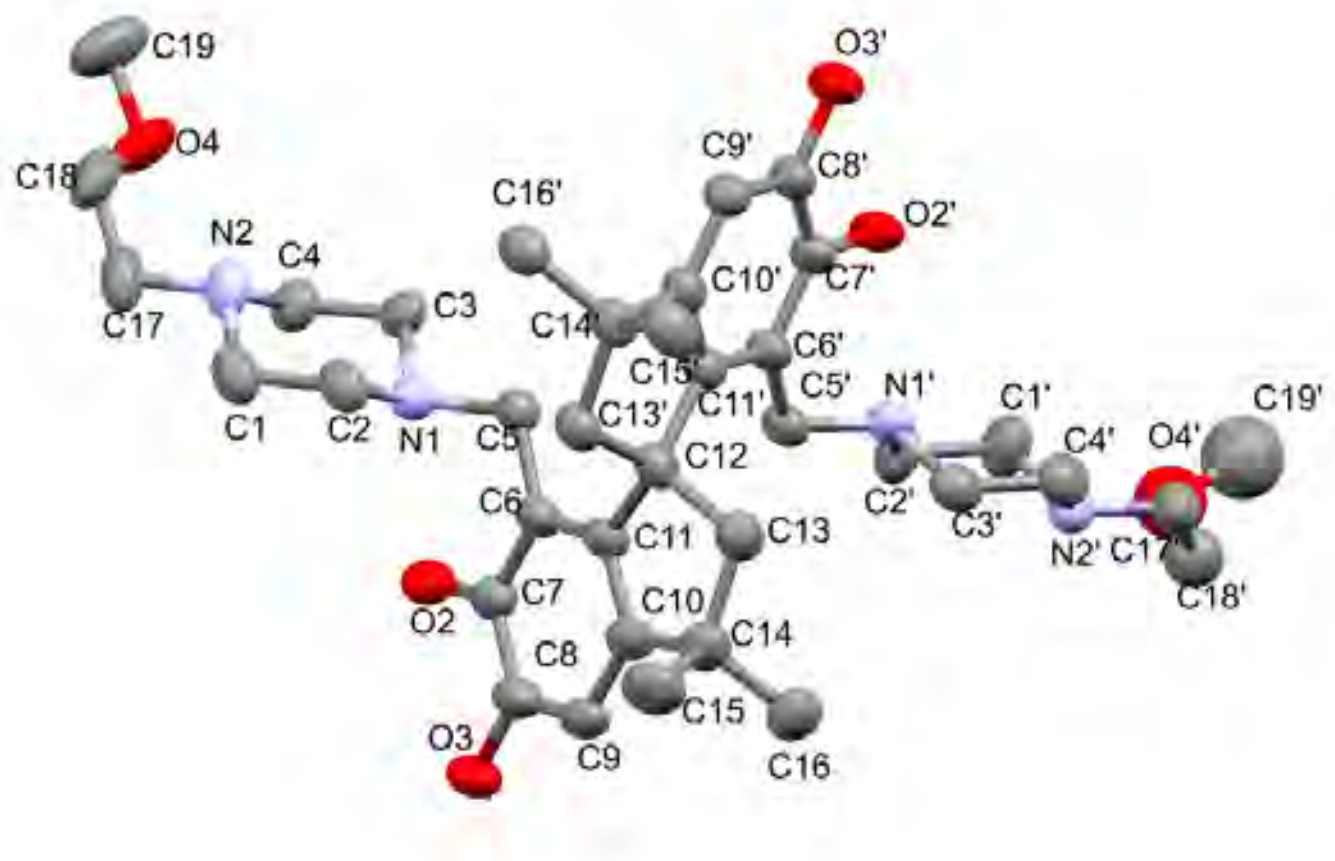

**Supplementary Figure 4 | Structure of compound 6 determined by single crystal X-ray diffraction.** Hydrogen atoms and minor component of the methoxy group disorder have been omitted for clarity, 50% probability ellipsoids, red oxygen, blue nitrogen and grey carbon.

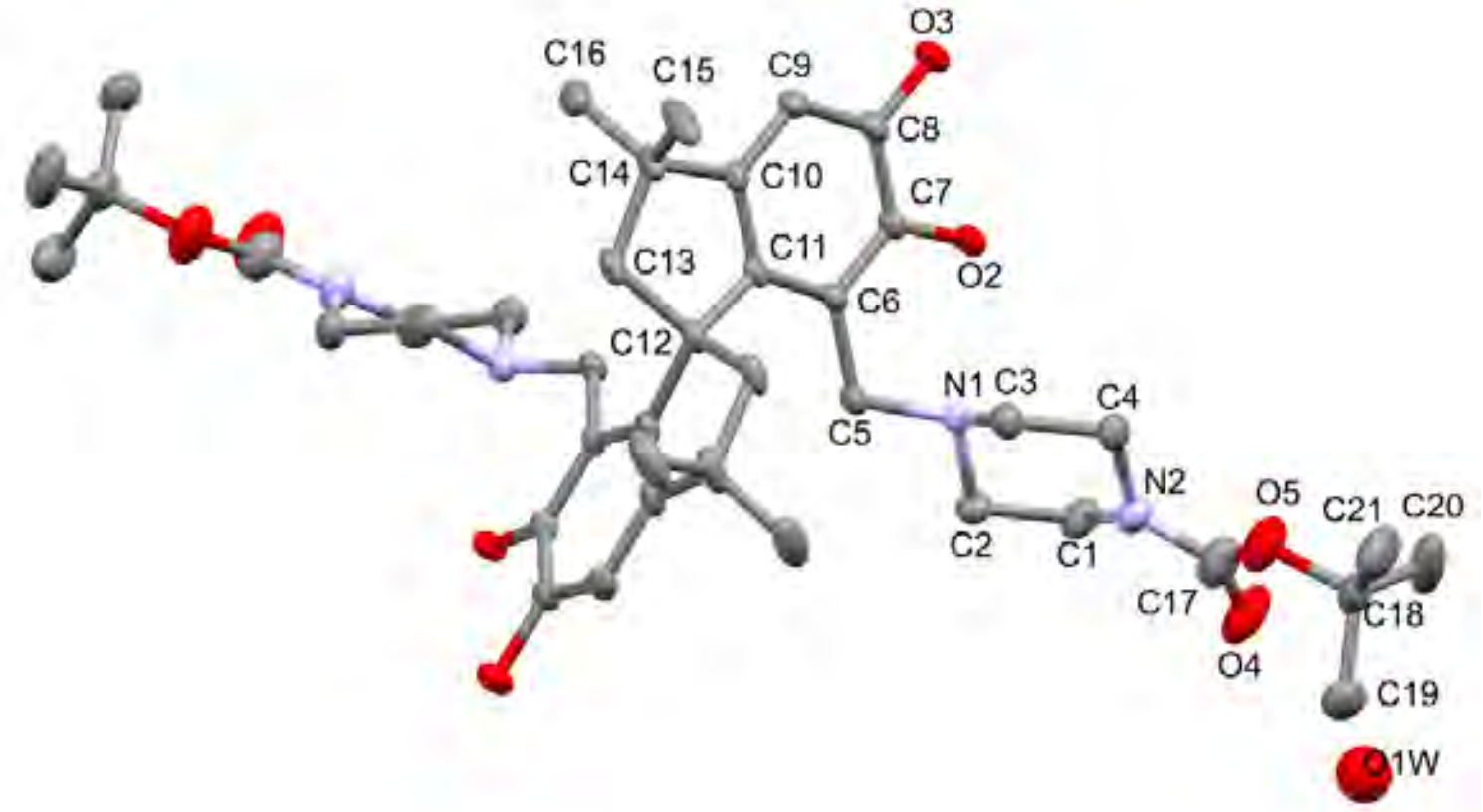

**Supplementary Figure 5 | Structure of compound 7 determined by single crystal X-ray diffraction.** Hydrogen atoms and minor component of the butoxy group disorder have been omitted for clarity, 50% probability ellipsoids, red oxygen, blue nitrogen and grey carbon. As half the molecule is generated by symmetry only the asymmetry unit has been labelled.

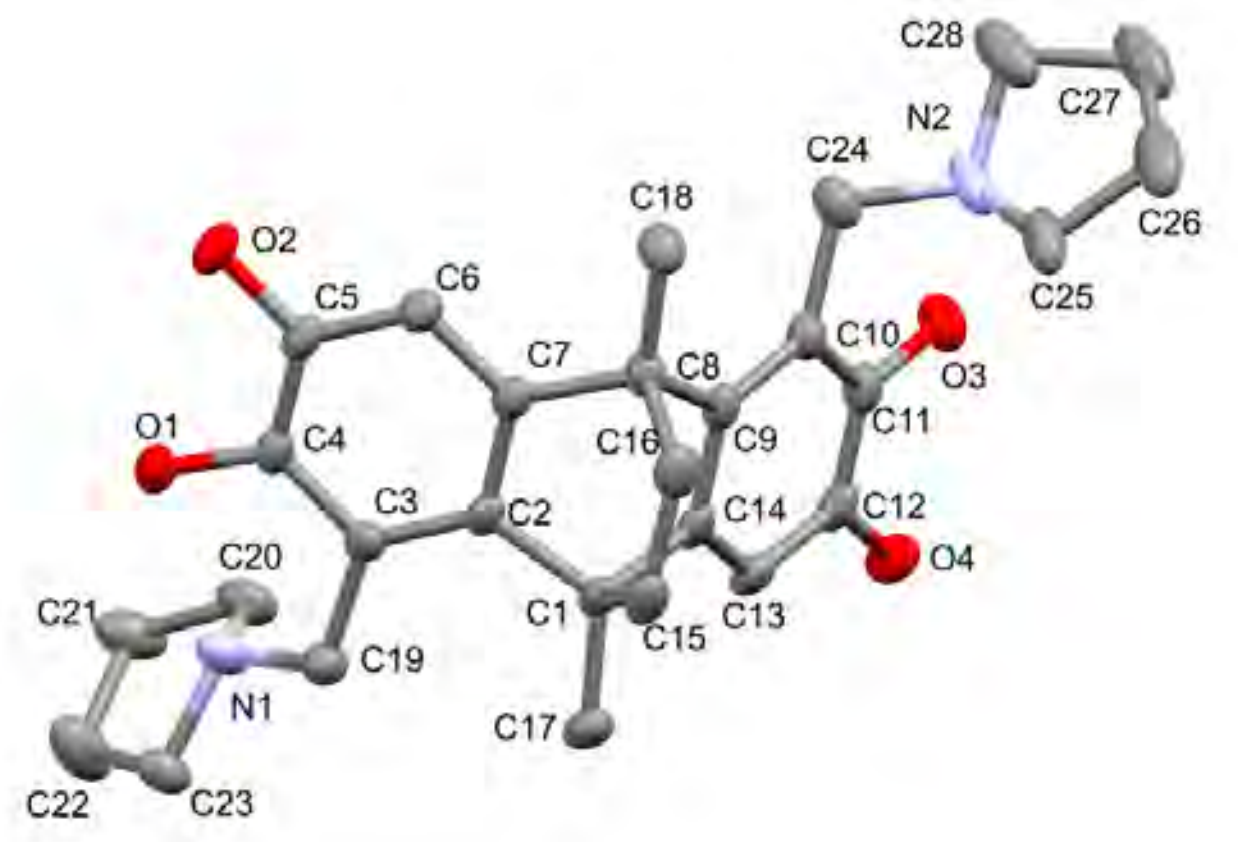

**Supplementary Figure 6 | Structure of compound 9 determined by single crystal X-ray diffraction.** Hydrogen atoms and minor component of the Pyrrolidine disorder have been omitted for clarity, 50% probability ellipsoids, red oxygen, blue nitrogen and grey carbon.

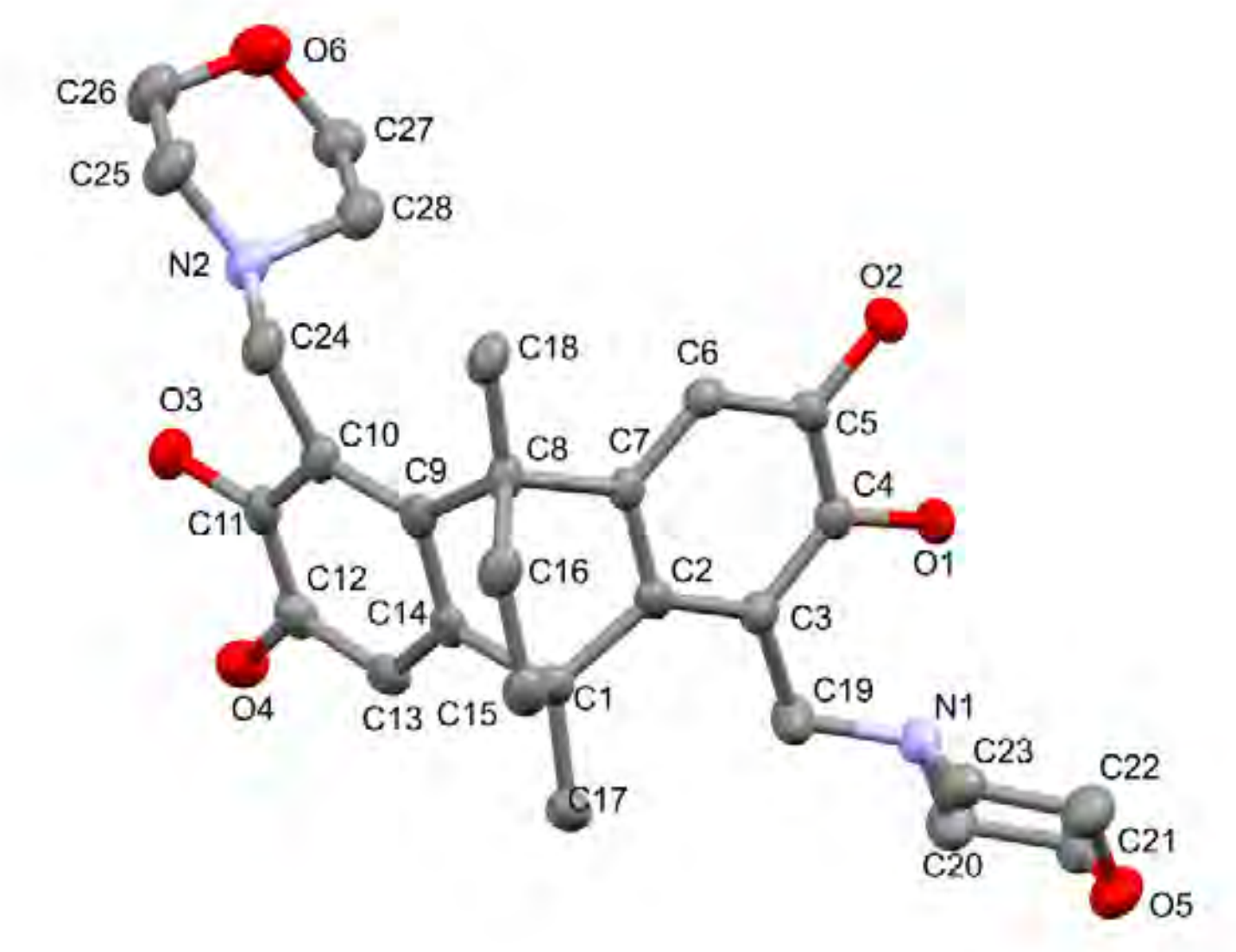

**Supplementary Figure 7 | Structure of compound 10 determined by single crystal X-ray diffraction.** Hydrogen atoms have been omitted for clarity, 50% probability ellipsoids, red oxygen, blue nitrogen and grey carbon.

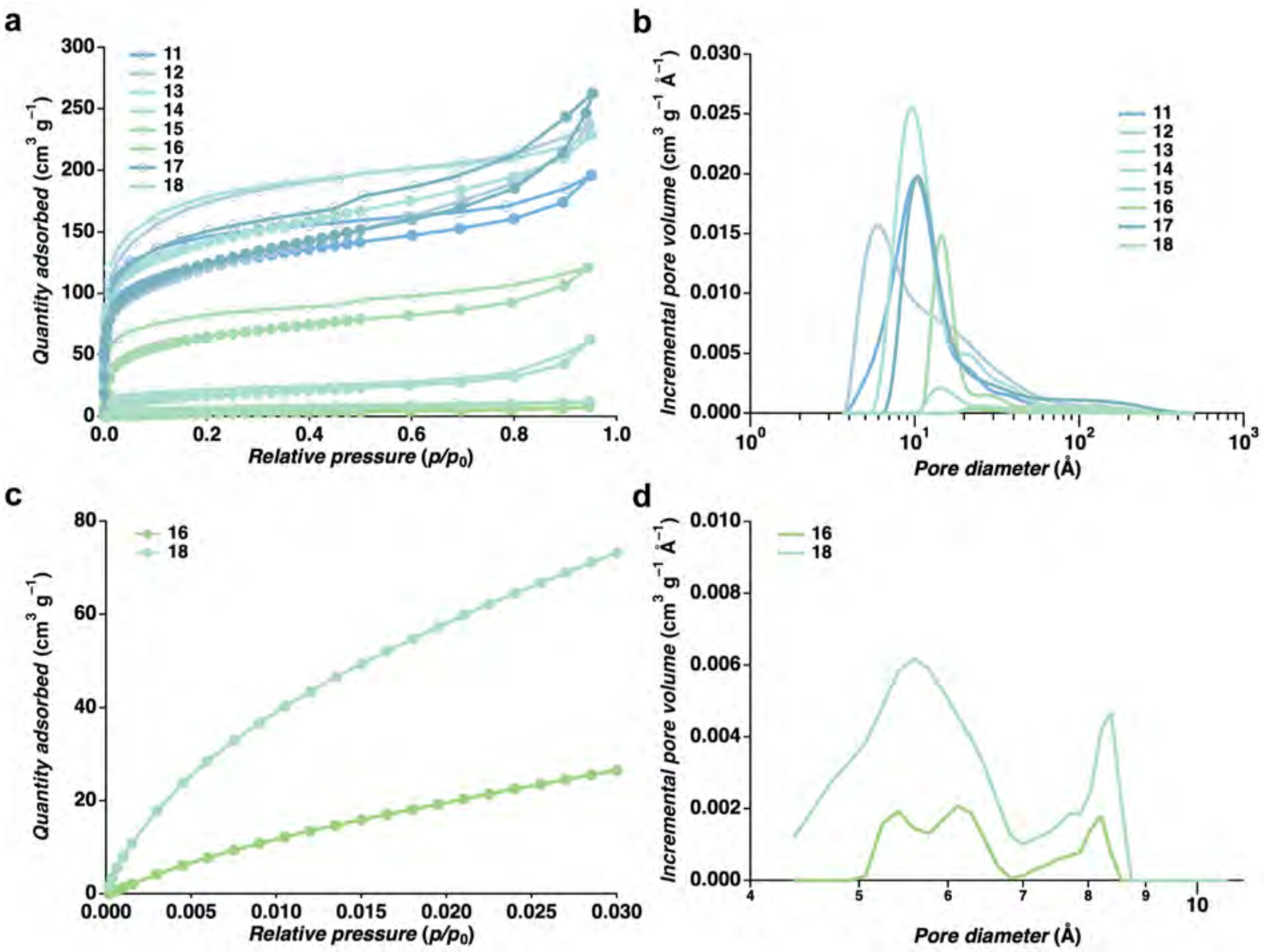

**Supplementary Figure 8 | PIM 11–18 gas adsorption. a,** $N_2$ adsorption isotherm and **b,** pore size distribution for PIMs **11–18**, and **c**, $CO_2$ adsorption and **d**, pore size distributions for those with low surface area (>30 $m^2g^{-1}$) by $N_2$.

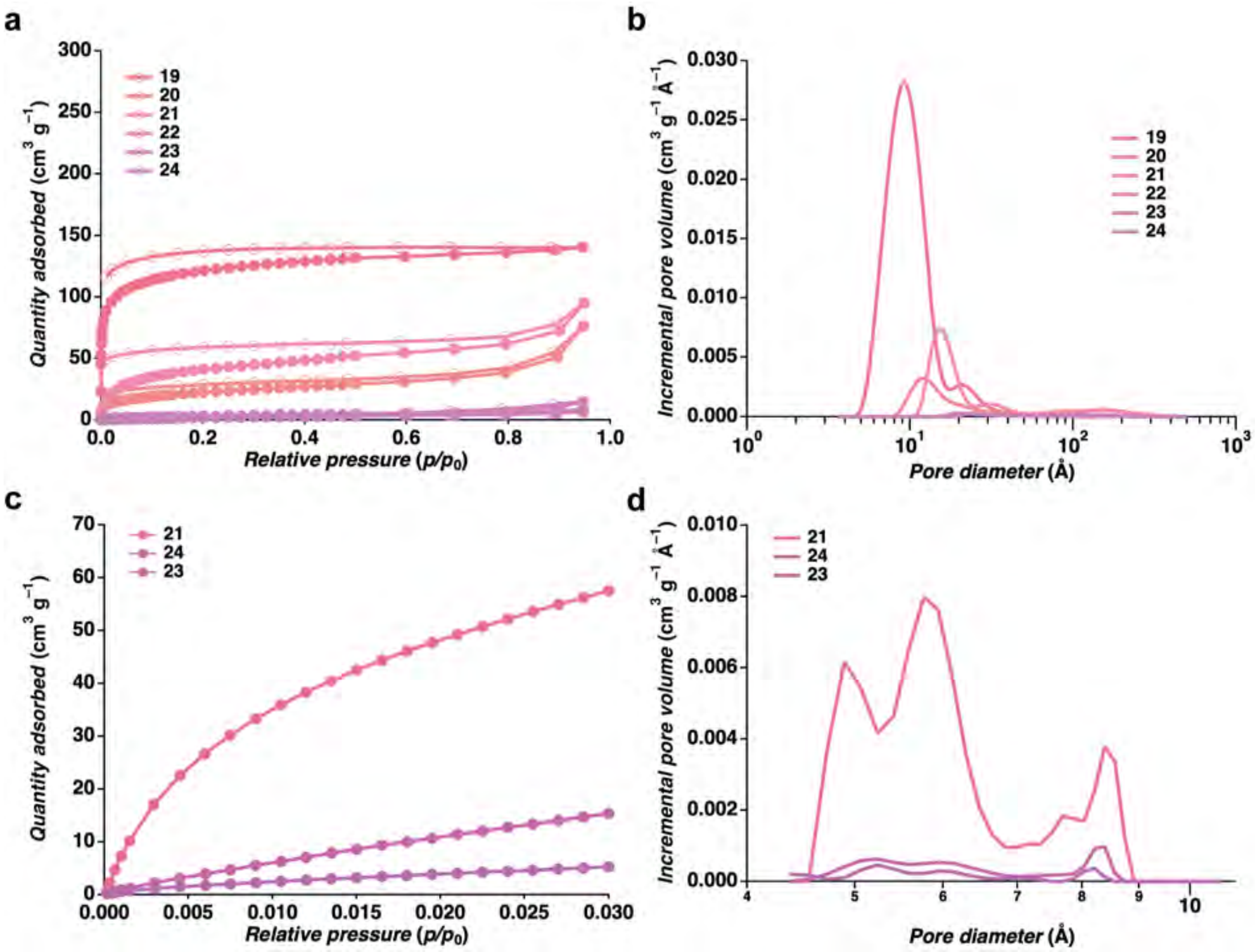

**Supplementary Figure 9 | PIM 19–24 gas adsorption. a,** $N_2$ adsorption isotherm and **b,** pore size distribution for PIMs **19–24**, and **c**, $CO_2$ adsorption and **d**, pore size distributions for those with low surface area (>30 $m^2g^{-1}$) by $N_2$.

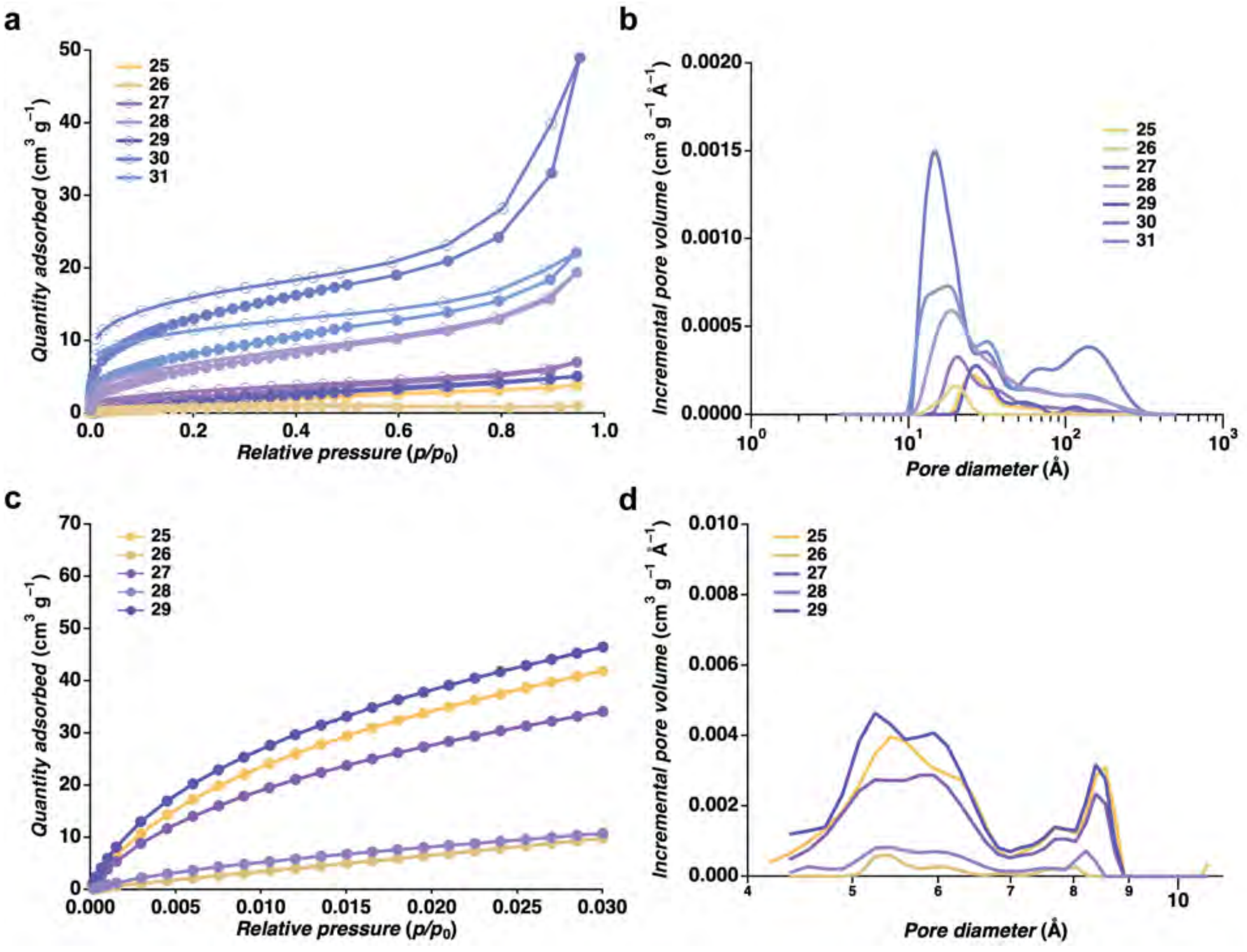

**Supplementary Figure 10 | PIM 25–31 gas adsorption. a,** $N_2$ adsorption isotherm and **b,** pore size distribution for PIMs **25–31**, and **c**, $CO_2$ adsorption and **d**, pore size distributions for those with low surface area (>30 $m^2g^{-1}$) by $N_2$.

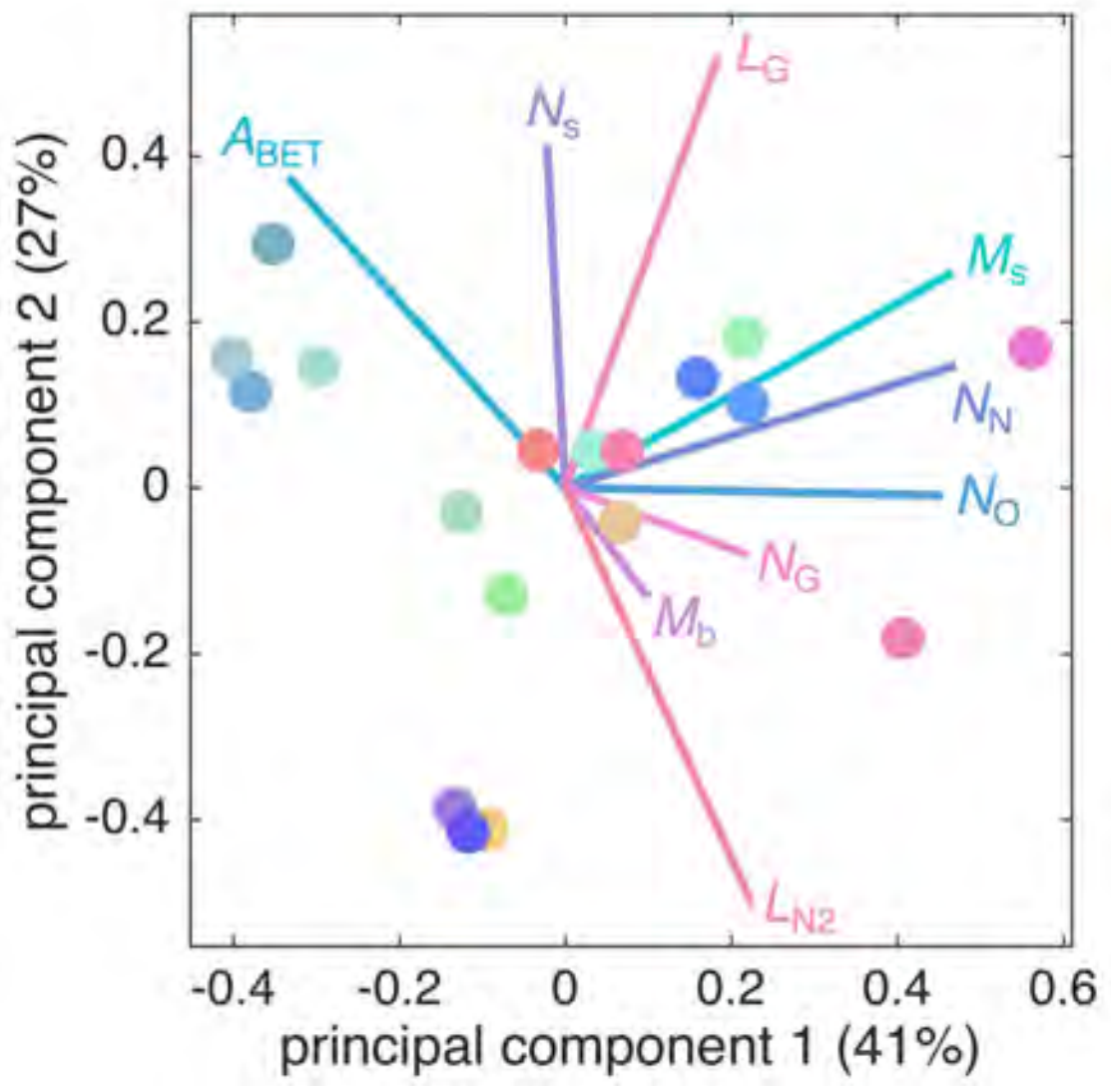

**Supplementary Figure 11 | Expanded PCA 2 for the DOS PIMs Library.**

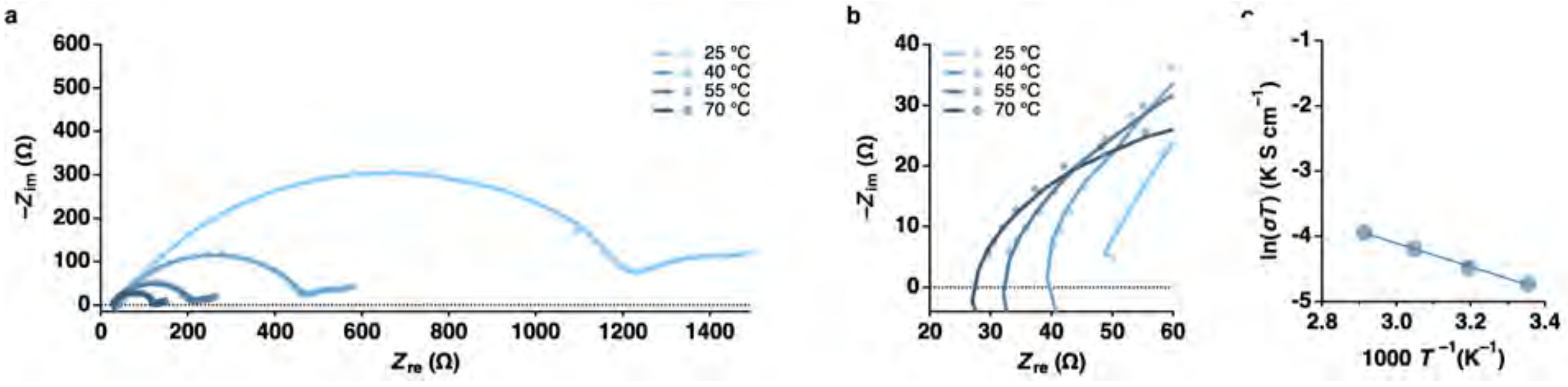

**Supplementary Figure 12 | PIM 11 variable temperature EIS. a,** Full Nyquist plot and **b,** zoom in on high frequency, and **c**, Arrhenius plot for at 25 °C, 40 °C, 55 °C, and 70 °C.

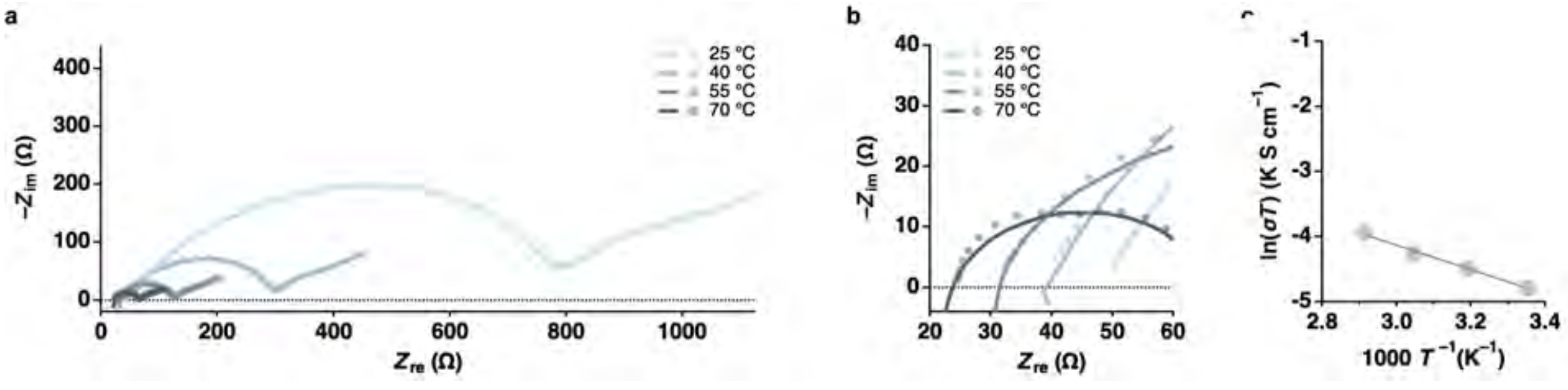

**Supplementary Figure 13 | PIM 12 variable temperature EIS. a,** Full Nyquist plot and **b,** zoom in on high frequency, and **c**, Arrhenius plot for at 25 °C, 40 °C, 55 °C, and 70 °C.

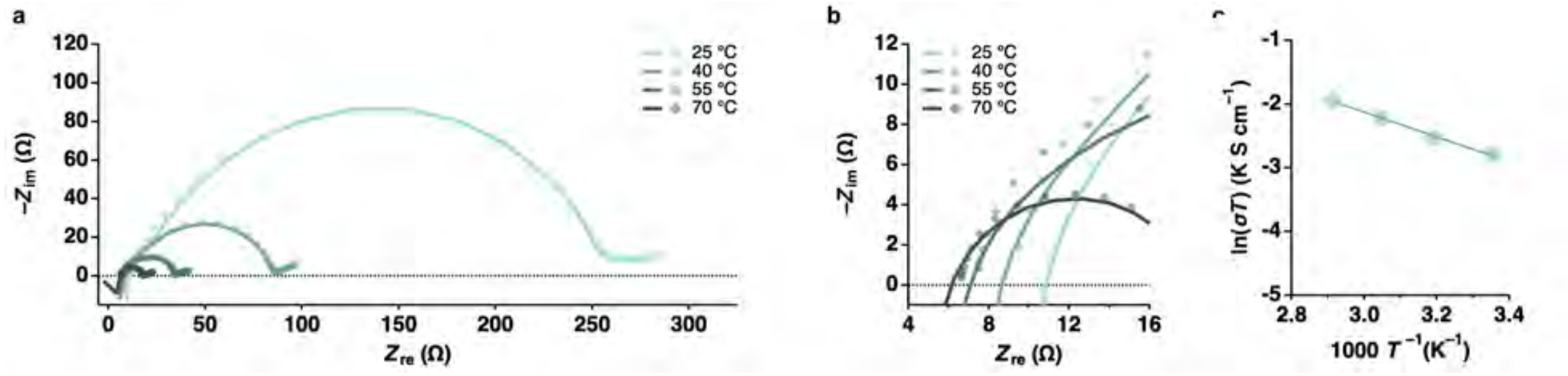

**Supplementary Figure 14 | PIM 13 variable temperature EIS. a,** Full Nyquist plot and **b,** zoom in on high frequency, and **c**, Arrhenius plot for at 25 °C, 40 °C, 55 °C, and 70 °C.

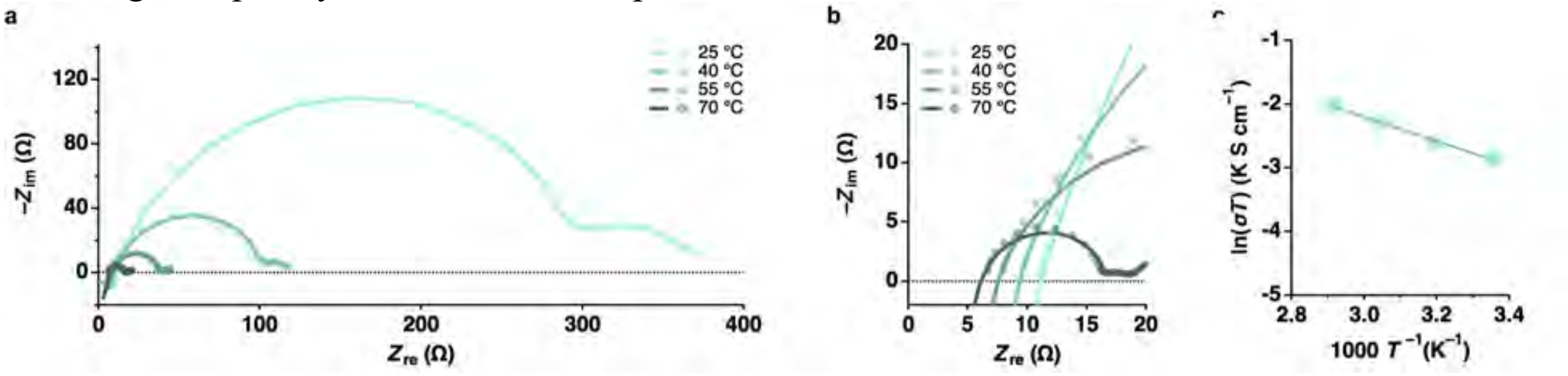

**Supplementary Figure 15 | PIM 14 variable temperature EIS. a,** Full Nyquist plot and **b,** zoom in on high frequency, and **c**, Arrhenius plot for at 25 °C, 40 °C, 55 °C, and 70 °C.

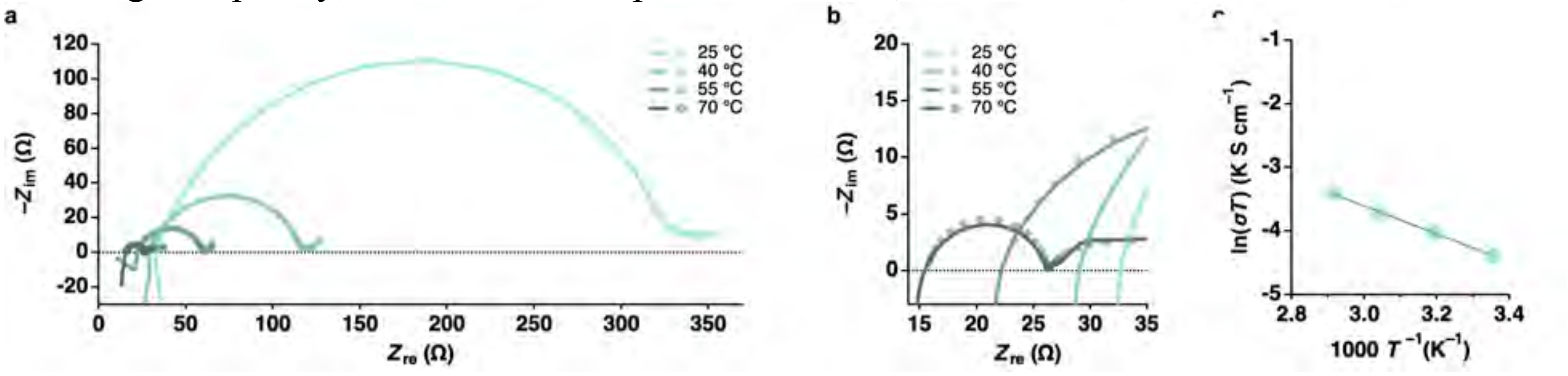

**Supplementary Figure 16 | PIM 15 variable temperature EIS. a,** Full Nyquist plot and **b,** zoom in on high frequency, and **c**, Arrhenius plot for at 25 °C, 40 °C, 55 °C, and 70 °C.

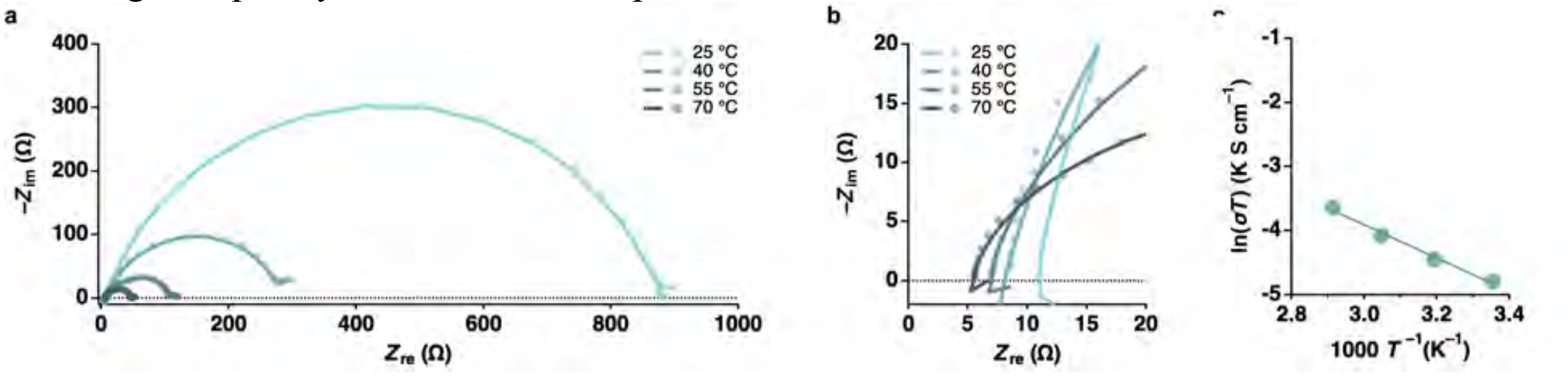

**Supplementary Figure 17 | PIM 17 variable temperature EIS. a,** Full Nyquist plot and **b,** zoom in on high frequency, and **c**, Arrhenius plot for at 25 °C, 40 °C, 55 °C, and 70 °C.

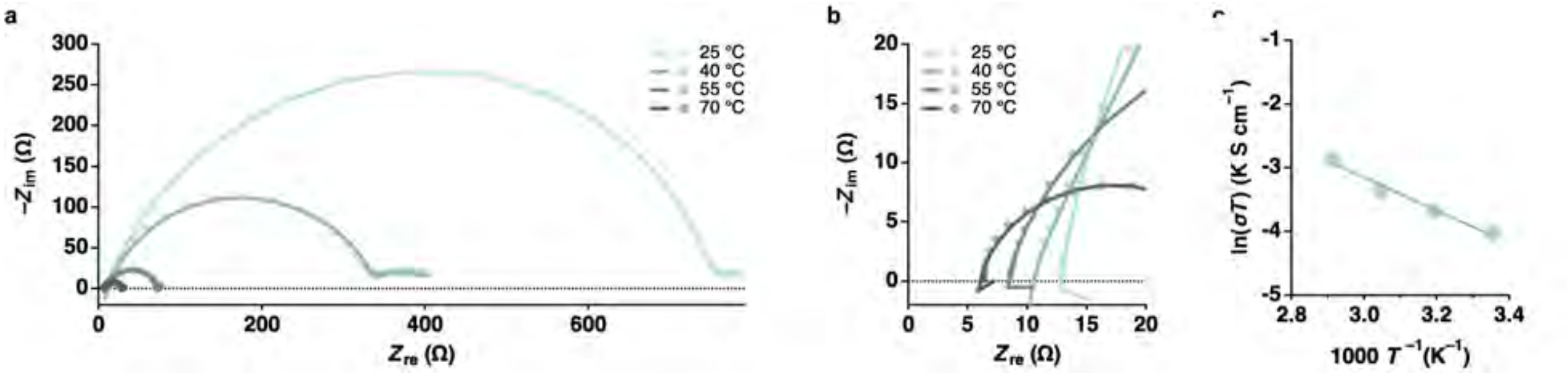

**Supplementary Figure 18 | PIM 18 variable temperature EIS. a,** Full Nyquist plot and **b,** zoom in on high frequency, and **c**, Arrhenius plot for at 25 °C, 40 °C, 55 °C, and 70 °C.

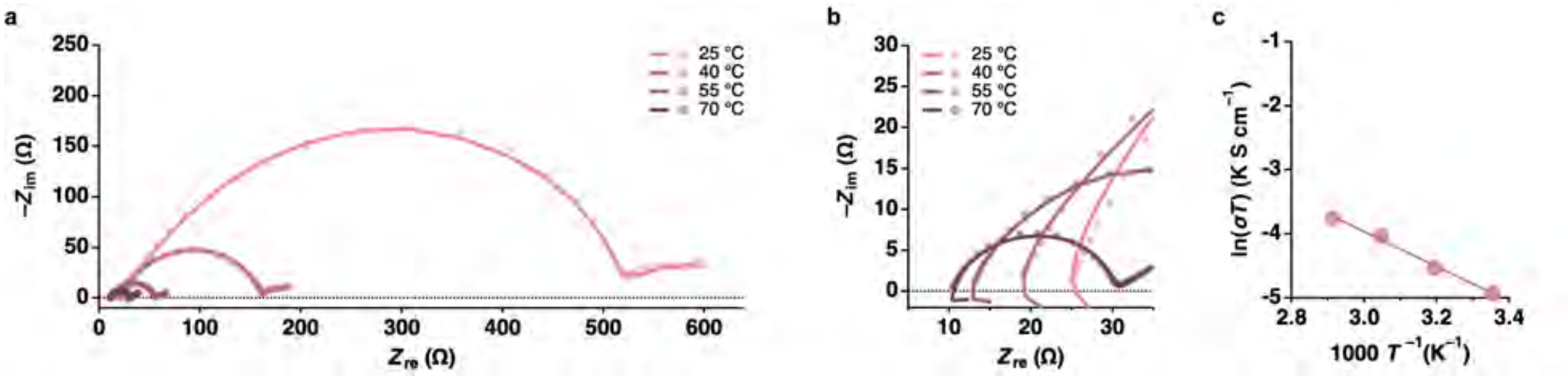

**Supplementary Figure 19 | PIM 21 variable temperature EIS. a,** Full Nyquist plot and **b,** zoom in on high frequency, and **c**, Arrhenius plot for at 25 °C, 40 °C, 55 °C, and 70 °C.

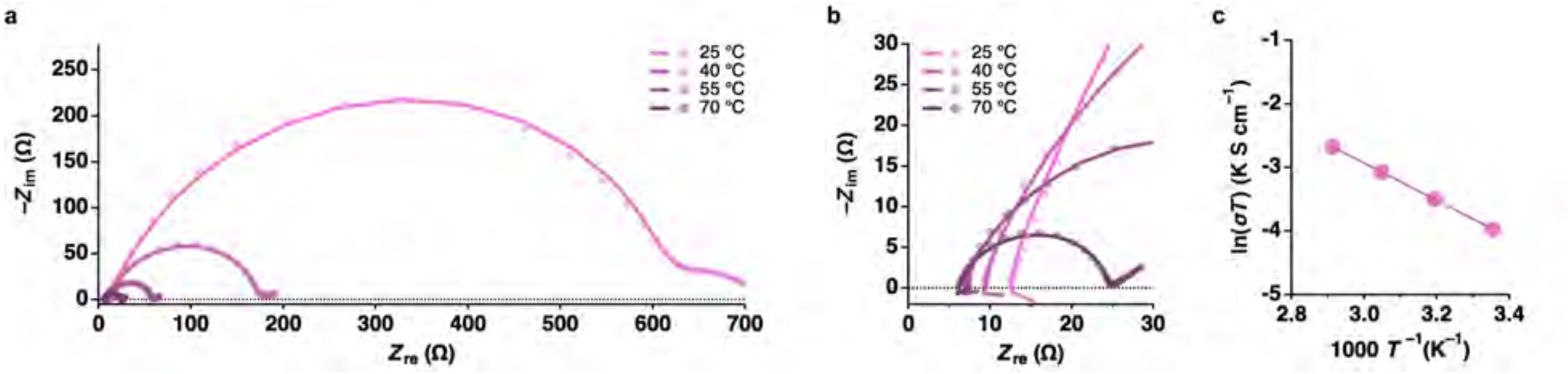

**Supplementary Figure 20 | PIM 23 variable temperature EIS. a,** Full Nyquist plot and **b,** zoom in on high frequency, and **c**, Arrhenius plot for at 25 °C, 40 °C, 55 °C, and 70 °C.

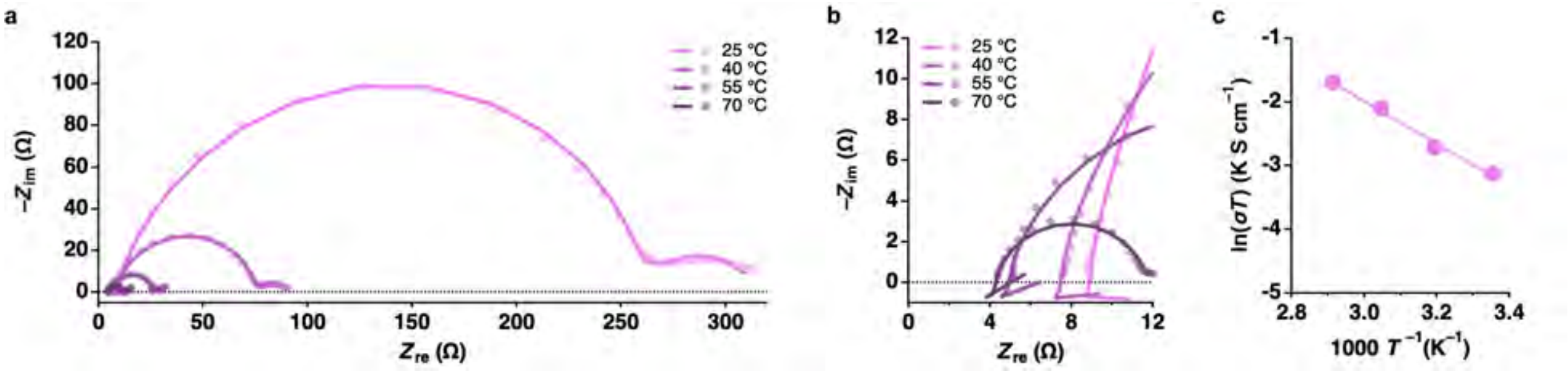

**Supplementary Figure 21 | PIM 24 variable temperature EIS. a,** Full Nyquist plot and **b,** zoom in on high frequency, and **c**, Arrhenius plot for at 25 °C, 40 °C, 55 °C, and 70 °C.

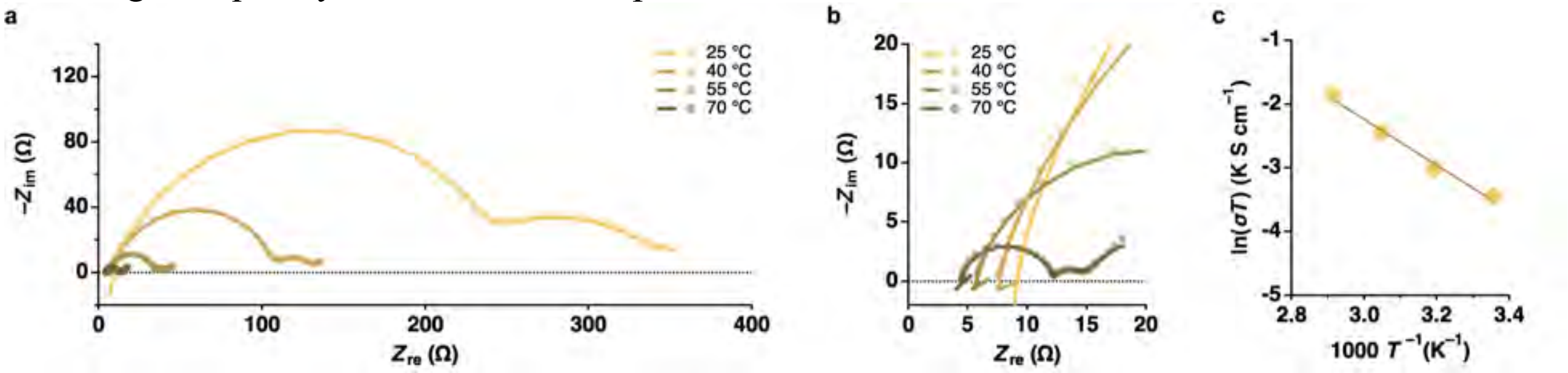

**Supplementary Figure 22 | PIM 25 variable temperature EIS. a,** Full Nyquist plot and **b,** zoom in on high frequency, and **c**, Arrhenius plot for at 25 °C, 40 °C, 55 °C, and 70 °C.

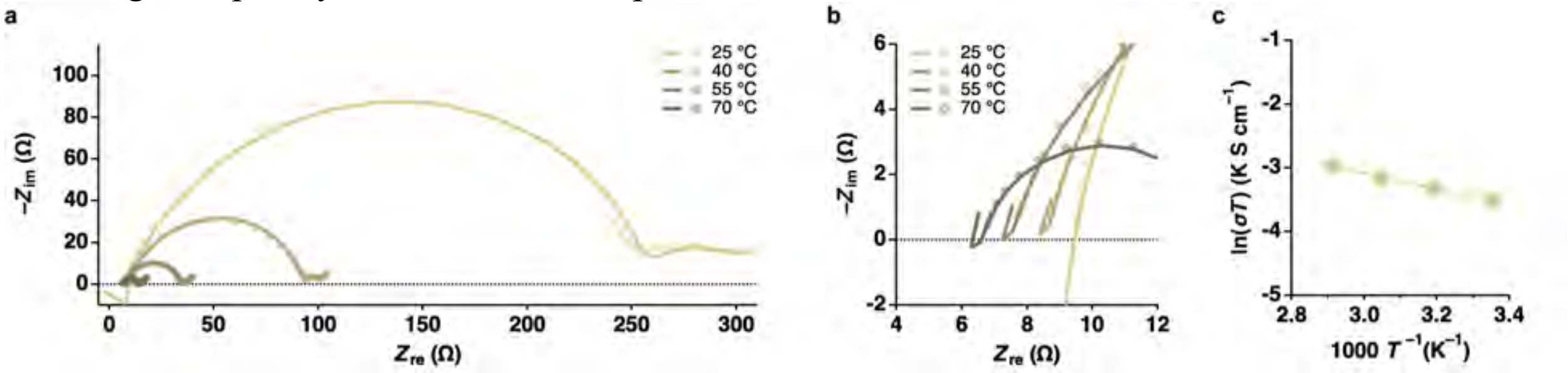

**Supplementary Figure 23 | PIM 26 variable temperature EIS. a,** Full Nyquist plot and **b,** zoom in on high frequency, and **c**, Arrhenius plot for at 25 °C, 40 °C, 55 °C, and 70 °C.

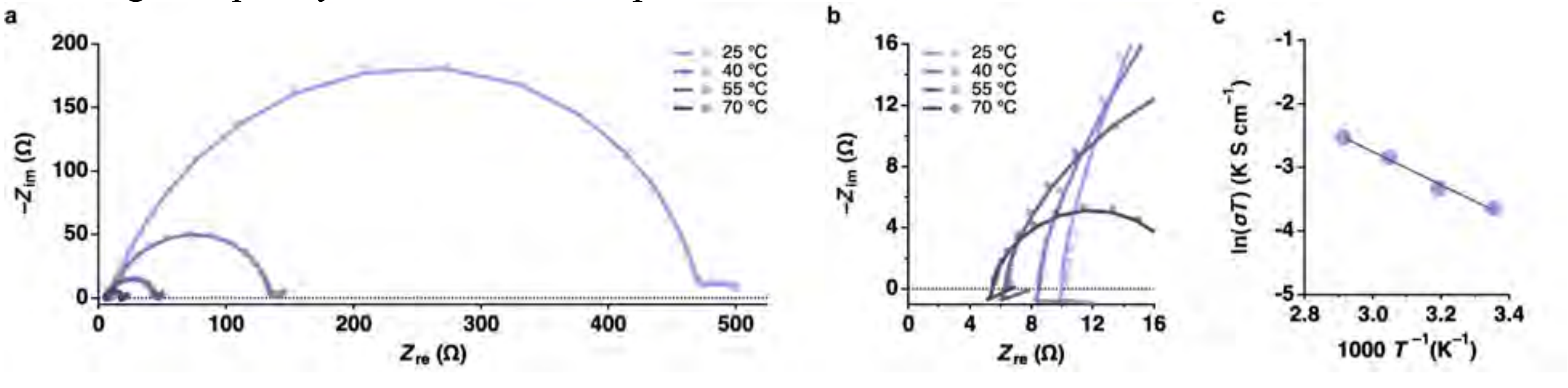

**Supplementary Figure 24 | PIM 28 variable temperature EIS. a,** Full Nyquist plot and **b,** zoom in on high frequency, and **c**, Arrhenius plot for at 25 °C, 40 °C, 55 °C, and 70 °C.

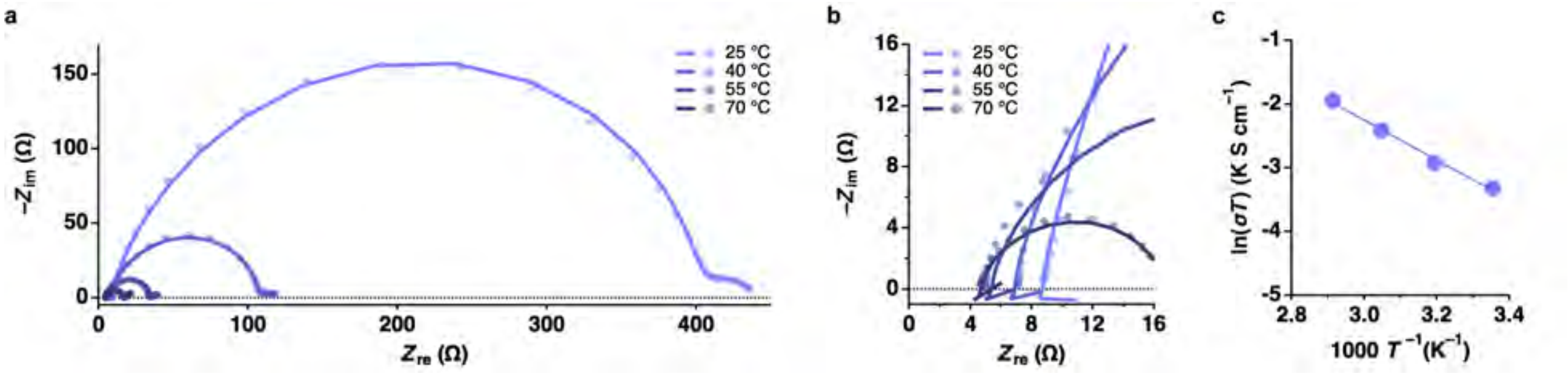

**Supplementary Figure 25 | PIM 30 variable temperature EIS. a,** Full Nyquist plot and **b,** zoom in on high frequency, and **c**, Arrhenius plot for at 25 °C, 40 °C, 55 °C, and 70 °C.

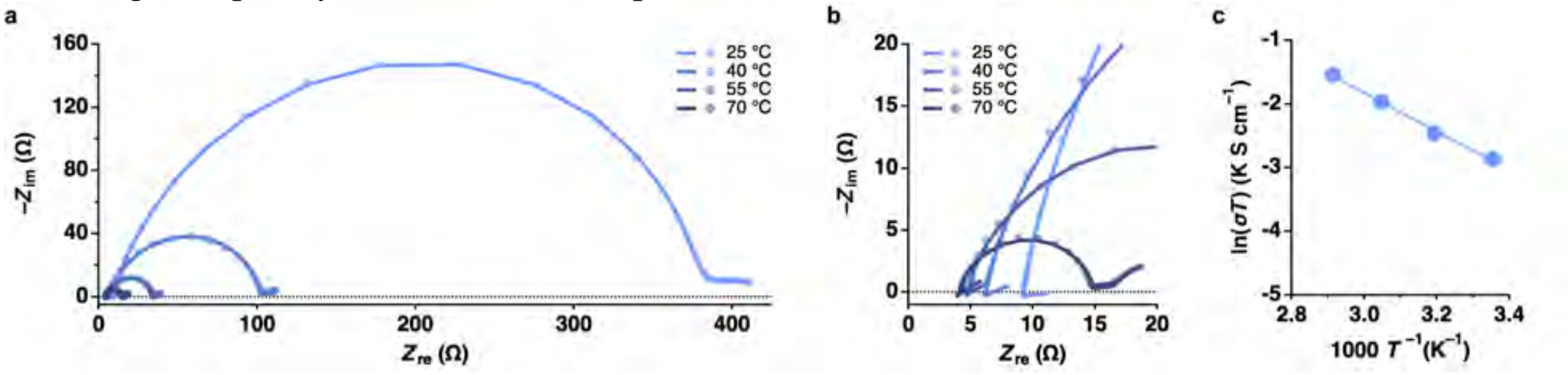

**Supplementary Figure 26 | PIM 31 variable temperature EIS. a,** Full Nyquist plot and **b,** zoom in on high frequency, and **c**, Arrhenius plot for at 25 °C, 40 °C, 55 °C, and 70 °C.

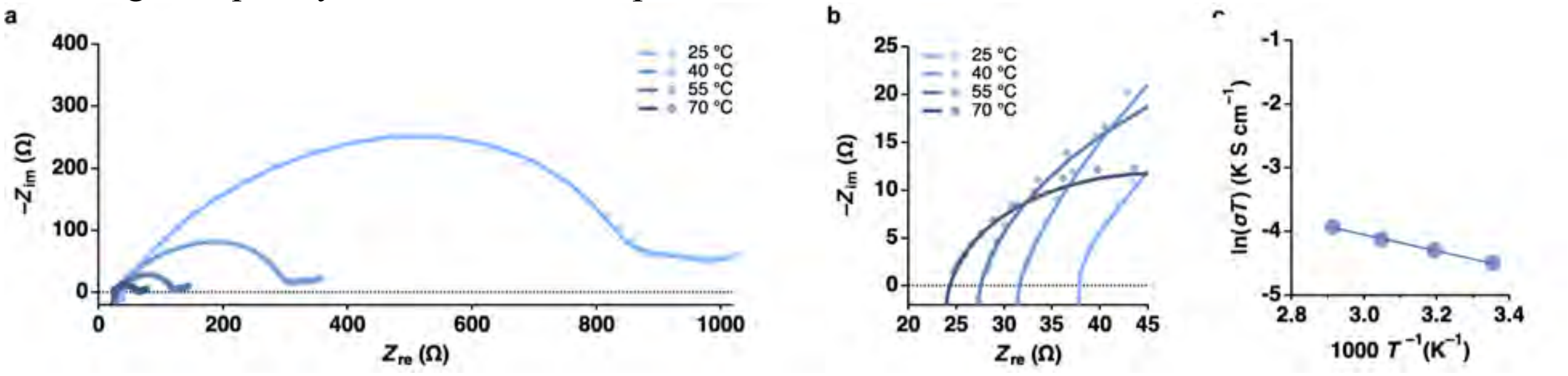

**Supplementary Figure 27 | PIM-1 variable temperature EIS. a,** Full Nyquist plot and **b,** zoom in on high frequency, and **c**, Arrhenius plot for at 25 °C, 40 °C, 55 °C, and 70 °C.

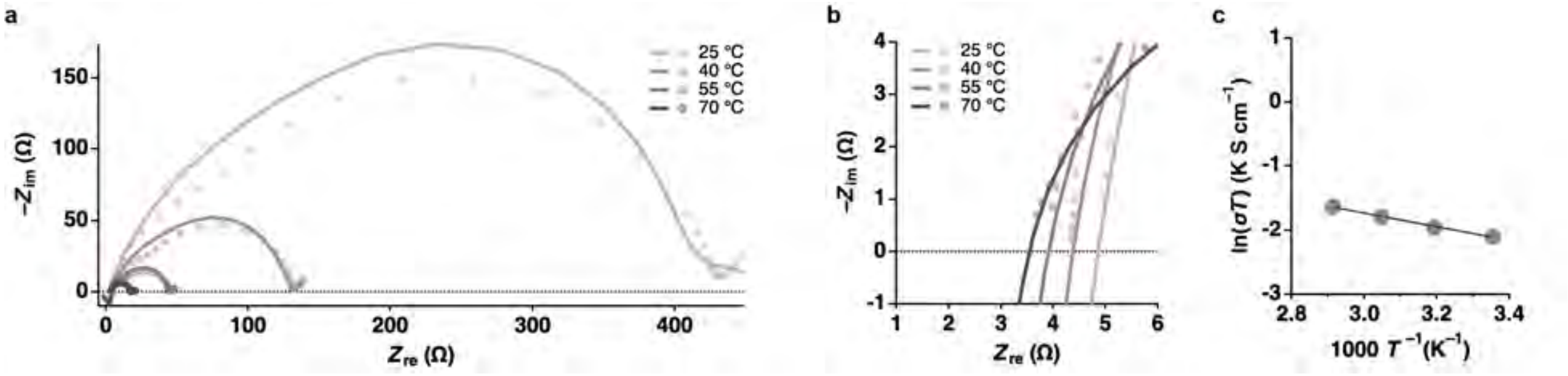

**Supplementary Figure 28 | Celgard variable temperature EIS. a,** Full Nyquist plot and **b,** zoom in on high frequency, and **c**, Arrhenius plot for at 25 °C, 40 °C, 55 °C, and 70 °C.

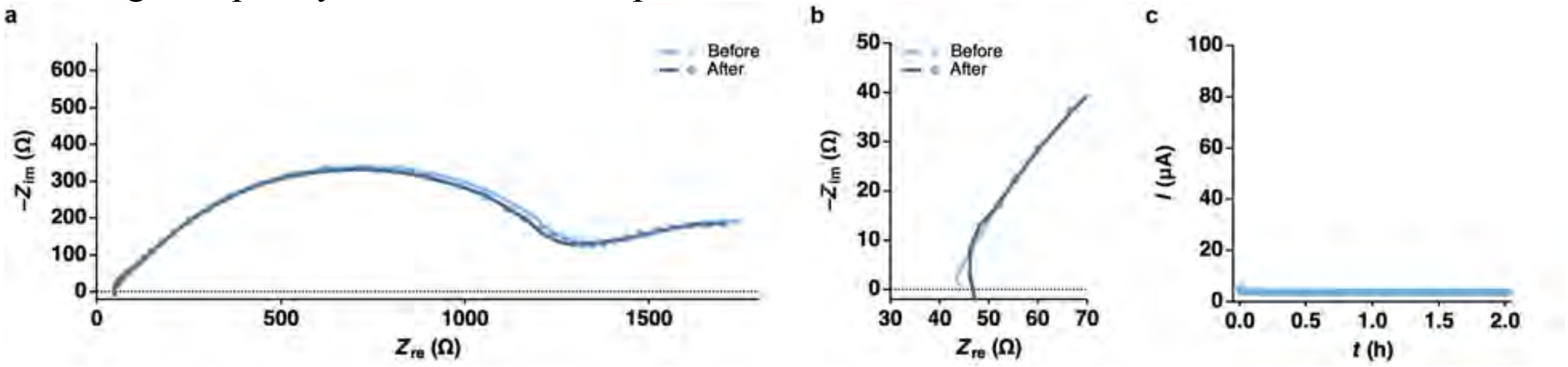

**Supplementary Figure 29 | Transference number determination for PIM 11. a**, Full Nyquist plot and **b**, zoom in on high frequency region before and after polarization, and **c**, current over time during potentiostatic polarization at 10 mV.

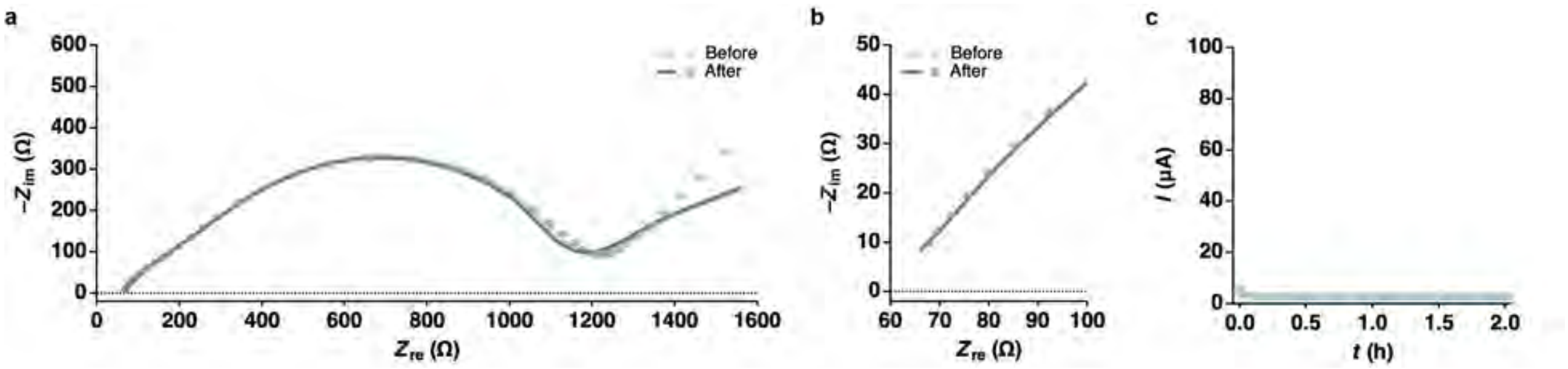

**Supplementary Figure 30 | Transference number determination for PIM 12. a**, Full Nyquist plot and **b**, zoom in on high frequency region before and after polarization, and **c**, current over time during potentiostatic polarization at 10 mV.

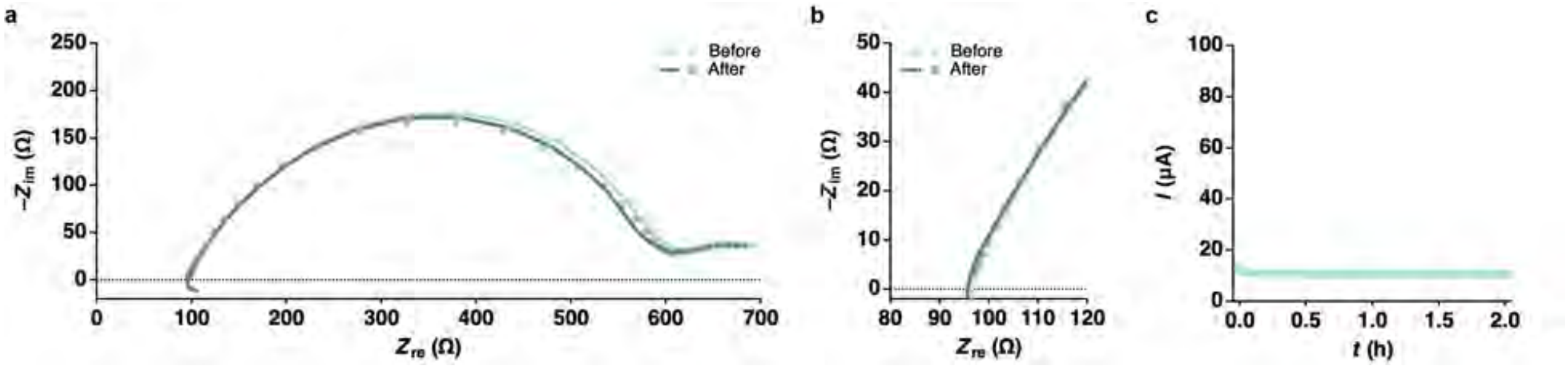

**Supplementary Figure 31 | Transference number determination for PIM 13. a**, Full Nyquist plot and **b**, zoom in on high frequency region before and after polarization, and **c**, current over time during potentiostatic polarization at 10 mV.

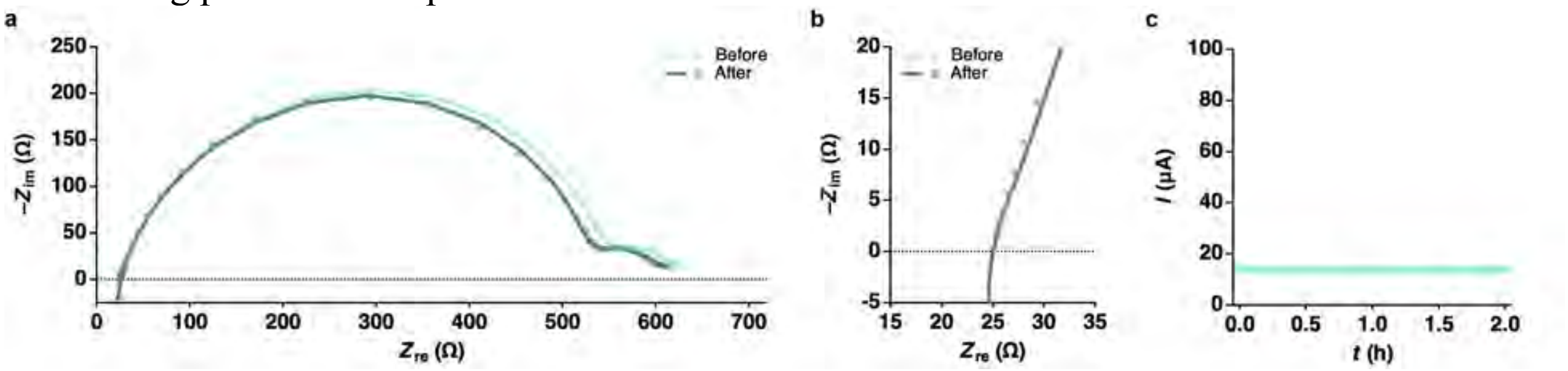

**Supplementary Figure 32 | Transference number determination for PIM 14. a**, Full Nyquist plot and **b**, zoom in on high frequency region before and after polarization, and **c**, current over time during potentiostatic polarization at 10 mV.

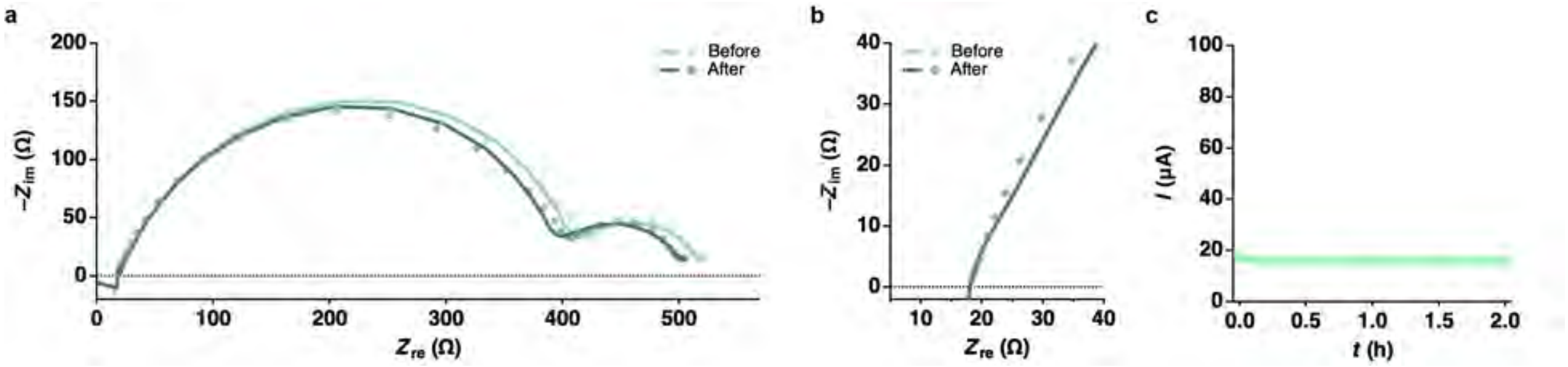

**Supplementary Figure 33 | Transference number determination for PIM 15. a**, Full Nyquist plot and **b**, zoom in on high frequency region before and after polarization, and **c**, current over time during potentiostatic polarization at 10 mV.

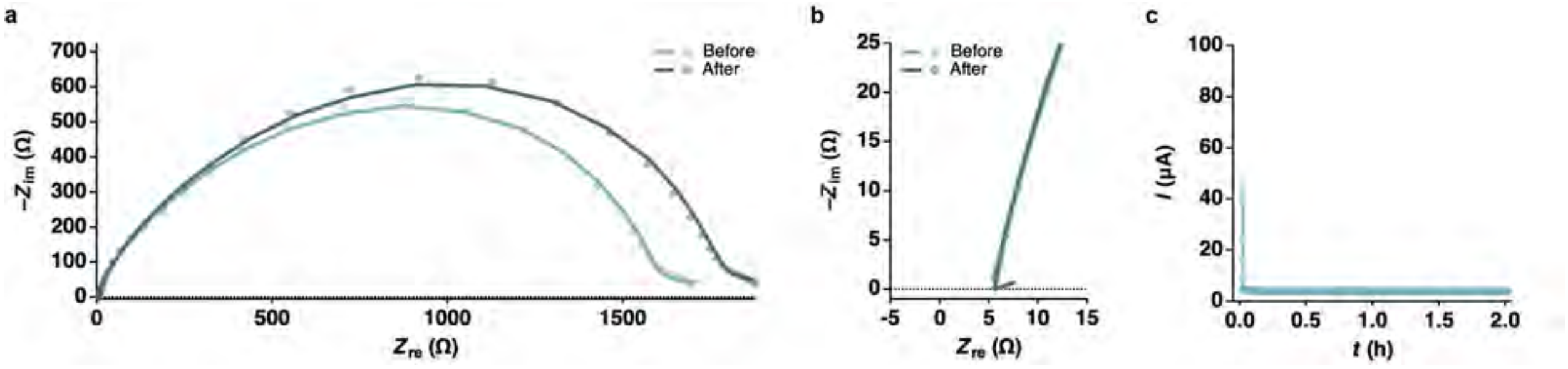

**Supplementary Figure 34 | Transference number determination for PIM 17. a**, Full Nyquist plot and **b**, zoom in on high frequency region before and after polarization, and **c**, current over time during potentiostatic polarization at 10 mV.

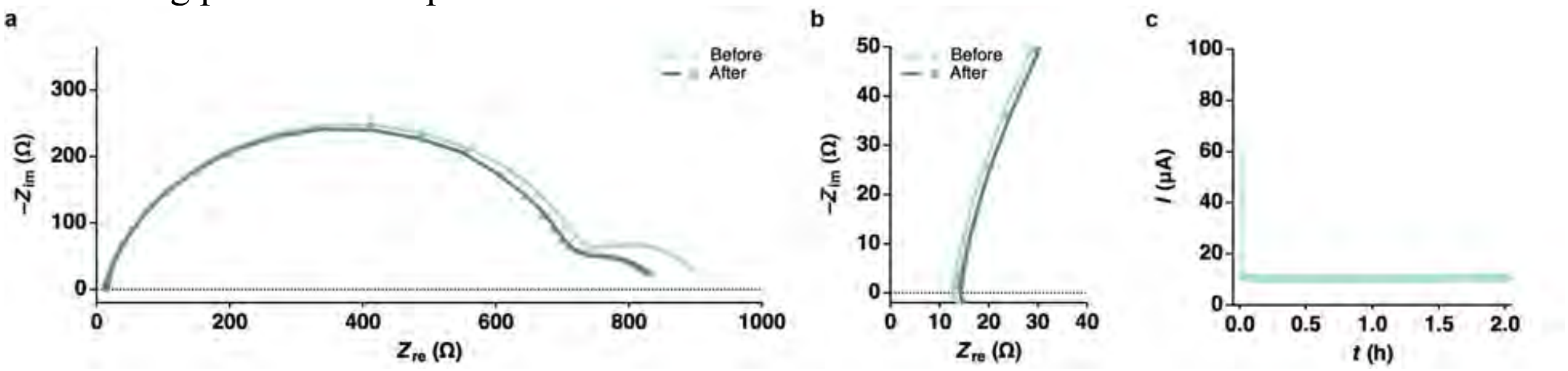

**Supplementary Figure 35 | Transference number determination for PIM 18. a**, Full Nyquist plot and **b**, zoom in on high frequency region before and after polarization, and **c**, current over time during potentiostatic polarization at 10 mV.

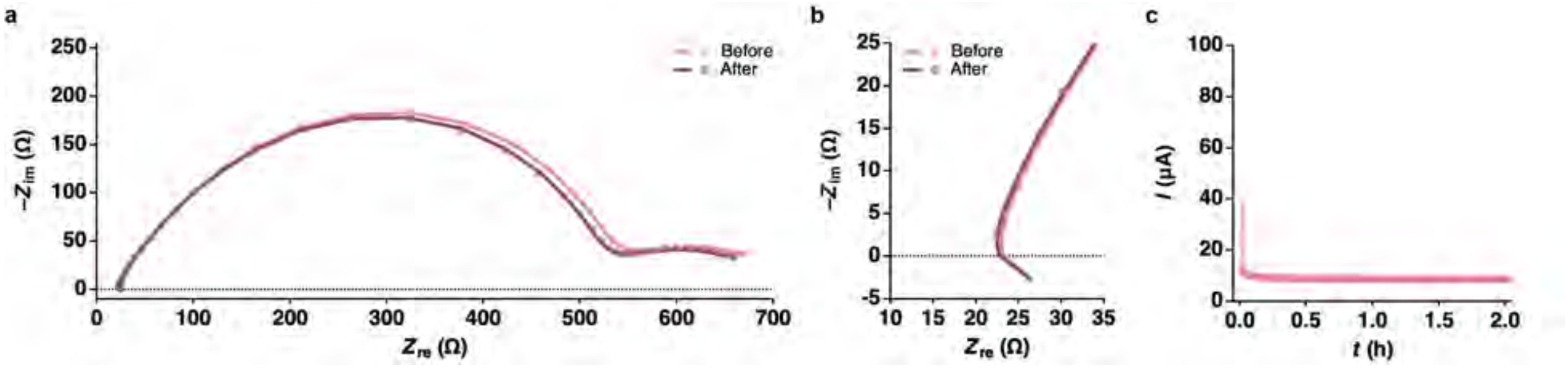

**Supplementary Figure 36 | Transference number determination for PIM 21. a**, Full Nyquist plot and **b**, zoom in on high frequency region before and after polarization, and **c**, current over time during potentiostatic polarization at 10 mV.

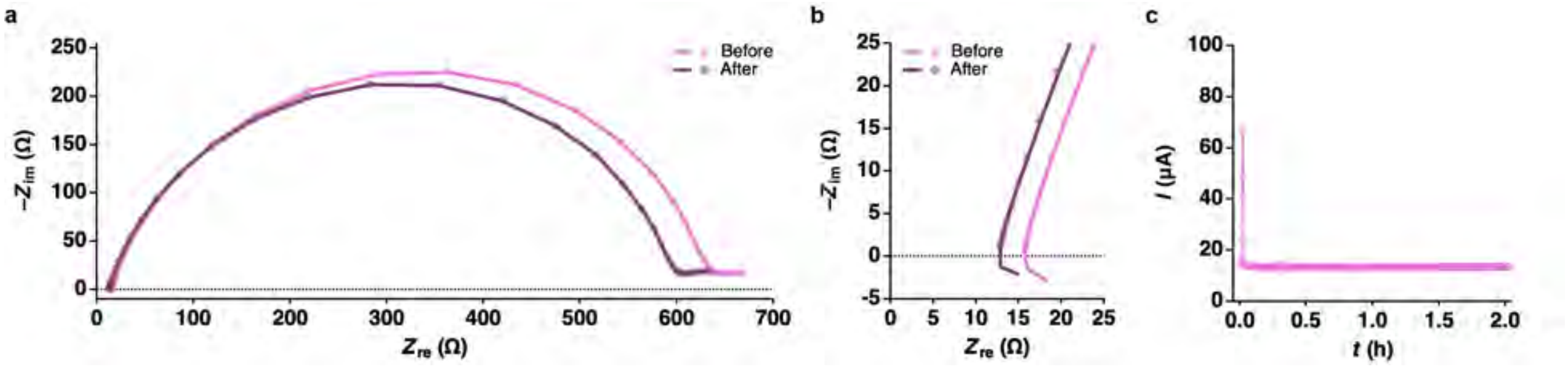

**Supplementary Figure 37 | Transference number determination for PIM 23. a**, Full Nyquist plot and **b**, zoom in on high frequency region before and after polarization, and **c**, current over time during potentiostatic polarization at 10 mV.

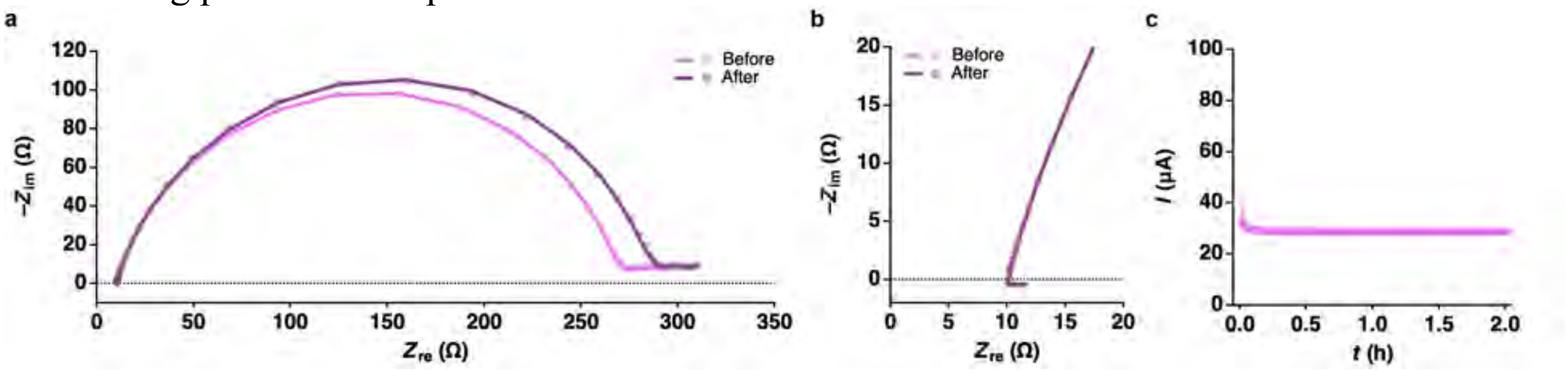

**Supplementary Figure 38 | Transference number determination for PIM 24. a**, Full Nyquist plot and **b**, zoom in on high frequency region before and after polarization, and **c**, current over time during potentiostatic polarization at 10 mV.

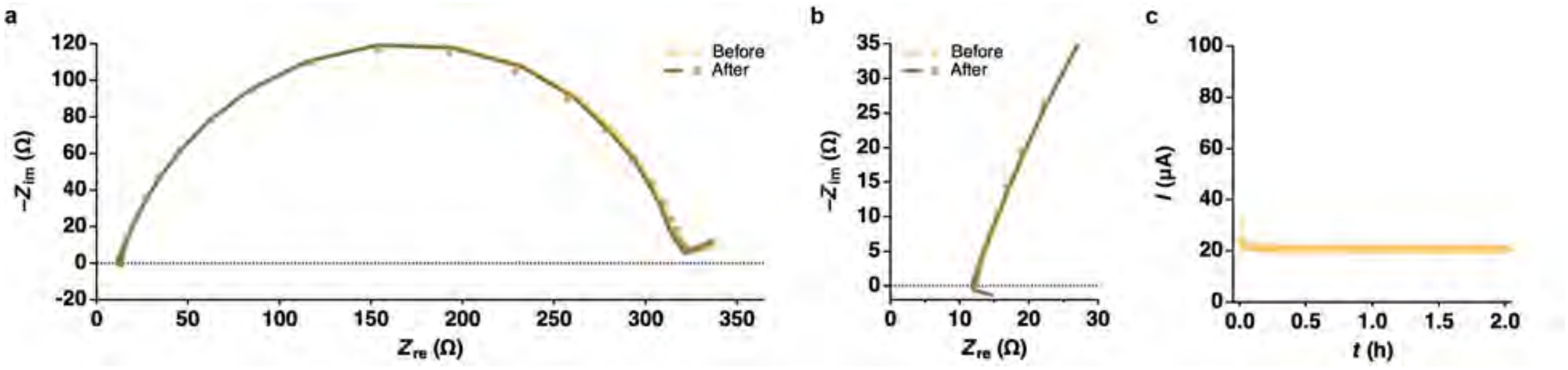

**Supplementary Figure 39 | Transference number determination for PIM 25. a**, Full Nyquist plot and **b**, zoom in on high frequency region before and after polarization, and **c**, current over time during potentiostatic polarization at 10 mV.

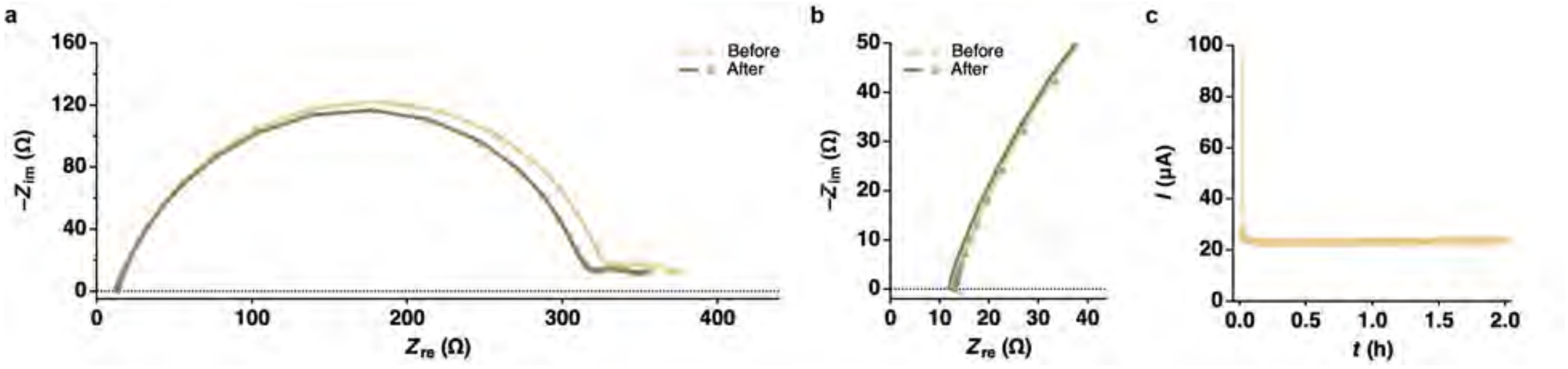

**Supplementary Figure 40 | Transference number determination for PIM 26. a**, Full Nyquist plot and **b**, zoom in on high frequency region before and after polarization, and **c**, current over time during potentiostatic polarization at 10 mV.

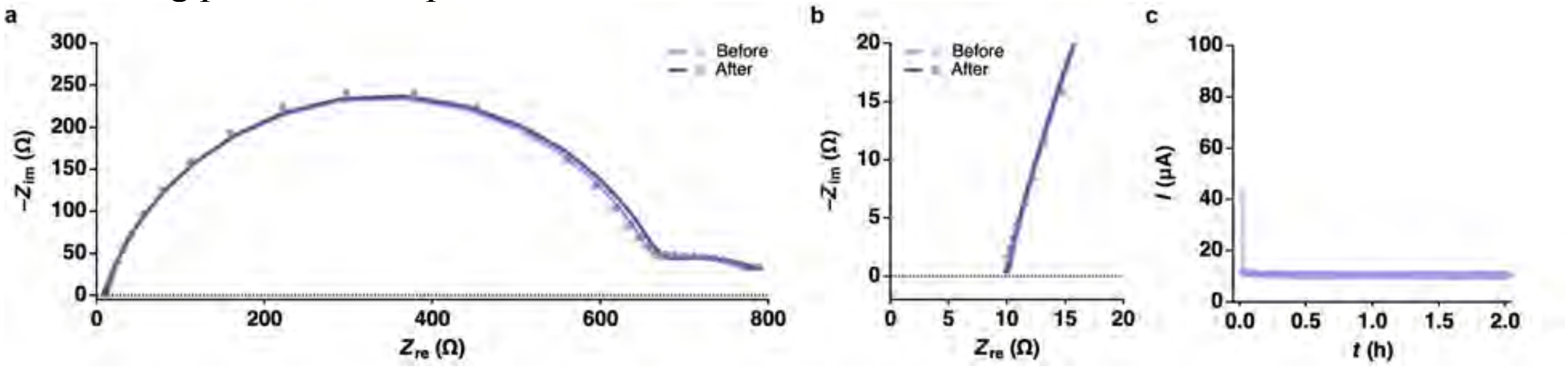

**Supplementary Figure 41 | Transference number determination for PIM 28. a**, Full Nyquist plot and **b**, zoom in on high frequency region before and after polarization, and **c**, current over time during potentiostatic polarization at 10 mV.

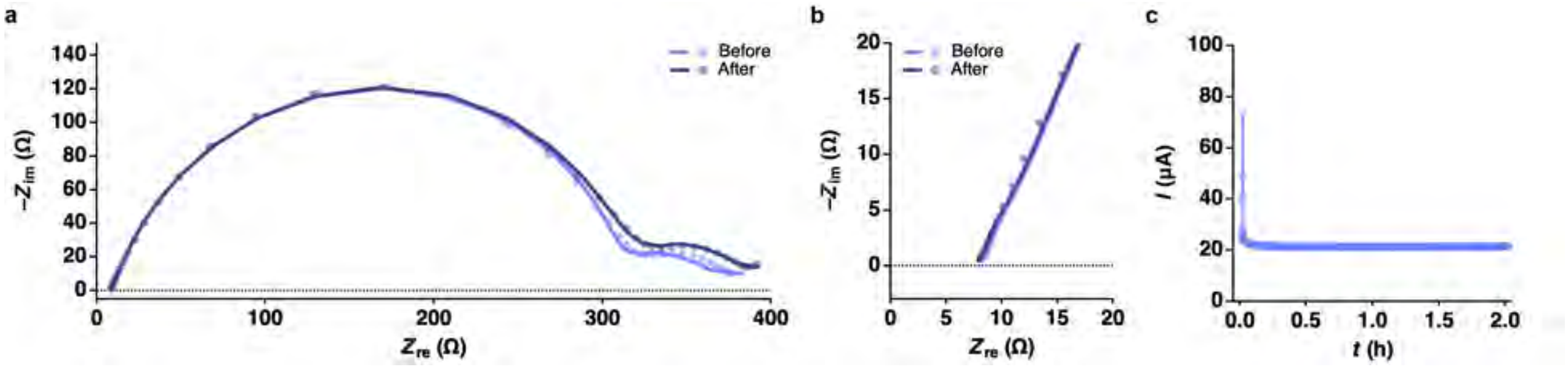

**Supplementary Figure 42 | Transference number determination for PIM 30. a**, Full Nyquist plot and **b**, zoom in on high frequency region before and after polarization, and **c**, current over time during potentiostatic polarization at 10 mV.

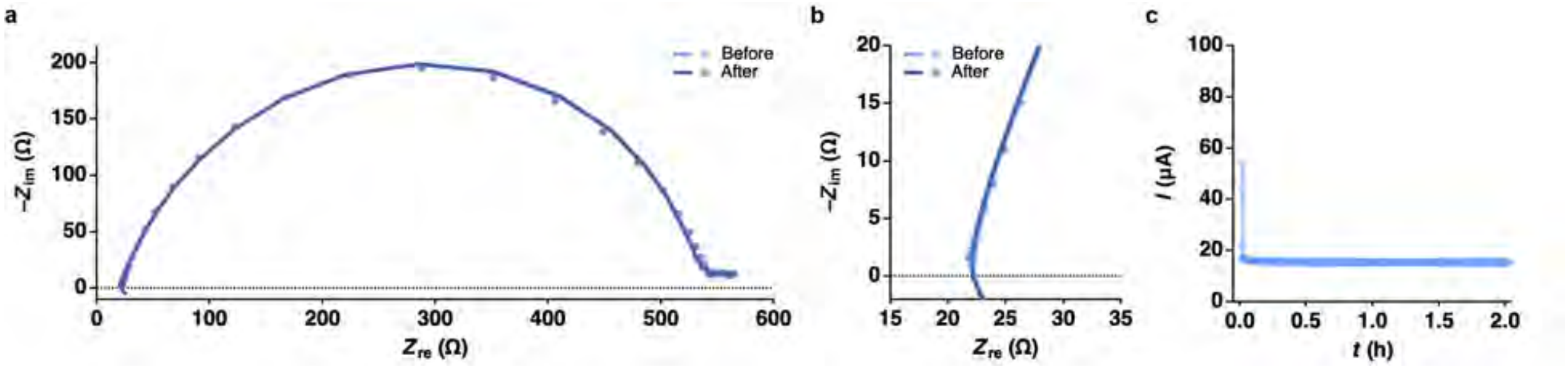

**Supplementary Figure 43 | Transference number determination for PIM 31. a**, Full Nyquist plot and **b**, zoom in on high frequency region before and after polarization, and **c**, current over time during potentiostatic polarization at 10 mV.

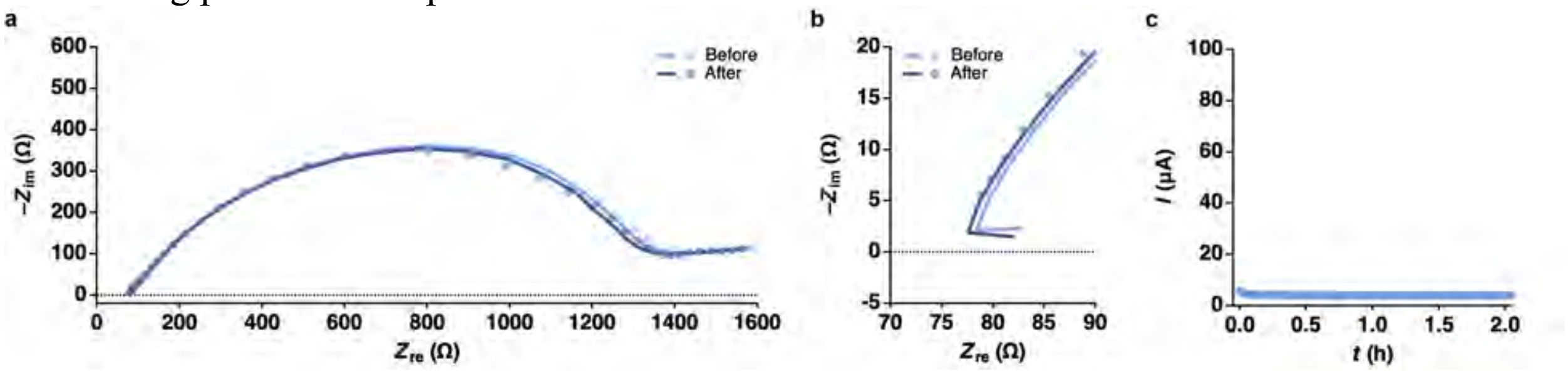

**Supplementary Figure 44 | Transference number determination for PIM-1. a**, Full Nyquist plot and **b**, zoom in on high frequency region before and after polarization, and **c**, current over time during potentiostatic polarization at 10 mV.

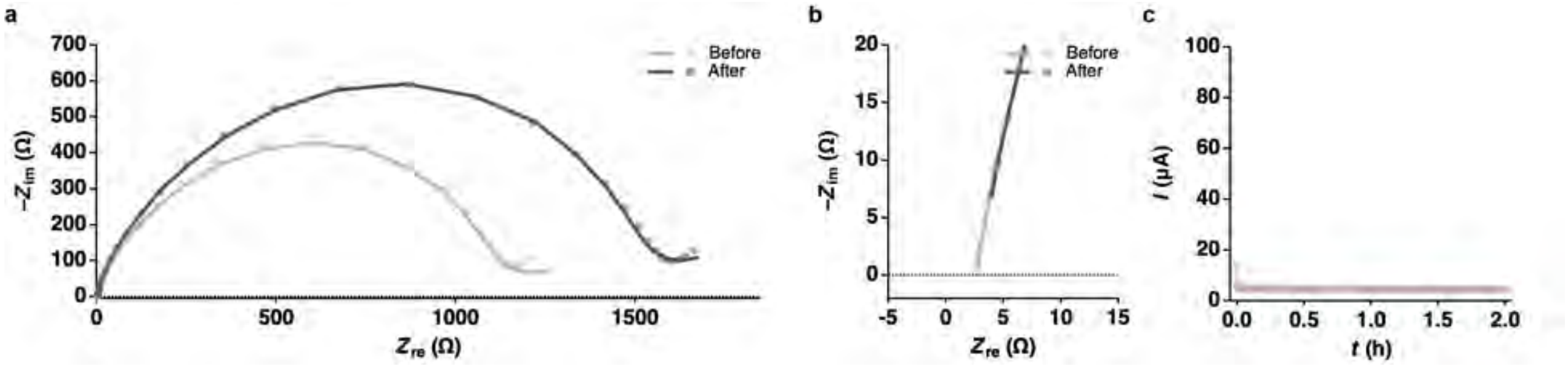

**Supplementary Figure 45 | Transference number determination for Celgard. a**, Full Nyquist plot and **b**, zoom in on high frequency region before and after polarization, and **c**, current over time during potentiostatic polarization at 10 mV.

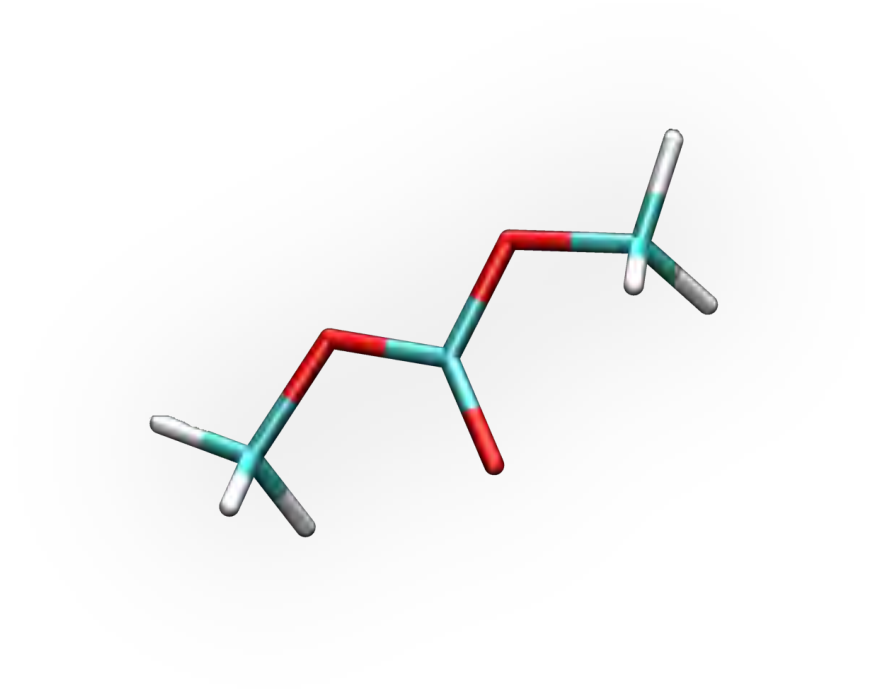

**Supplementary Figure 46 | Equilibrium geometry of DMC.**

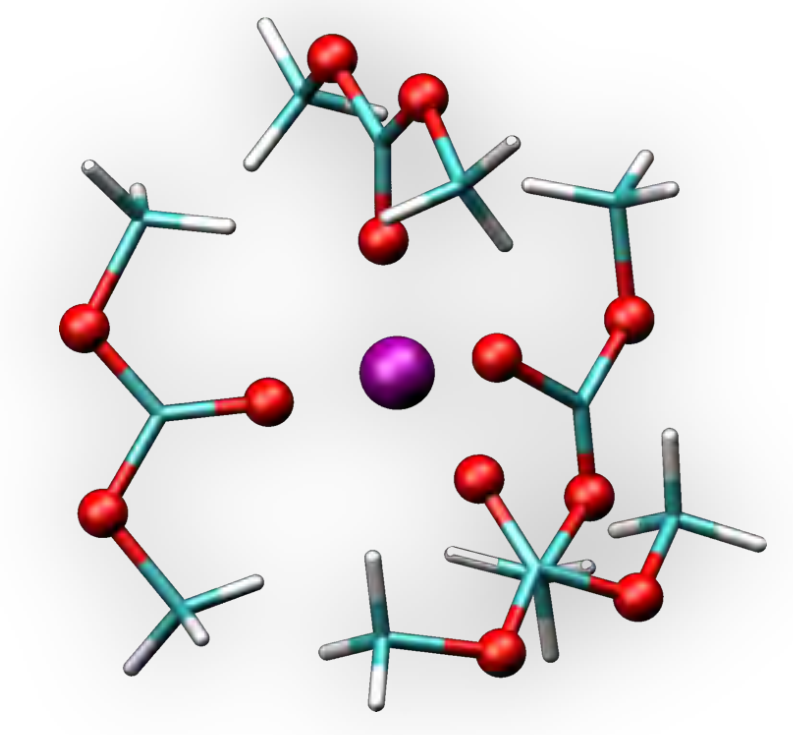

**Supplementary Figure 47 | Equilibrium geometry of a $Li^+$/4DMC molecular cluster.**

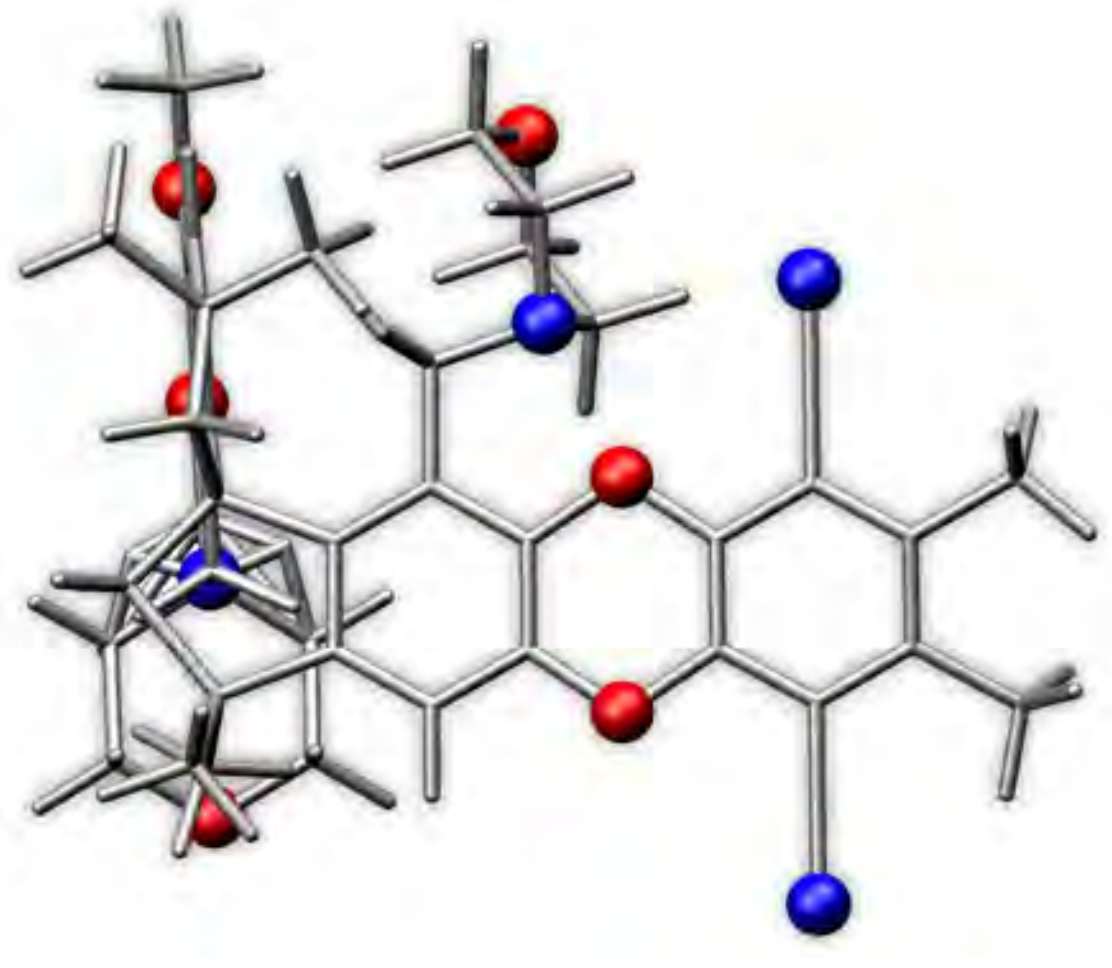

**Supplementary Figure 48 | Equilibrium geometry of the PIM 13 fragment.**

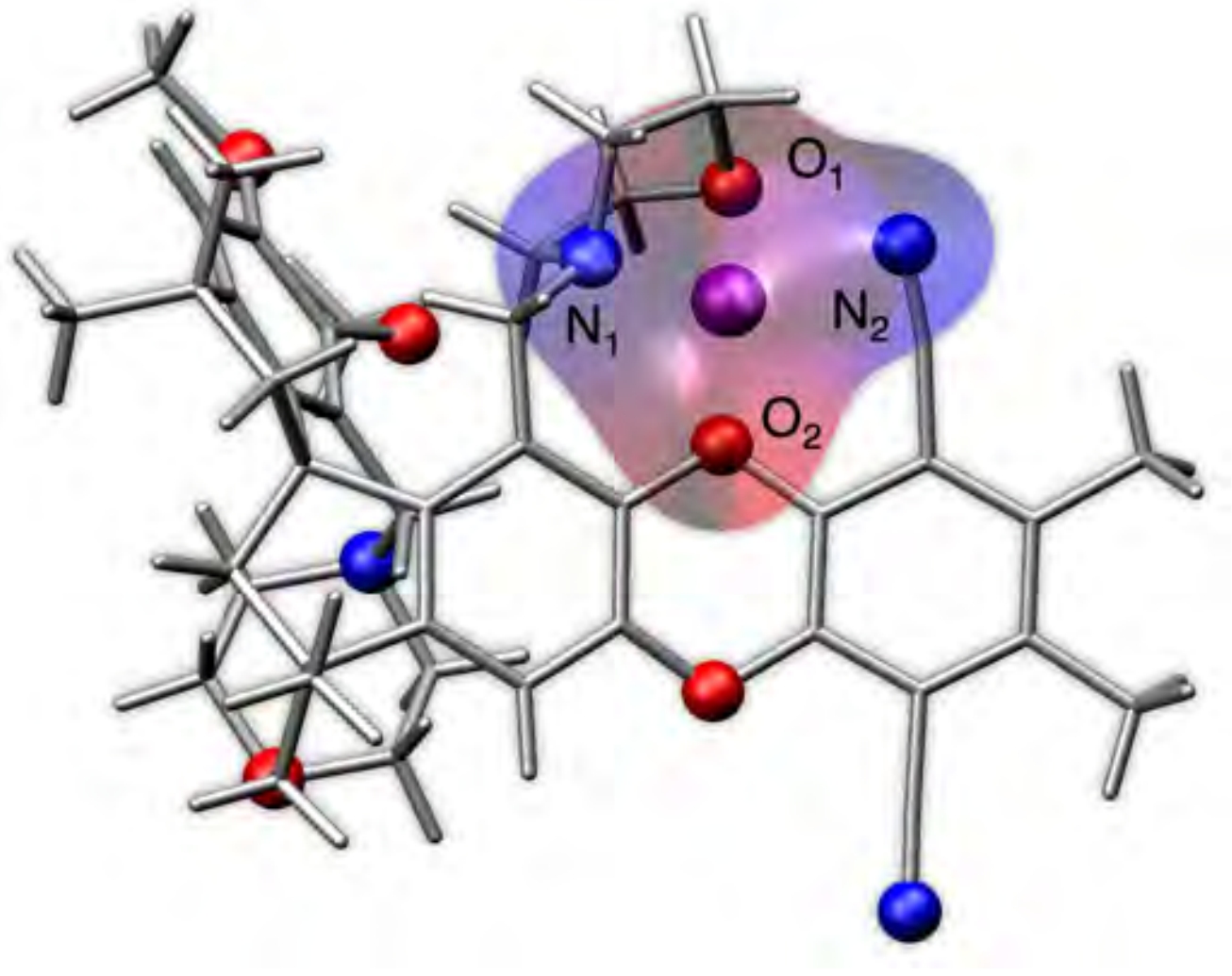

**Supplementary Figure 49 | Equilibrium geometry of the PIM 13 fragment and the $Li^+$ ion in the cage.**

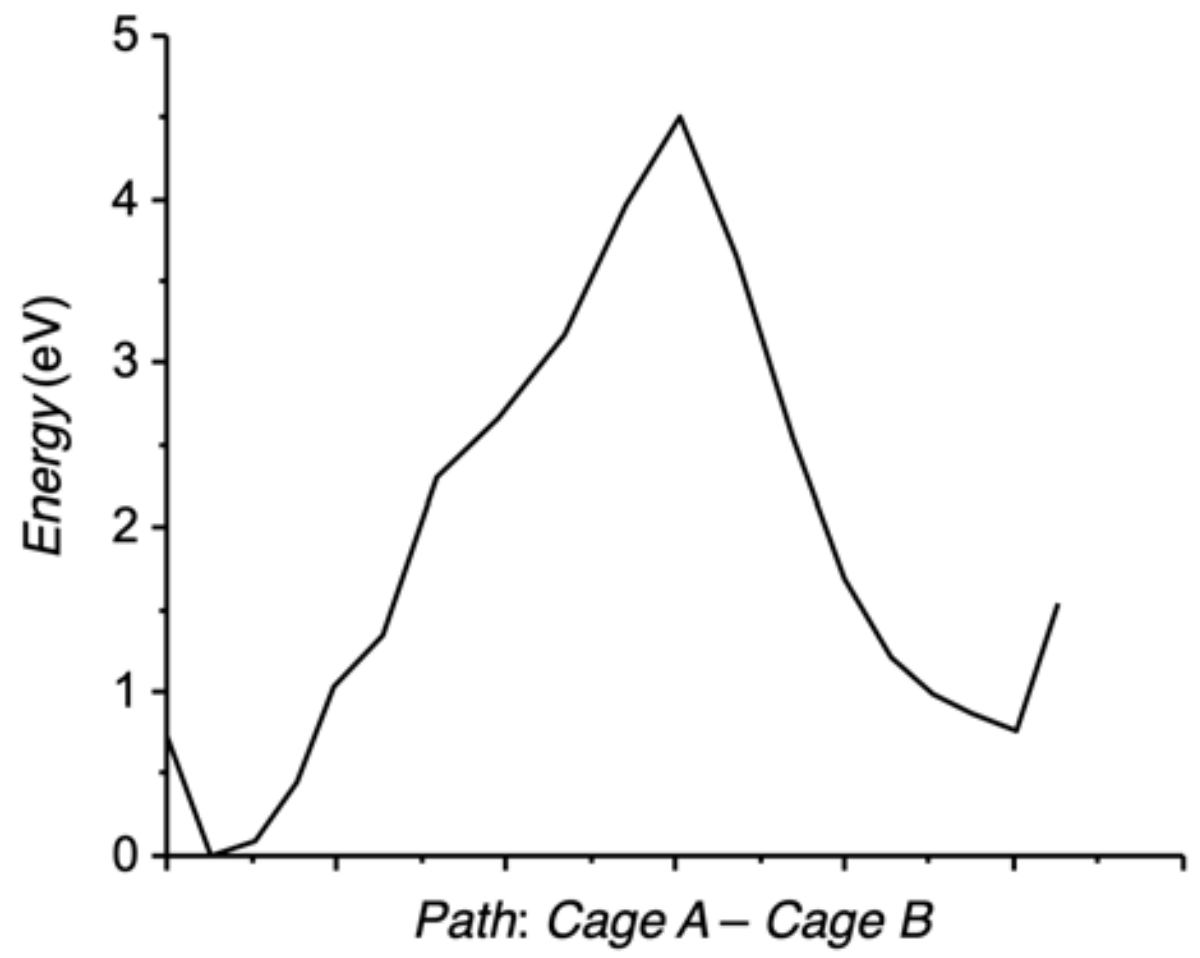

**Supplementary Figure 50 | NEB energies along the optimal pathway of the $Li^+$ ion from one cage to another in the PIM 13 unit used in quantum chemistry calculations (Supplementary Figures 48–49).**

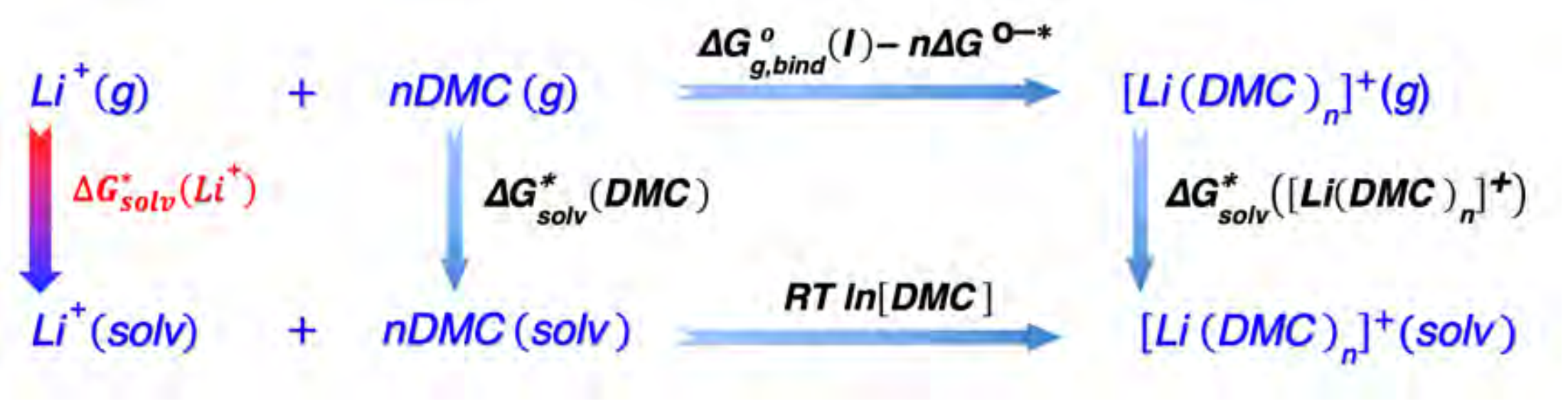

**Supplementary Figure 51 | Monomer thermodynamic cycle for calculation of the solvation free energy of the $Li^+$ ion in DMC.**

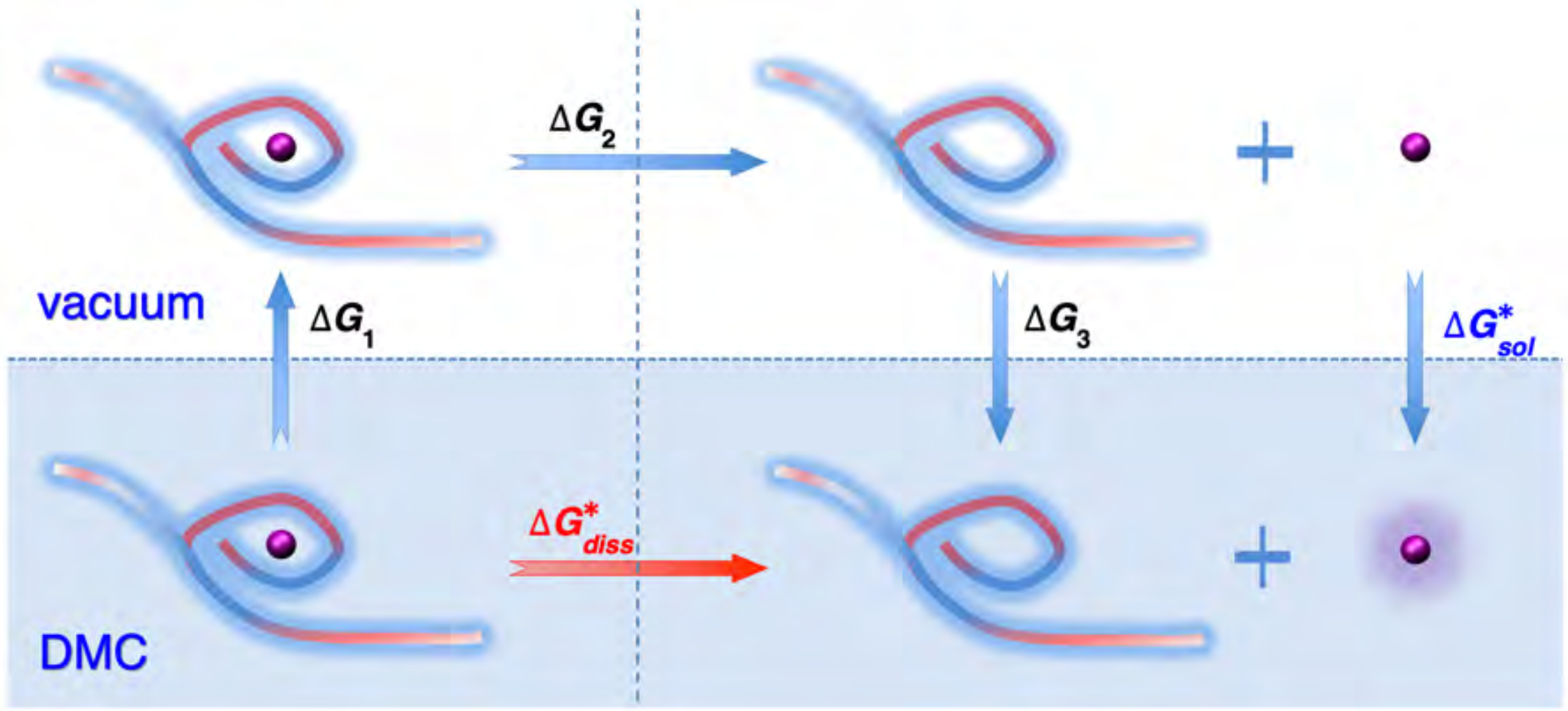

**Supplementary Figure 52 | Thermodynamic cycle for calculations of free energy of dissociation of the $Li^+$ ion in the bulk DMC solvent.**

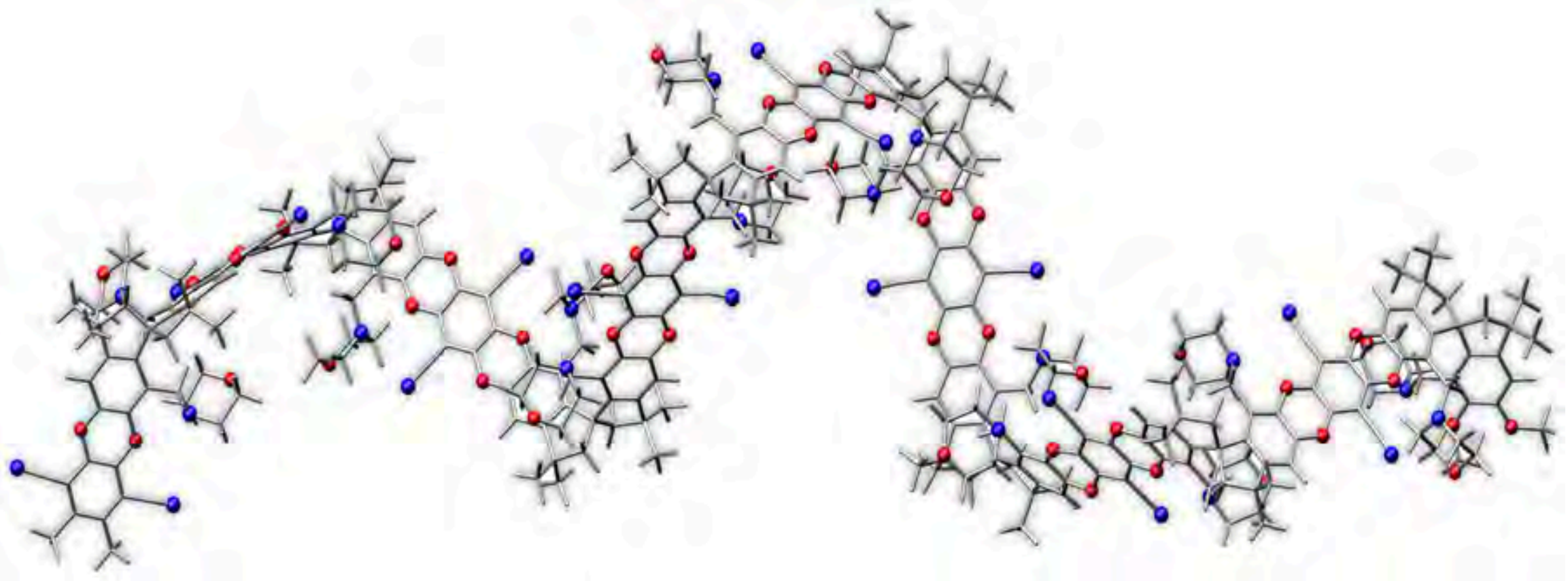

**Supplementary Figure 53 | A strand of PIM 13 (8 units) used in MD simulations.**

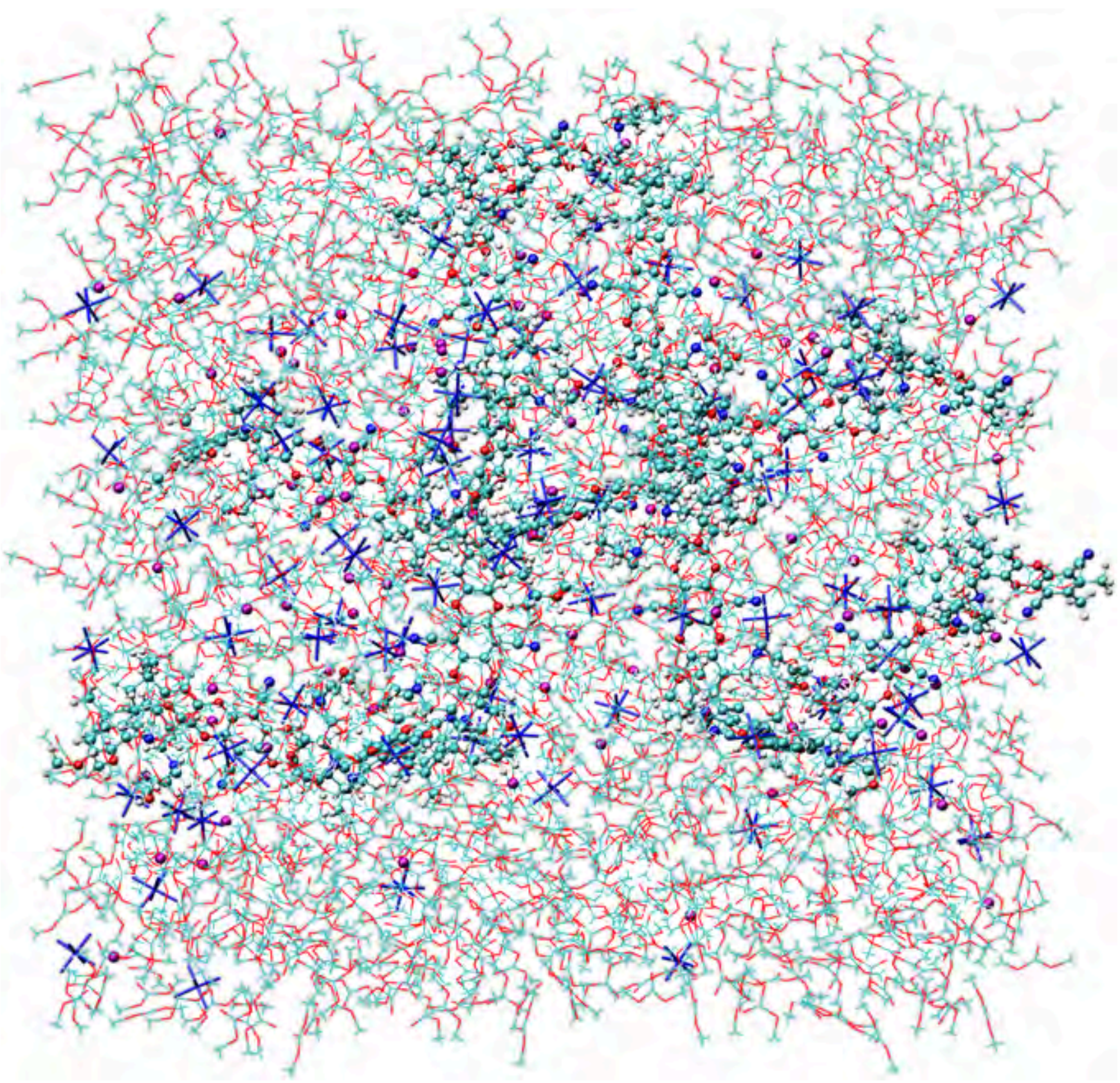

**Supplementary Figure 54 | A typical simulation box containing two strands of PIM 13 (8 units each) and 84 pairs of $Li^+/PF_6^-$ in DMC (0.7 M) used in MD simulations.**

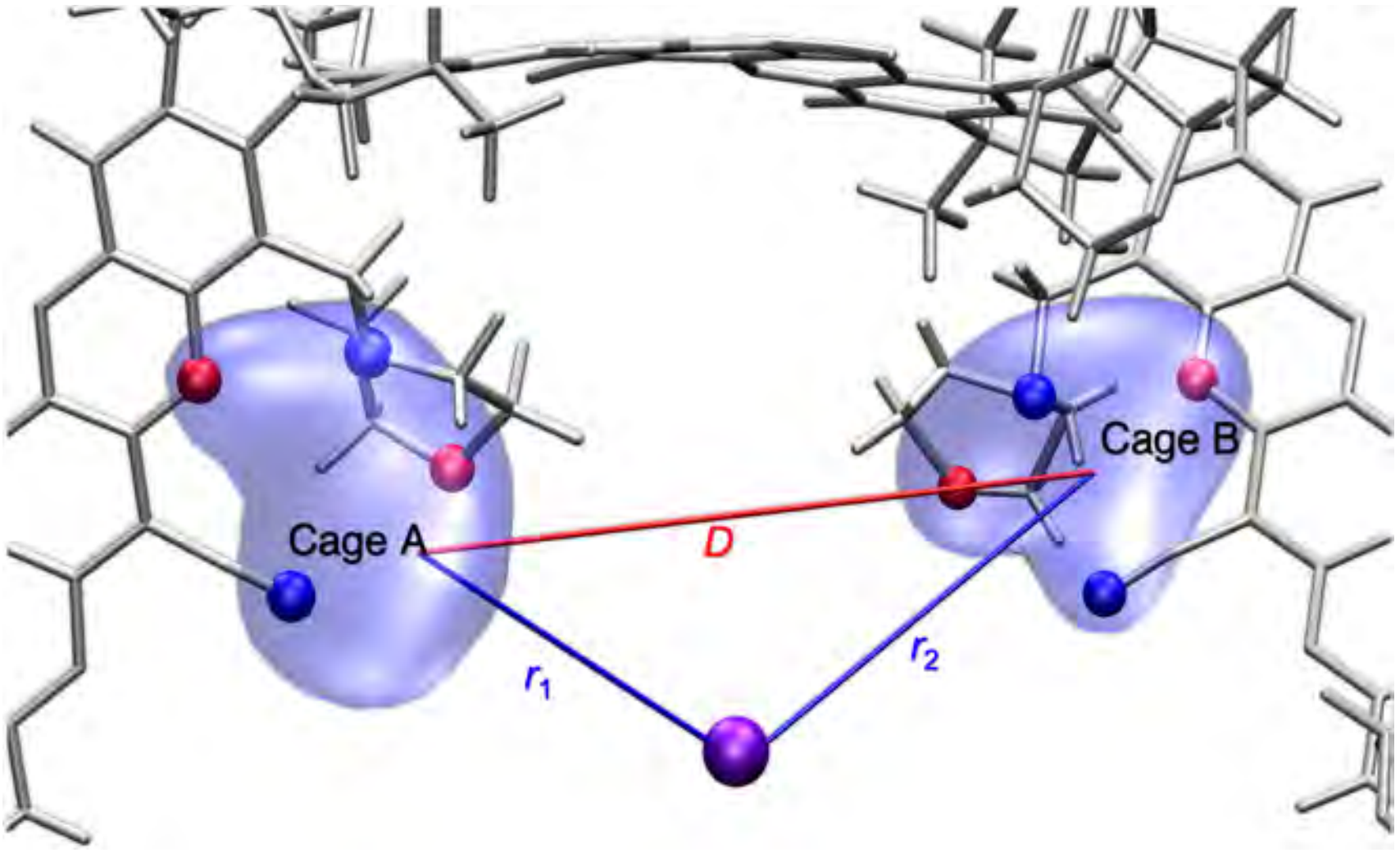

**Supplementary Figure 55 | Definitions of collective variables.** $\boldsymbol{r_1}$, $\boldsymbol{d} = \boldsymbol{r_1} - \boldsymbol{r_2}$ and $\boldsymbol{s} = \boldsymbol{r_1} + \boldsymbol{r_2}$. $D$ is the distance between centers of masses of two cages.

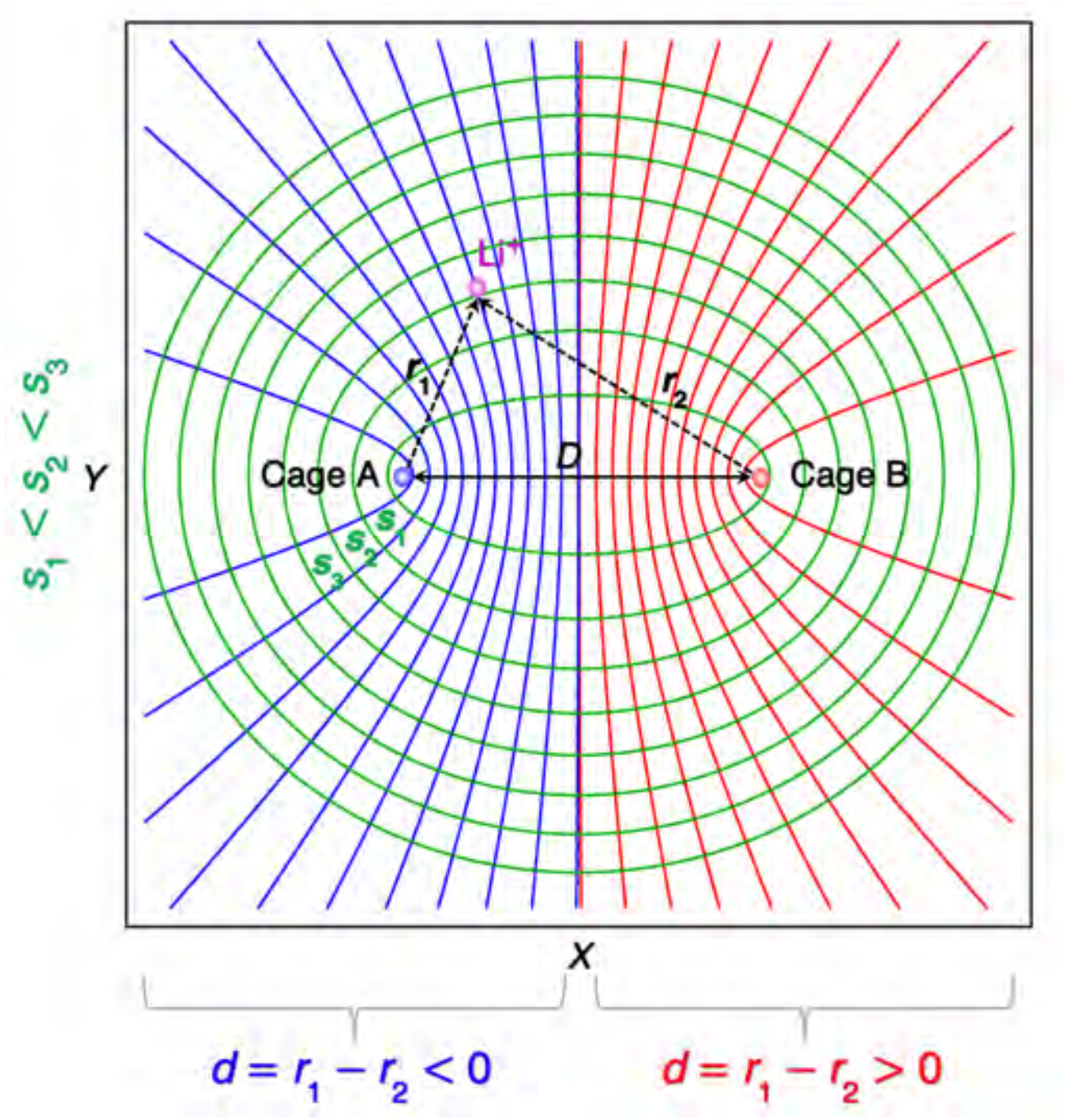

**Supplementary Figure 56 | Meaning of constraints on collective variables.** $\boldsymbol{r_1}$, $\boldsymbol{d = r_1 - r_2}$ and $\boldsymbol{s}$ $\boldsymbol{= r_1 + r_2}$. $D$ is the distance between centers of masses of two cages.

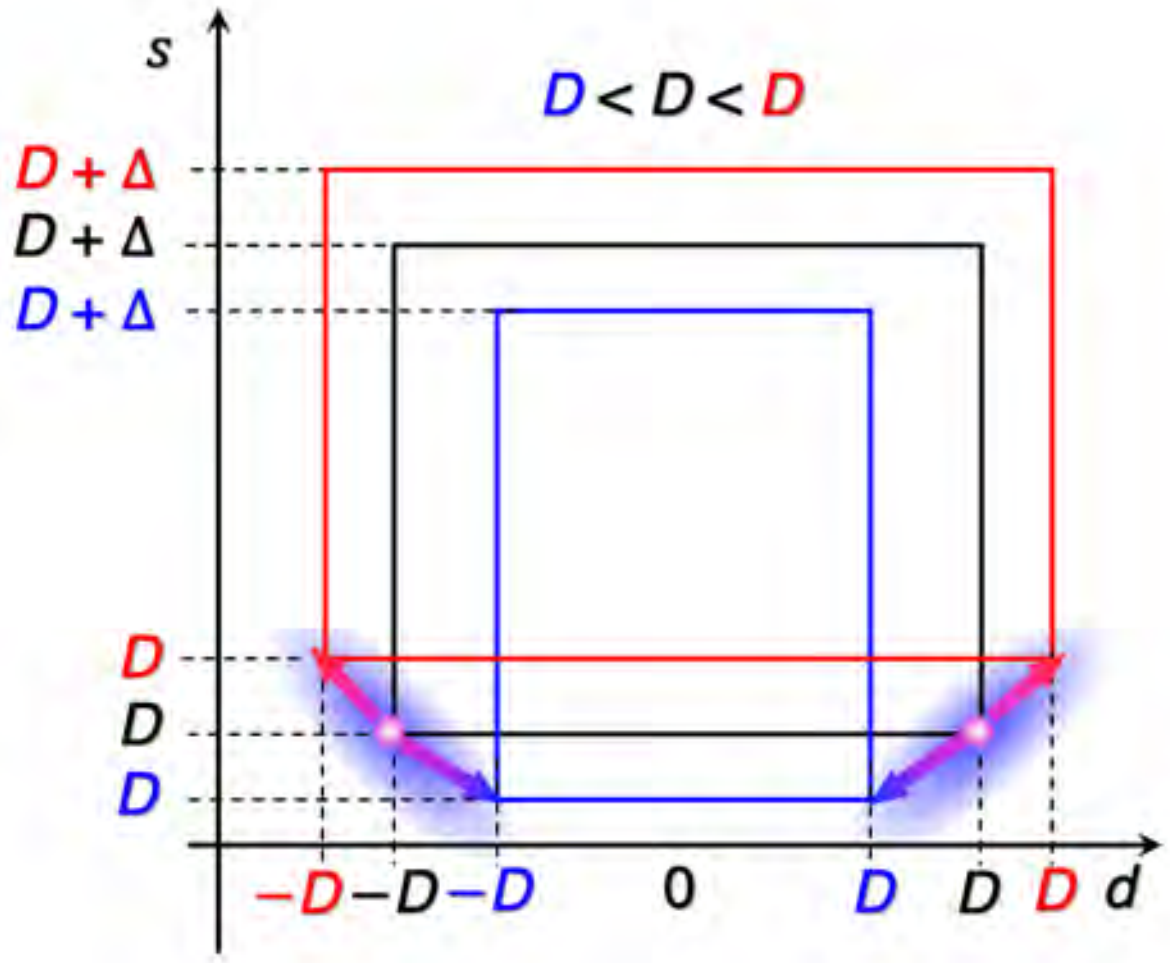

**Supplementary Figure 57 | Meaning of the phase space in $d = r_1 - r_2$ and $s = r_1 + r_2$ coordinates and the constraint Δ.**

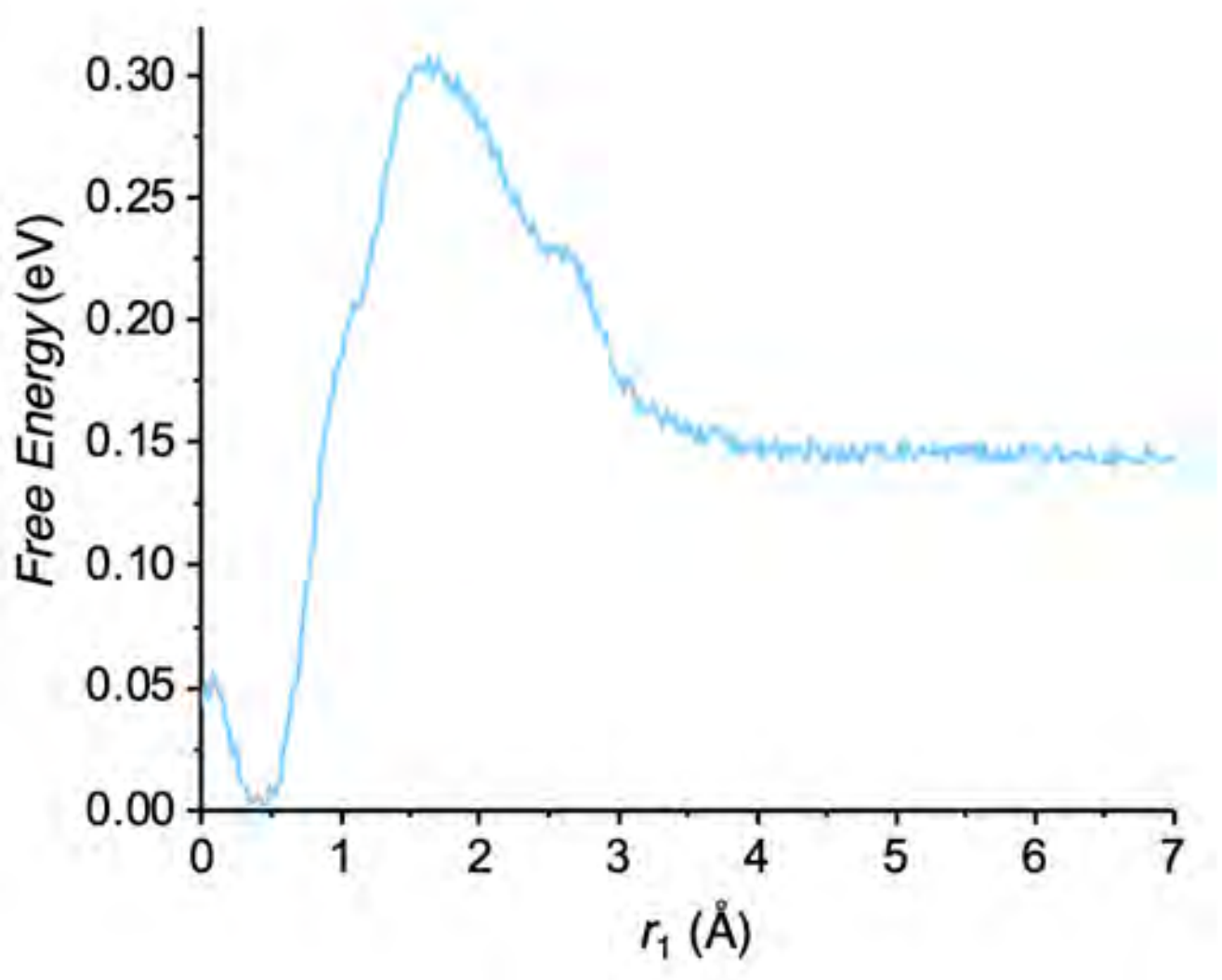

**Supplementary Figure 58 | Free energy profile as a function of $r_1$ – distance between the $Li^+$ and the center of mass of a cage.**

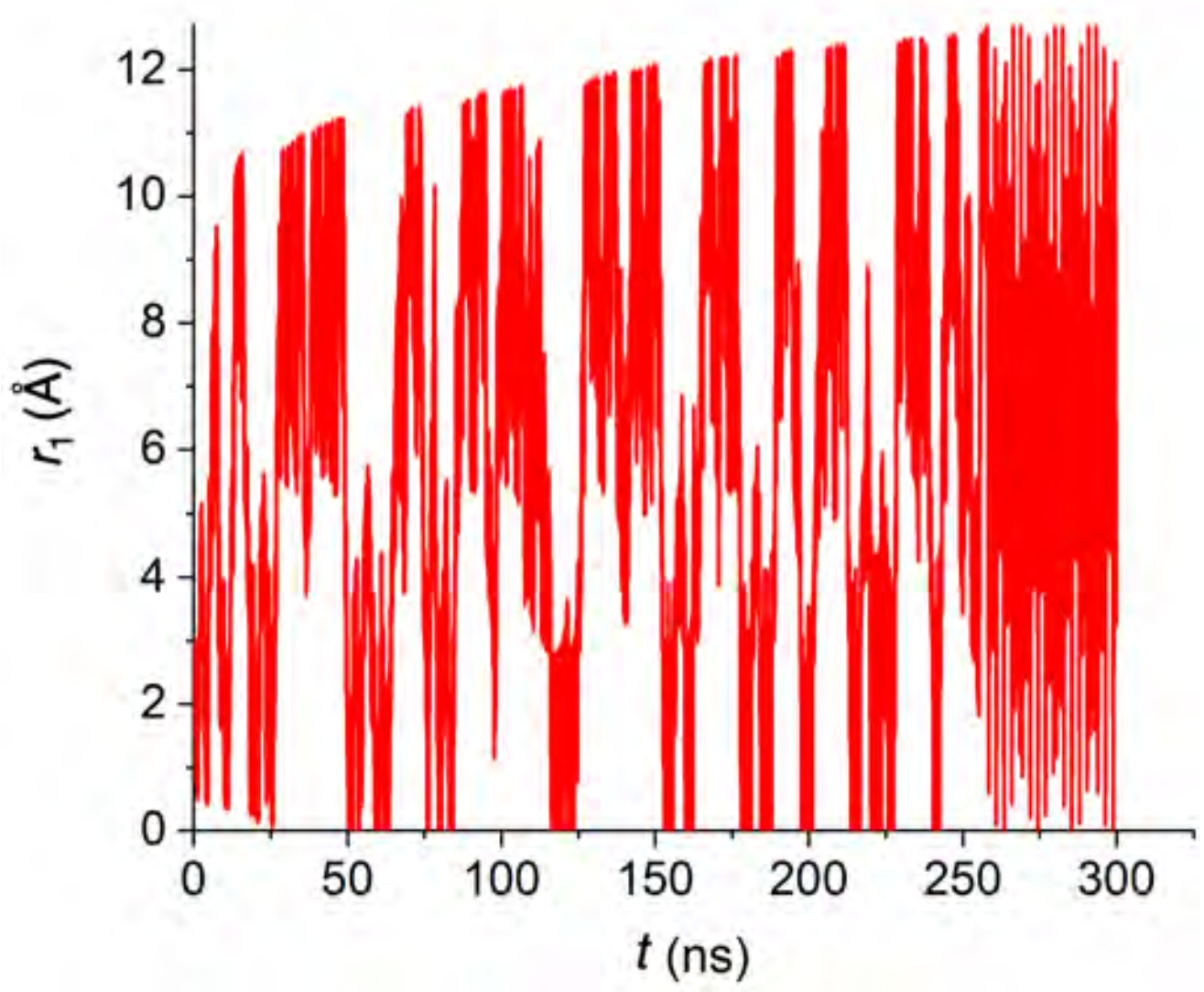

**Supplementary Figure 59 | Time evolution of $r_1$ in the metadynamics protocol.**

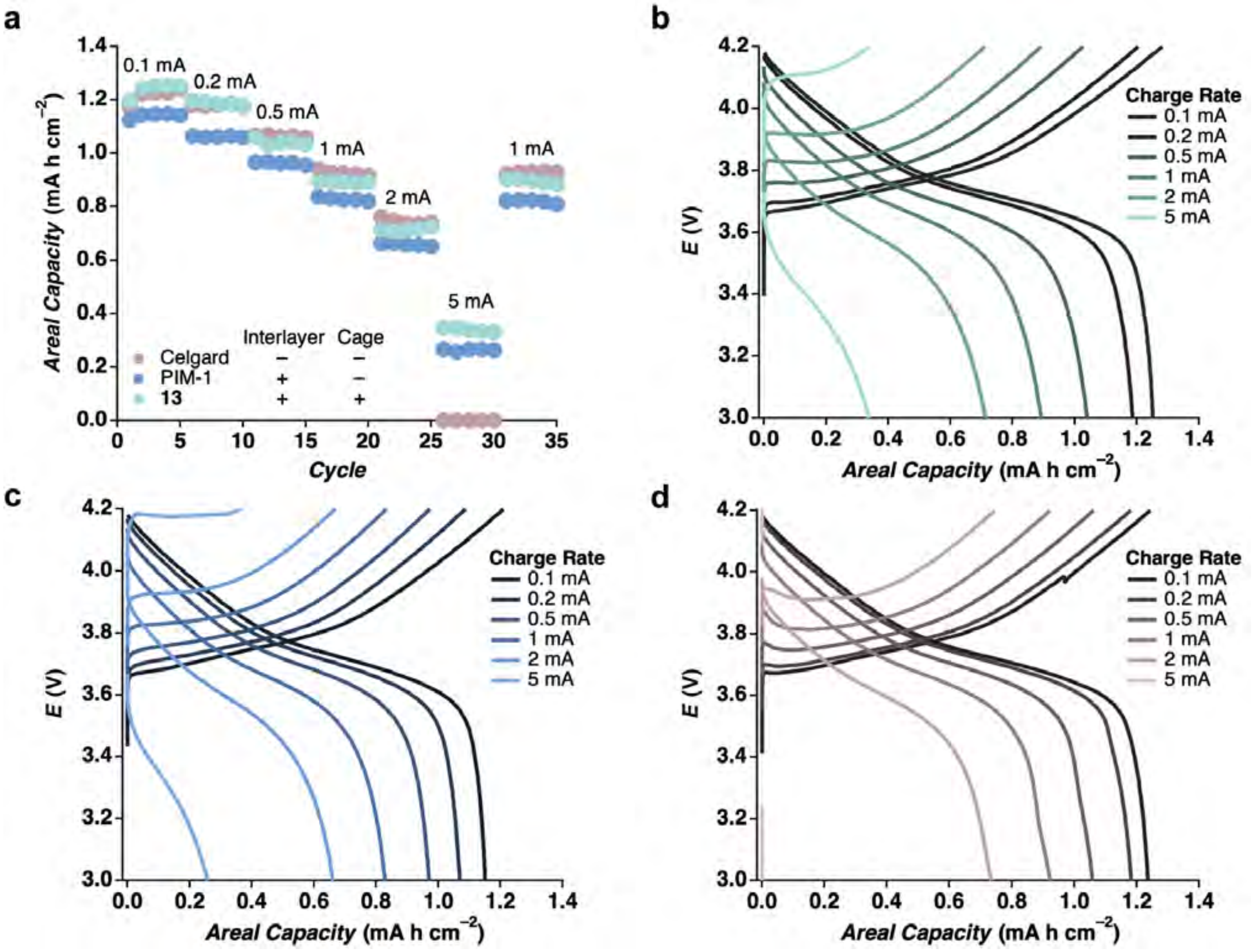

**Supplementary Figure 60 | Full cell rate performance**. **a**, Discharge capacity and **b–c** charge-discharge curves for the cycling of PIM **13**, PIM-1, and Celgard, respectively, at a variety of charge rates.

**Supplementary References**

1. Fritsch, D., Bengtson, G., Carta, M. & McKeown, N. B. Synthesis and gas permeation properties of Spirobischromane-Based polymers of intrinsic microporosity. *Macromol. Chem. Phys.* **212**, 1137–1146 (2011).
2. SAINT Software for CCD Diffractometers, Bruker AXS Inc., Madison, WI (2014).
3. Sheldrick, G. M. SADABS, Bruker Analytical X-ray Systems, Inc., Madison, WI (2000).
4. Sheldrick, G. M. SHELXT–Integrated space-group and crystal-structure determination. *Acta Crystallographica Section A: Foundations and Advances* **71**, 3–8 (2015).
5. Sheldrick, G. M. A short history of SHELX. *Acta Crystallographica Section A: Foundations of Crystallography* **64**, 112–122 (2008).
6. Hexemer, A., Bras, W., Glossinger, J., Schaible, E.,Gann, E., Kirian, R., MacDowell, A., Church, M., Rude, B. & Padmore, H. A SAXS/WAXS/GISAXS Beamline with Multilayer Monochromator. *J. Phys. Conf. Ser.* **247**, 012007 (2010).
7. Ilvasky, J. Nika: software for two-dimensional data reduction. *J. Appl. Crystallogr.* **45**, 324–328 (2012).
8. McDermott, A. G., Larsen, G. S., Budd, P. M., Colina, C. M. & Runt, J. Structural Characterization of a Polymer of Intrinsic Microporosity: X-ray Scattering with Interpretation Enhanced by Molecular Dynamics Simulations. *Macromolecules* **44**, 14–16 (2011).
9. McKeown, N. B. The synthesis of polymers of intrinsic microporosity (PIMs). *Sci. China Chem.* **60**, 1023–1032 (2017).
10. McKeown, N. B. Polymers of intrinsic microporosity. *ISRN Materials Science* **2012**, 513986 (2012).

11. Low, Z.-X., Budd, P. M, McKeown, N. B. & Patterson, D. A. Gas permeation properties, physical aging, and its mitigation in high free volume glassy polymers. *Chem. Rev.* **118**, 5871–5911 (2018).

12. Ufimtsev, I. S. & Martinez, T. J. Quantum Chemistry on Graphical Processing Units. 3. Analytical Energy Gradients, Geometry Optimization, and First Principles Molecular Dynamics**.** *J. Chem. Theory Comput.* **5**, 2619 (2009).

13. Jensen, F. Atomic orbital basis sets. *Comput. Mol. Sci.* **3**, 273–295 (2013).

14. Grimme, S., Antony, J., Ehrlich, S. & Krieg, H. A consistent and accurate *ab initio* parametrization of density functional dispersion correction (DFT-D) for the 94 elements H-Pu. *J. Chem. Phys.* **132**, 154104 (2010).

15. Kästner, J., Carr, J. M., Keal, T. W., Thiel, W., Wander, A. & Sherwood, P. DL-FIND: An Open-Source Geometry Optimizer for Atomistic Simulations. *J. Phys. Chem. A* **113**, 11856 (2009).

16. Boys, S. F. & Bernardi, F. The Calculation of Small Molecular Interactions by the Differences of Separate Total Energies. Some Procedures with Reduced Errors. *Mol. Phys*. **19**, 552–566 (1970).

17. Ponnuchamy, V., Mossa, S. & Skarmoutsos, I. Solvent and Salt Effect on Lithium Ion Solvation and Contact Ion Pair Formation in Organic Carbonates: A Quantum Chemical Perspective. *J. Phys. Chem. C* **122**, 25930−25939 (2018).

18. Han, S. Structure and dynamics in the lithium solvation shell of nonaqueous electrolytes. *Sci. Rep*. **9**, 5555 (2019).

19. Shim, Y. Computer simulation study of the solvation of lithium ions in ternary mixed carbonate electrolytes: free energetics, dynamics, and ion transport. *Phys. Chem. Chem. Phys.* **20**, 28649–28657 (2018).

20. Fulfer, K. D. & Kuroda, D. G. A comparison of the solvation structure and dynamics of the lithium ion in linear organic carbonates with different alkyl chain lengths. *Phys. Chem. Chem. Phys.* **19**, 25140–25150 (2017).

21. Mills, H. & Jónsson, G. Quantum and thermal effects in $H_2$ dissociative adsorption: Evaluation of free energy barriers in multidimensional quantum systems. *Phys. Rev. Lett.* **72**, 1124 (1994).

22. Klamt, A. & Schüürmann, G. COSMO: a new approach to dielectric screening in solvents with explicit expressions for the screening energy and its gradient. *J. Chem. Soc., Perkin Trans.* **2**, 799–805 (1993).

23. Bondi, A. van der Waals Volumes and Radii. *J. Phys. Chem.* **68**, 441−451(1964).

24. Bryantsev, V. S., Diallo, M. S. & Goddard III, W. A. Calculation of Solvation Free Energies of Charged Solutes Using Mixed Cluster/Continuum Models. *J. Phys. Chem.* **112**, 9709 (2008).

25. Fawcett, W. R. Thermodynamic Parameters for the Solvation of Monatomic Ions in Water. *J. Phys. Chem. B* **103**, 11181–11185 (1999).

26. Kelly, C. P., Cramer, C. J. & Truhlar, D. G. Single-Ion Solvation Free Energies and the Normal Hydrogen Electrode Potential in Methanol, Acetonitrile, and Dimethyl Sulfoxide. *J. Phys. Chem. B* **111**, 408−422 (2007).

27. Plimpton, S. Fast Parallel Algorithms for Short-Range Molecular Dynamics. *J. Comp. Phys.* **117**, 1 (1995).

28. Nosé, S. A. A unified formulation of the constant temperature molecular dynamics methods. *J. Chem. Phys.* **81**, 511 (1984).

29. Hoover, W. G. Canonical dynamics: Equilibrium phase-space distributions. *Phys. Rev. A: At. Mol. Opt. Phys.* **31**, 1965 (1985).

30. Wang, J., Wolf, R. M., Caldwell, J. W., Kollman, P. A. & Case, D. A. Development and testing of a general amber force field. *J. Comput. Chem.* **25**, 1157 (2004).

31. Aqvist, J. Ion–Water Interaction Potentials Derived from Free Energy Perturbation Simulations. *J. Phys. Chem.* **94**, 8021–8024 (1990).

32. Laio, A. & Gervasio, F. L. Metadynamics - A Method to Stimulate Rare Events and Reconstruct the Free Energy in Biophysics. *Rep. Prog. Phys*. **71**, 126601 (2008).

33. Barducci, A., Bonomi, M. & Parrinello, M. Metadynamics. *Wiley Inter. Rev.: Comp. Mol. Sci*. **1**, 826–843 (2011).